\documentclass[aps,prd,twocolumn,superscriptaddress,preprintnumbers,floatfix,longbibliography,nofootinbib]{revtex4-1}

\usepackage{dcolumn}
\usepackage{bm}

\usepackage{graphicx}
\usepackage{amsmath}
\usepackage{multirow}
\usepackage{tensor}
\usepackage[caption=false]{subfig}
\usepackage{siunitx}
\usepackage{placeins}
\usepackage{braket}
\usepackage{color}
\usepackage{float}
\usepackage{standalone}
\usepackage{dcolumn}
\usepackage{bm}
\usepackage{comment}
\usepackage{microtype}
\usepackage{etoolbox}
\usepackage{amssymb}
\usepackage{mathrsfs}
\usepackage{accents}
\usepackage[normalem]{ulem}
\usepackage[dvipsnames]{xcolor}
\usepackage[colorlinks,urlcolor=NavyBlue,citecolor=NavyBlue,linkcolor=NavyBlue,pdfusetitle]{hyperref}
\usepackage{gensymb}
\AtBeginDocument{\usepackage{booktabs}}
\graphicspath{{./figs/}}
\usepackage[colorlinks,urlcolor=NavyBlue,citecolor=NavyBlue,linkcolor=NavyBlue,pdfusetitle]{hyperref}
\usepackage{diagbox}
\usepackage{enumitem}
\DeclareMathAlphabet{\mathpzc}{OT1}{pzc}{m}{it}

\newcommand{\etal}{\textit{et al.\ }}

\newlist{todolist}{itemize}{2}
\setlist[todolist]{label=$\square$}

\begin{document}

\title{Using rational filters to uncover the first ringdown overtone in GW150914}

\newcommand{\Cornell}{\affiliation{Cornell Center for Astrophysics
    and Planetary Science, Cornell University, Ithaca, New York 14853, USA}}
\newcommand\CornellPhys{\affiliation{Department of Physics, Cornell
    University, Ithaca, New York 14853, USA}}
\newcommand\Caltech{\affiliation{TAPIR 350-17, California Institute of
    Technology, 1200 E California Boulevard, Pasadena, CA 91125, USA}}
\newcommand{\AEI}{\affiliation{Max Planck Institute for Gravitational Physics
    (Albert Einstein Institute), Am M\"uhlenberg 1, Potsdam 14476, Germany}} %
\newcommand{\UMassD}{\affiliation{Department of Mathematics,
    Center for Scientific Computing and Visualization Research,
    University of Massachusetts, Dartmouth, MA 02747, USA}}
\newcommand\Olemiss{\affiliation{Department of Physics and Astronomy,
    The University of Mississippi, University, MS 38677, USA}}
\newcommand{\Bham}{\affiliation{School of Physics and Astronomy and Institute
    for Gravitational Wave Astronomy, University of Birmingham, Birmingham, B15
    2TT, UK}}
\newcommand{\ANU}{\affiliation{OzGrav-ANU, Centre for Gravitational Astrophysics, College of Science,
The Australian National University, ACT 2601, Australia}}

\author{Sizheng Ma}
\email{sma@caltech.edu}
\Caltech

\author{Ling Sun}
\ANU

\author{Yanbei Chen}
\Caltech

\hypersetup{pdfauthor={Ma et al.}}

\date{\today}

\begin{abstract}
There have been debates in the literature about the existence of the first overtone in the ringdown of GW150914. We develop a novel Bayesian framework to reanalyze the data of this event, by incorporating a new technique, the ``rational filter'' that can clean particular modes from the ringdown signal. We examine the existence of the first overtone in GW150914 from multiple novel perspectives. First, we confirm that the estimates of the remnant black hole mass and spin are more consistent with those obtained from the full inspiral-merger-ringdown signal when including the first overtone at an early stage of the ringdown (right after the inferred signal peak); such improvement fades away at later times. Second, we formulate a new way to compare the ringdown models with and without the first overtone by calculating the Bayes factor at different times during the ringdown. We obtain a Bayes factor of 600 at the time when the signal amplitude reaches its peak. The Bayes factor decreases sharply when moving away from the peak time and eventually oscillates around a small value when the overtone signal is expected to have decayed.
Third, we clean the fundamental mode from the ringdown of GW150914 and estimate the amplitudes of the modes using the filtered data with Markov chain Monte Carlo (MCMC). The inferred amplitude of the fundamental mode is $\sim 0$ whereas the amplitude of the first overtone remains almost unchanged, implying that the filtered data is consistent with a first-overtone-only template. Similarly, if we remove the first overtone from the GW150914 data, the filtered data are consistent with a fundamental-mode-only template. Finally, after removing the fundamental mode, we use MCMC to infer the remnant black hole mass and spin from the first overtone alone. We find the posteriors are still informative and consistent with those inferred from the fundamental mode. The conclusions are also verified through simulations in Gaussian noise using a GW150914-like numerical relativity waveform. 
\end{abstract}

\maketitle


\section{Introduction}
\label{sec:introduction}
The ringdown stage of a gravitational wave (GW) emitted by a binary black hole (BBH) corresponds to the oscillation of the remnant BH, which encodes rich information about the system. At the linear order, a ringdown waveform is given by a superposition of a set of complex-valued quasinormal modes (QNMs) \cite{Kokkotas:1999bd,Nollert_1999,Cardoso:2016ryw,Berti:2009kk}, labeled by two angular numbers $(l,m)$ and one overtone index $n$. Within the general theory of relativity, they are fully determined by the mass and spin of the corresponding BH due to the no-hair theorem \cite{penrose2002golden,Chrusciel:2012jk,PhysRevLett.26.331,PhysRev.164.1776}. Thus measuring the frequency and decay rate of a QNM from a ringdown signal can lead to the estimates of the mass and spin of the remnant BH \cite{PhysRevD.40.3194}. Alternatively, if multiple modes are detected at the same time, we can use them to test the no-hair theorem \cite{Dreyer:2003bv,Berti:2005ys,Berti:2007zu}. This method is known as \textit{BH spectroscopy} \cite{Gossan:2011ha,Caudill:2011kv,Meidam:2014jpa,Bhagwat:2016ntk,Berti:2016lat,Baibhav:2017jhs,Maselli:2017kvl,Yang:2017zxs,DaSilvaCosta:2017njq,Baibhav:2018rfk,Carullo:2018sfu,Brito:2018rfr,Nakano:2018vay,Cabero:2019zyt,Bhagwat:2019bwv,Ota:2019bzl,Bustillo:2020buq,JimenezForteza:2020cve,Isi:2021iql,Finch:2021qph,Isi:2022mhy,Finch:2022ynt,Cotesta:2022pci,TheLIGOScientific:2016src,Isi:2019aib,Capano:2021etf,Capano:2022zqm}.

An important topic of BH spectroscopy is to understand which QNMs are present in the ringdown of a numerical relativity (NR) waveform \cite{Buonanno:2006ui,Berti:2007fi,Berti:2007dg,Kamaretsos:2011um,London:2014cma,Thrane:2017lqn,Giesler:2019uxc,Cook:2020otn,Dhani:2020nik,Dhani:2021vac,Forteza:2021wfq,Finch:2021iip,Li:2021wgz,MaganaZertuche:2021syq,Ma:2021znq,Mitman:2022qdl,Cheung:2022rbm,Lagos:2022otp} and when they start \cite{Bhagwat:2017tkm,Bhagwat:2019dtm,Okounkova:2020vwu}. To address these questions, a common method is to fit the waveform after the merger using a ringdown template that consists of a group of QNMs, and explore when the mismatch between the two can be minimized by varying the QNMs and the fitting start time. In particular, Giesler \etal \cite{Giesler:2019uxc} demonstrated that the ringdown of a GW150914-like NR waveform \cite{Lovelace:2016uwp} starts as early as when the strain amplitude reaches its peak, if seven overtones are included. Motivated by this result, Isi \etal \cite{Isi:2019aib} (and also \cite{Isi:2022mhy}) extended the initial ringdown analyses \cite{TheLIGOScientific:2016src,Carullo:2019flw} of GW150914 \cite{LIGOScientific:2016aoc} and explored earlier start times for fitting. A significance of $3.6\sigma$ was found for the existence of the first overtone.
However, the conclusion was later challenged by Cotesta \etal \cite{Cotesta:2022pci}, who argued that the early (ringdown) signal could be noise dominated, and thus the existence of the first overtone might not be reliable. The claim by Cotesta \etal \cite{Cotesta:2022pci} was then disputed by a subsequent response by Isi \etal \cite{Isi:2022mhy} who found the impact of noise was not reproducible; Finch and Moore \cite{Finch:2022ynt} also showed that the noise fluctuations might be overestimated. On the other hand, Bustillo \etal \cite{Bustillo:2020buq} and Finch \etal \cite{Finch:2022ynt} tackled the problem via different approaches; tentative evidence was found. 

The lack of a definitive conclusion over the ringdown modes of GW150914 leaves unresolved issues for BH spectroscopy, posting questions for the ringdown analysis in the upcoming LIGO-Virgo-KAGRA fourth observing run (O4). Here we propose a new framework for BH spectroscopy and revisit the issues from a different perspective. 
Recently, we proposed a new methodology, the so-called ``QNM filters'' \cite{Ma:2022wpv}. The method includes the use of two filter classes: a rational filter and a full filter. The rational filter is constructed via a QNM frequency, whereas the full filter is constructed from the BH transmissivity, based on the \textit{hybrid approach} \cite{Nichols:2010qi,Nichols:2011ih,Ma:2022xmp}. They were originally designed to remove QNM(s) from the ringdown when studying NR waveforms. After filtering out some dominant modes, we were able to show the existence of subdominant effects confidently, such as the mixing of modes, retrograde modes, and also the second-order QNMs \cite{Mitman:2022qdl,Cheung:2022rbm,Lagos:2022otp}. In our companion paper \cite{Ma_prl}, a novel framework for BH spectroscopy is outlined: by incorporating the rational filter into Bayesian inference, a new scheme is developed to analyze the ringdown of real GW events, independent of the usual Markov chain Monte Carlo (MCMC) method. In this paper, we extend the discussions therein and provide full details to demonstrate the efficacy and efficiency of this framework. In particular, we demonstrate the existence of the first overtone in the ringdown of GW150914 with detailed evidence.

This paper is organized as follows. In Sec.~\ref{sec:ration_filter}, we introduce the properties of the rational filter. In Sec.~\ref{sec:filter_bayes}, we use the filter to construct a two-dimensional (2D) ringdown likelihood function in the time domain that depends only on the mass and spin of the remnant BH (independent of mode amplitudes and phases). Based on the likelihood function, we define a new method to compute model evidence and Bayes factor for QNM(s). Detailed case studies of a simulated signal using a NR waveform and the real event GW150914 are given in Secs.~\ref{subsec:injection_likelihood} and \ref{subsec:GW150914_likelihood}, respectively. Next, in Sec.~\ref{sec:mixed_appraoch}, we carry out a mixed BH spectroscopy analysis by combining our new approach with the usual MCMC treatment. Again, the NR simulation (Sec.~\ref{sec:mixed_injection}) and GW150914 (Sec.~\ref{sec:mixed_gw150914}) are discussed as detailed examples. Finally, we summarize
the results in Sec.~\ref{sec:conclusion}.

Throughout this paper, we use geometric units with
$G = c = 1$. We always use the notation $\omega_{lmn}=2\pi f_{lmn}-i/\tau_{lmn}$ to refer to the $(l,m,n)$ QNM, with $2\pi f_{lmn}$ and $-1/\tau_{lmn}$ being its real and imaginary parts. All of our analyses are in the detector frame.









\section{The rational filter}
\label{sec:ration_filter}
To start with, let us consider two complex GW harmonics $h_{l,\pm m}(t)$. Below, we always assume $m>0$; thus $h_{l,m}$ ($h_{l,-m}$) represents the harmonic component that is emitted towards the north (south) with respect to the system. Within the ringdown regime, their time evolution reads
\begin{subequations}
\label{eq:hlm_ringdown}
    \begin{align}
    &h_{lm}(t)=\sum_{n}A_{lmn}e^{-i\omega_{lmn}(t-t_0)+i\phi_{lmn}},\\
    &h_{l,-m}(t)=\sum_{n}A_{lmn}^{\prime}e^{i\omega_{lmn}^*(t-t_0)+i\phi^\prime_{lmn}},
    \end{align}
\end{subequations}
where $n$ stands for the overtone index, and $A_{lmn}$'s and $\phi_{lmn}$'s are the amplitudes and phases of the QNMs, respectively. Note that $\omega_{lmn}$ on the right-hand side of Eq.~\eqref{eq:hlm_ringdown} refers to prograde modes, and we always neglect the contribution of retrograde modes in the rest of this paper. 

As discussed in Ref.~\cite{Ma:2022wpv}, to clean a mode $\omega_{lmn}$ from $h_{lm}(t)$, we can apply a rational filter:
\begin{align}
    \frac{\omega-\omega_{lmn}}{\omega-\omega_{lmn}^*}. \notag
\end{align}
Similarly, we need to apply another filter:
\begin{align}
    \frac{\omega+\omega_{lmn}^*}{\omega+\omega_{lmn}}, \notag
\end{align}
to eliminate the same mode $\omega_{lmn}$ from $h_{l,- m}(t)$. For an actual GW event, its time-domain real strain $h_t$ consists of both the complex harmonics $h_{l,\pm m}(t)$. In consequence, the final form of the filter $\mathcal{F}_{lmn}$ that is associated with the QNM $\omega_{lmn}$ is given by
\begin{align}
    &\mathcal{F}_{lmn}(f;M_f,\chi_f)=\frac{\omega-\omega_{lmn}}{\omega-\omega_{lmn}^*}\frac{\omega+\omega_{lmn}^*}{\omega+\omega_{lmn}} \notag \\
    &=\frac{f-f_{lmn}+\frac{i}{2\pi\tau_{lmn}}}{f- f_{lmn}-\frac{i}{2\pi\tau_{lmn}}}\times\frac{f+f_{lmn}+\frac{i}{2\pi\tau_{lmn}}}{f+ f_{lmn}-\frac{i}{2\pi\tau_{lmn}}}, \label{eq:rational_filter}
\end{align}
where $\omega=2\pi f$; $f_{lmn}$ and $\tau_{lmn}$ corresponds to the real and imaginary parts of $\omega_{lmn}$:
\begin{align}
    \omega_{lmn}=2\pi f_{lmn}-\frac{i}{\tau_{lmn}}.
\end{align}
According to the no-hair theorem \cite{penrose2002golden,Chrusciel:2012jk,PhysRevLett.26.331,PhysRev.164.1776}, $f_{lmn}$ and $\tau_{lmn}$ are fully determined by the mass $M_f$ and spin $\chi_f$ of the remnant BH. We obtain $f_{lmn}$ and $\tau_{lmn}$ using the Python package $\textsf{qnm}$ \cite{Stein:2019mop}. To apply the filter $\mathcal{F}_{lmn}$ to real GW data, we first transform the time-domain data $d_t$ to the frequency domain via fast Fourier transform (FFT)\footnote{More technical details are provided in Sec.~\ref{subsec:injection_likelihood}.}
\begin{align}
    &\tilde{d}_f=\int d_t e^{2\pi i f t}dt.
\end{align}
Note that the length of $d_t$ needs to be at least comparable to the entire inspiral-merger-ringdown (IMR) signal to avoid spectral leakage. Then the filtered data read
\begin{align}
    d^F_t=\int df \tilde{d}_f\mathcal{F}_{lmn}(f)e^{-i2\pi f t}.
\end{align}
In practice, multiple filters could be applied simultaneously via a total filter
\begin{align}
  \mathcal{F}_{\rm tot}=\prod_{lmn} \mathcal{F}_{lmn}. \label{eq:total_filter}
\end{align}
Since each filter $\mathcal{F}_{lmn}(f)$ satisfies 
\begin{align}
    \mathcal{F}_{lmn}(-f)=\left[\mathcal{F}_{lmn}(f)\right]^*,
\end{align}
the filtered data $d^F_t$ is still real-valued.

Because we apply the filter to the entire IMR signal, different portions of the signal have distinct responses. As discussed in Ref.~\cite{Ma:2022wpv}, the early low-frequency inspiral signal is shifted backward in time by the filter, which does not impact the ringdown analysis. Here we continue the discussion therein and investigate the impact of the filter on two other aspects. In Sec.~\ref{subsec:impact_irre_qnm}, we study the effect of $\mathcal{F}_{lmn}$ on a different QNM $\omega_{l^\prime m^\prime n^\prime}$, namely $(l\neq l^\prime\, {\rm or}\, m\neq m^\prime\, {\rm or}\,n\neq n^\prime)$. Then in Sec.~\ref{subsec:impact_detector_noise}, the impact of the filter on detector noise is discussed.

\begin{figure}[htb]
        \includegraphics[width=\columnwidth,clip=true]{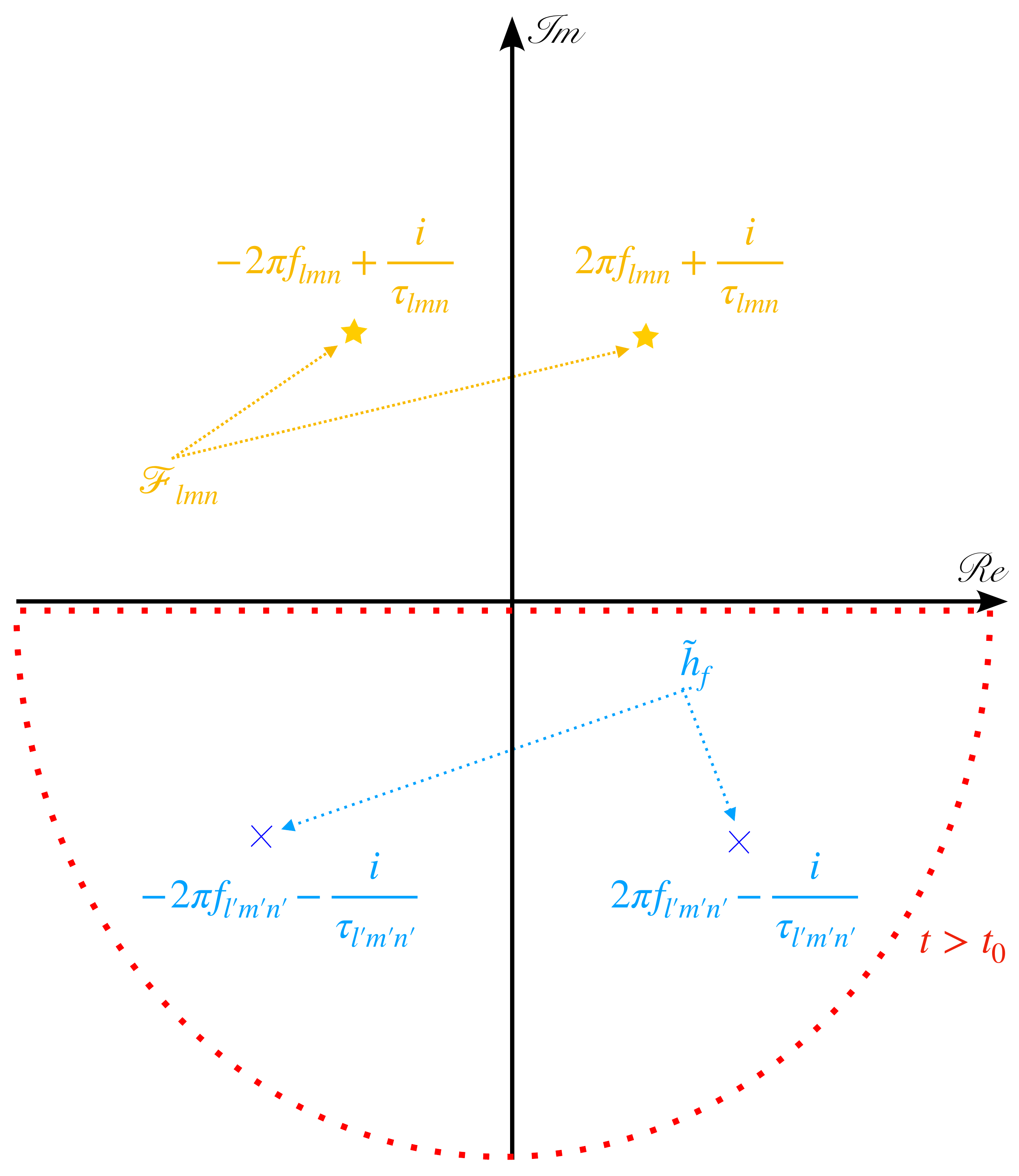}
  \caption{Poles of the filtered waveform [Eq.~\eqref{eq:filtered_h_toy_model}] on the complex plane. Two poles are in the lower half-plane (blue crosses), contributed by the original waveform in Eq.~\eqref{eq:toy_model_original_model}. The other two (yellow stars) are in the upper half-plane, coming from the filter $\mathcal{F}_{lmn}$. The red dashed curve corresponds to the time regime of $t>t_0$. Before $t_0$, the time-domain signal is contributed by the two ring-up modes $f=\pm f_{lmn}+i/(2\pi \tau_{lmn})$ outside the closed region. After $t_0$, the two ringdown modes $f=\pm f_{l^\prime m^\prime n^\prime}-i/(2\pi \tau_{l^\prime m^\prime n^\prime})$ contribute.}
 \label{fig:filter_poles}
\end{figure}

\subsection{Impact on a different QNM}
\label{subsec:impact_irre_qnm}
To investigate the impact of $\mathcal{F}_{lmn}$ on $\omega_{l^\prime m^\prime n^\prime}$, let us consider a toy model for the excitation of $\omega_{l^\prime m^\prime n^\prime}$
\begin{align}
    h_t=&e^{-(t-t_0)/\tau_{l^\prime m^\prime n^\prime}}\notag \\
    &\times\cos \left[2\pi f_{l^\prime m^\prime n^\prime}(t-t_0)+\phi_{l^\prime m^\prime n^\prime}\right]\Theta(t-t_0), \label{eq:filter_other_qnm_formula}
\end{align}
where $\Theta(t-t_0)$ is the Heaviside step function, meaning that the QNM is excited at $t_0$. The Fourier transformation of $h_t$ reads 
\begin{align}
    \tilde{h}_f=&\frac{i}{2}e^{i\omega t_0}\left[\frac{e^{i\phi_{l^\prime m^\prime n^\prime}}}{2\pi f-2\pi f_{l^\prime m^\prime n^\prime}+i/\tau_{l^\prime m^\prime n^\prime}}\right. \notag \\
    &+\left.\frac{e^{-i\phi_{l^\prime m^\prime n^\prime}}}{2\pi f+2\pi f_{l^\prime m^\prime n^\prime}+i/\tau_{l^\prime m^\prime n^\prime}}\right]. \label{eq:toy_model_original_model}
\end{align}
The two poles $f=\pm f_{l^\prime m^\prime n^\prime}-i/(2\pi \tau_{l^\prime m^\prime n^\prime})$ of $\tilde{h}_f$ are plotted in Fig.~\ref{fig:filter_poles}. Both of them lie in the lower half of the complex plane, indicating the fact that there is no $\omega_{l^\prime m^\prime n^\prime}$ signal before $t=t_0$. After applying the filter $\mathcal{F}_{lmn}$, two new poles $f=\pm f_{lmn}+i/(2\pi \tau_{lmn})$ appear in the upper half of the plane. This implies that the $\omega_{lmn}$ component of the filtered waveform exists before $t_0$. On the other hand, the two original poles in the lower plane remain unchanged, and the start time of the $\omega_{l^\prime m^\prime n^\prime}$ component remains at $t_0$. This is different from the situation where the early inspiral signal is shifted to an earlier time \cite{Ma:2022wpv}. However, the amplitude and phase of the $\omega_{l^\prime m^\prime n^\prime}$ component in the filtered waveform are changed. We can calculate the changes quantitatively by computing the following integral 
\begin{align}
    h^F_t=\int df \tilde{h}_f\mathcal{F}_{lmn}(f)e^{-i2\pi f t}, \label{eq:filtered_h_toy_model}
\end{align}
and obtain
\begin{align}
    &h^F_t=B_{lmn}^{l^\prime m^\prime n^\prime} e^{-(t-t_0)/\tau_{l^\prime m^\prime n^\prime}}\notag \\
    &\times \cos \left[2\pi f_{l^\prime m^\prime n^\prime}(t-t_0)+\phi_{l^\prime m^\prime n^\prime}+\varphi_{lmn}^{l^\prime m^\prime n^\prime}\right], \quad t>t_0, \label{eq:filter_on_other_qnm}
\end{align}
where
\begin{align}
    B_{lmn}^{l^\prime m^\prime n^\prime} e^{i\varphi_{lmn}^{l^\prime m^\prime n^\prime}}\equiv\mathcal{F}_{lmn}(\omega_{l^\prime m^\prime n^\prime}).
\end{align}
It is straightforward to show that
\begin{align}
    B_{lmn}^{l^\prime m^\prime n^\prime}=B^{lmn}_{l^\prime m^\prime n^\prime}.
\end{align}
Eq.~\eqref{eq:filter_on_other_qnm} shows that the amplitude of the $\omega_{l^\prime m^\prime n^\prime}$ mode is reduced by a factor of $B_{lmn}^{l^\prime m^\prime n^\prime}$. As for a Kerr BH with $\chi_f=0.692$, we have $B_{220}^{221}=0.487$, meaning that the amplitude of the first overtone (fundamental mode) is reduced by a factor of two after applying the filter that cleans the fundamental mode (first overtone).
For completeness, we also provide the expression of $h^F_t$ when $t<t_0$,
\begin{align}
    &h^F_t=\frac{1}{\pi f_{lmn}\tau_{lmn}}e^{(t-t_0)/\tau_{lmn}}\Phi_{lmn}^{l^\prime m^\prime n^\prime}(t), \quad t<t_0, \label{eq:h_t_ana_ringup}
\end{align}
with
\begin{align}
    &\Phi_{lmn}^{l^\prime m^\prime n^\prime}(t)=-{\rm Im} \,\, e^{2\pi i f_{lmn}(t-t_0)} \notag \\
    &\times\left(\frac{e^{i\phi_{l^\prime m^\prime n^\prime}}}{1+\omega_{l^\prime m^\prime n^\prime}/\omega_{lmn}}+\frac{e^{-i\phi_{l^\prime m^\prime n^\prime}}}{1-\omega^*_{l^\prime m^\prime n^\prime}/\omega_{lmn}}\right).
\end{align}
The term $e^{(t-t_0)/\tau_{lmn}}$ in Eq.~\eqref{eq:h_t_ana_ringup} shows that $h^F_t$ is a ``ring-up'' component at $t<t_0$.

\begin{figure}[htb]
        \includegraphics[width=\columnwidth,clip=true]{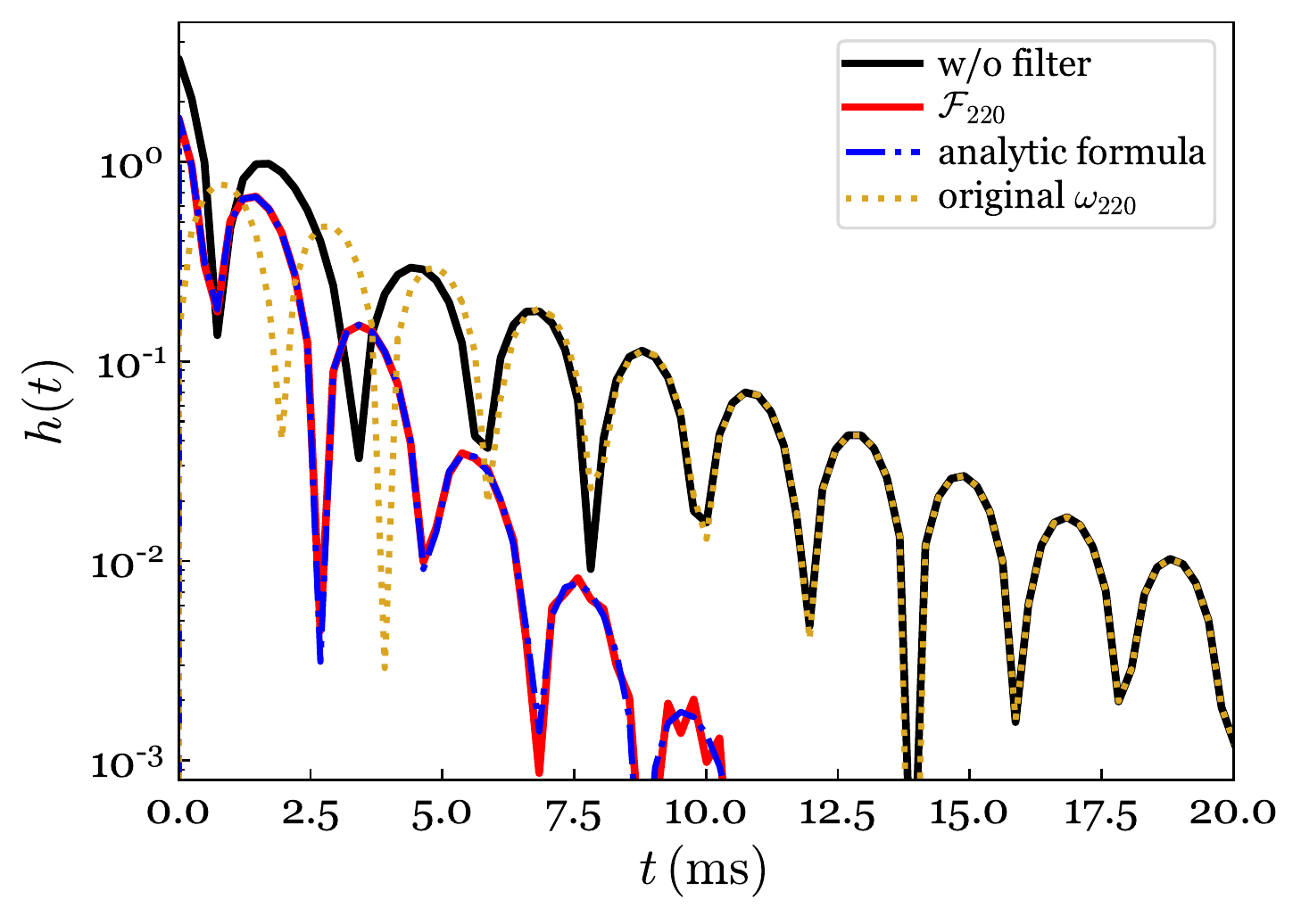}
  \caption{Validation of the filtered waveform in Eq.~\eqref{eq:filter_on_other_qnm}. We construct a ringdown waveform, consisting of two QNMs (black curve). The amplitudes and phases of these two modes are consistent with a GW150914-like NR waveform (with a remnant BH spin of 0.692). After applying the filter $\mathcal{F}_{220}$ to remove the fundamental mode (yellow dashed curve), the filtered waveform (red curve) agrees with the analytic formula in Eq.~\eqref{eq:filter_on_other_qnm} (blue curve). }
 \label{fig:filter_ana_reduction}
\end{figure}


To verify the analytical filtered waveform described by Eq.~\eqref{eq:filter_on_other_qnm}, we consider a remnant BH of a GW150914-like system, with a final spin of $0.692$. Then we construct a two-QNM waveform that consists of the fundamental mode $\omega_{220}$ and the first overtone $\omega_{221}$, given by
\begin{align}
    &h_t=\sum_{n^\prime=0}^{1}A_{n^\prime}e^{-|t|/\tau_{22n^\prime}}\cos \left[2\pi f_{22n^\prime}|t|+\phi_{22 n^\prime}\right],
\end{align}
where $A_0=0.96,A_1=4.15,\varphi_{220}=1.43,\varphi_{221}=-0.71$ are chosen to be consistent with those of the GW150914-like NR simulation \cite{Lovelace:2016uwp,Giesler:2019uxc}. Both modes are assumed to start at $t=0$. In Fig.~\ref{fig:filter_ana_reduction}, $h_t$ is shown as the black curve. Comparing the two-QNM $h_t$ to the fundamental-mode-only ($\omega_{220}$) evolution (yellow dashed curve), we can see that the two curves agree with each other at late times because the first overtone ($\omega_{221}$) has decayed. After applying the filter $\mathcal{F}_{220}$ to clean the fundamental mode, the filtered waveform, indicated by the red curve, is consistent with the analytic result from Eq.~\eqref{eq:filter_on_other_qnm} (blue dash-dotted curve). 
Here we verify Eq.~\eqref{eq:filter_on_other_qnm} using this two-QNM simplified waveform. In Sec.~\ref{sec:mixed_gw150914}, we show that Eq.~\eqref{eq:filter_on_other_qnm} also applies to real GW data (e.g. GW150914).

\begin{figure*}[htb]
        \subfloat[Bandlimited White Noise\label{fig:psds_filter_white_noise}]{\includegraphics[width=\columnwidth,clip=true]{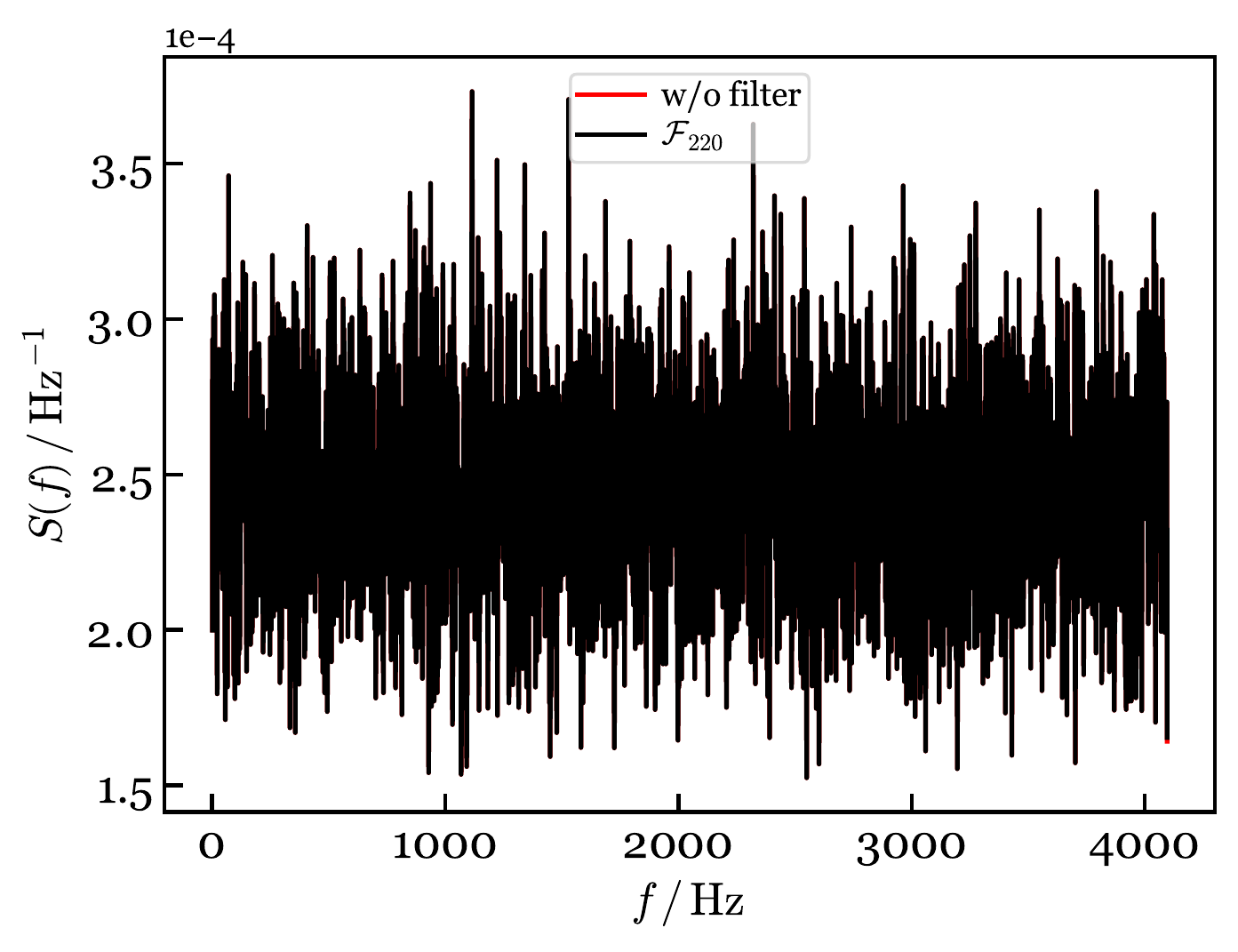}}
        \subfloat[Simulated Hanford\label{fig:psds_filter_sim_han}]{\includegraphics[width=\columnwidth,clip=true]{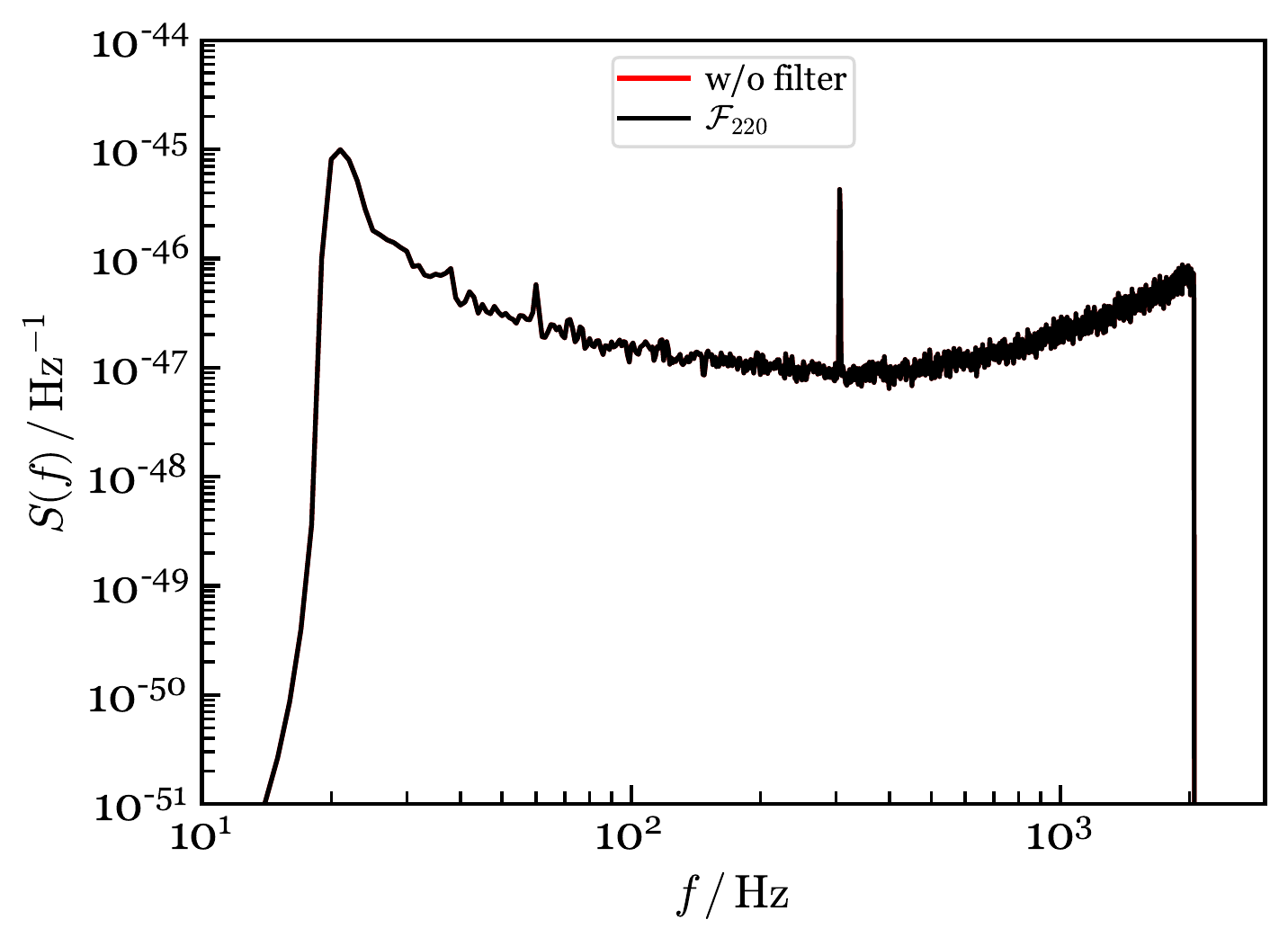}}\\
        \subfloat[GW150914 Hanford\label{fig:psds_filter_gw150914_han}]{\includegraphics[width=\columnwidth,clip=true]{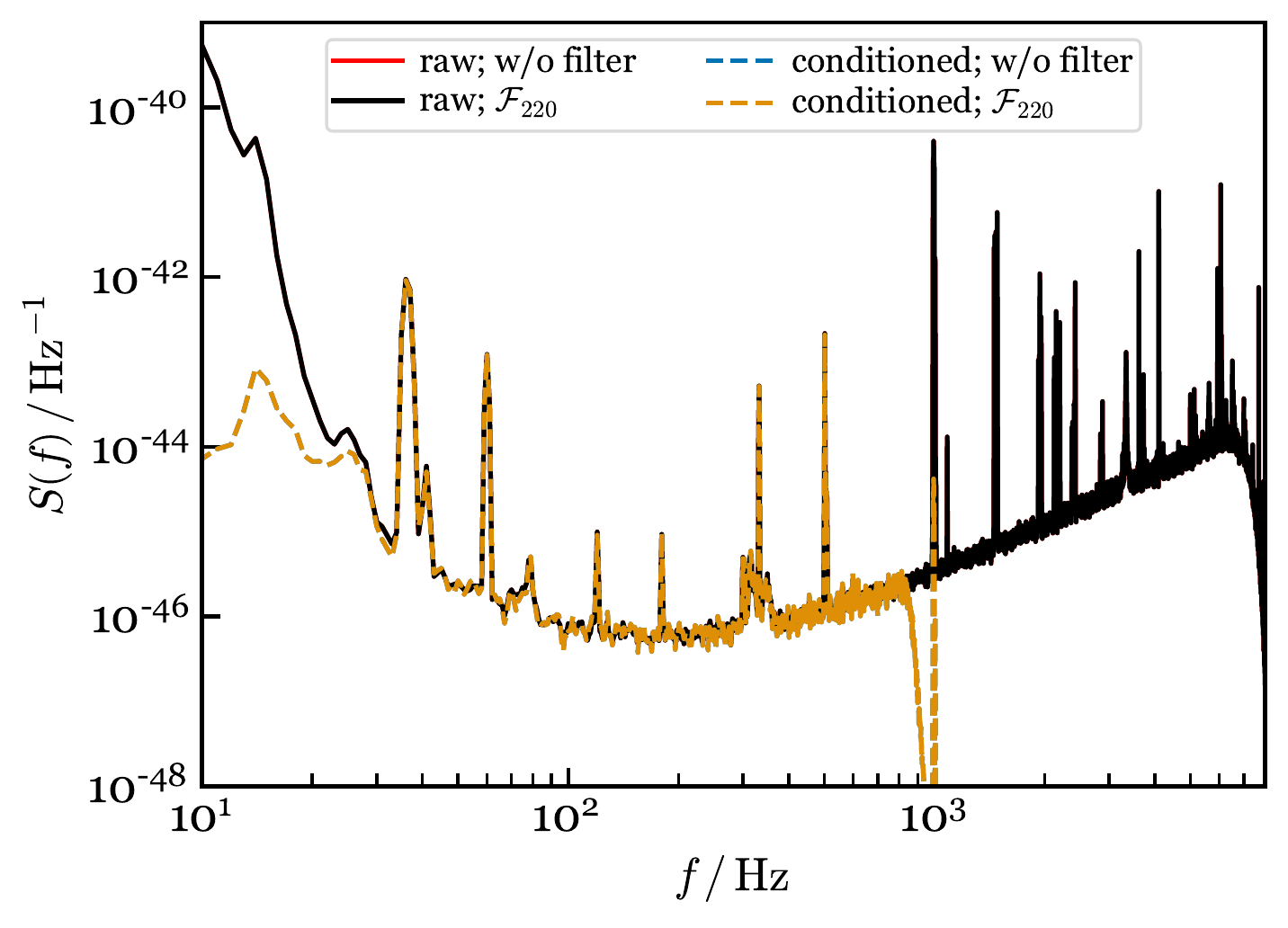}}
        \subfloat[GW150914 Livingston\label{fig:psds_filter_gw150914_liv}]{\includegraphics[width=\columnwidth,clip=true]{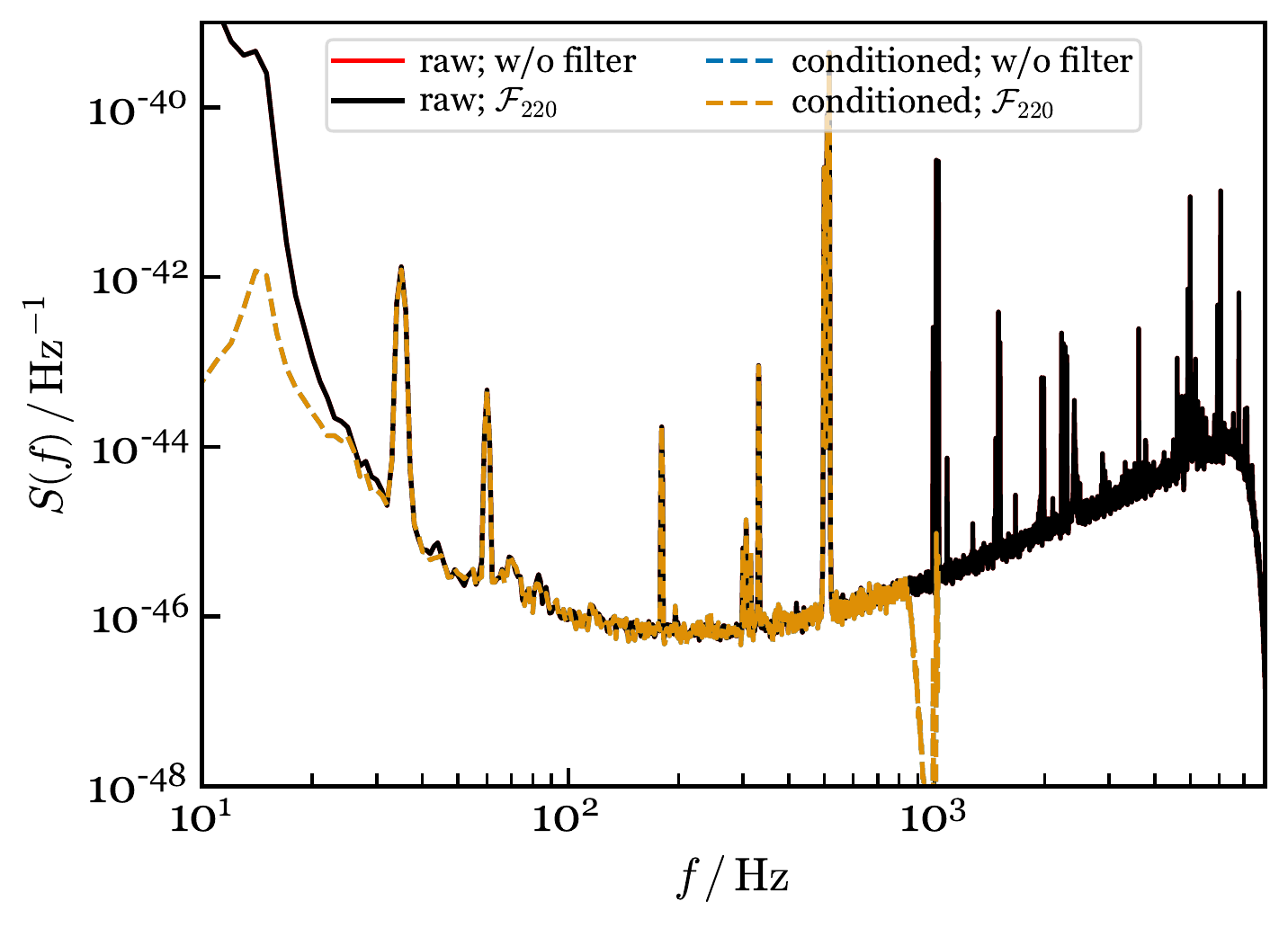}}
  \caption{Comparison of the one-sided PSDs of the filtered and unfiltered noise. The top panels show results in (a) band-limited white noise $\sim \mathcal{N}(0,1)$ and (b) simulated Hanford noise. The filtered and unfiltered PSDs of the raw data are shown in black and red, respectively. The bottom panels show results in (c) LIGO Hanford and (d) LIGO Livingston data around the event time of GW150914. The filtered (black) and unfiltered (red) PSDs are in perfect agreement. After conditioning the raw data, the filtered (dashed orange) and unfiltered (dashed blue) PSDs also overlap.}
 \label{fig:psds_filter}
\end{figure*}

\subsection{Impact on detector noise}
\label{subsec:impact_detector_noise}
For a noise series $n_t$, its covariance matrix is given by
\begin{align}
    C_{t,t+\tau}=E[n_tn_{t+\tau}]-E[n_t]E[n_{t+\tau}],
\end{align}
where $E$ denotes expectation. By assuming stationarity and  $E[n_t]=0$, $C_{t,t+\tau}$ takes a simple form
\begin{align}
    C_{t,t+\tau}=\rho(\tau), \label{eq:autocovariance}
\end{align}
where $\rho(\tau)$ is the autocovariance function (ACF). In the frequency domain, Eq.~\eqref{eq:autocovariance} becomes (known as the Wiener–Khinchin theorem)
\begin{align}
    E[\tilde{n}_f\tilde{n}_{f^\prime}^*]=\frac{1}{2}\delta(f-f^\prime)S(f), \label{eq:expectation_psd}
\end{align}
with $S(f)$ being the one-sided noise power spectral density (PSD)
\begin{align}
    &S(f)=2\int e^{2\pi i f\tau}\rho(\tau)d\tau. \label{eq:PSD_vs_rho}
\end{align}
Then we apply the rational filter $\mathcal{F}_{lmn}$ to $\tilde{n}_f$. Since the modulus of $\mathcal{F}_{lmn}$ is unity, we can write 
\begin{align}
    \mathcal{F}_{lmn}(f)=e^{i\delta_{lmn}(f)},
\end{align}
and the filtered noise writes
\begin{align}
    \tilde{n}_f^F=e^{i\delta_{lmn}(f)}\tilde{n}_f, \label{eq:noise_filter}
\end{align}
which still remains stationary. In addition, 
after plugging Eq.~\eqref{eq:noise_filter} into Eq.~\eqref{eq:expectation_psd}, we find that the one-sided PSD is not impacted by the filter.

To support our conclusion, we generate two noise series with a sampling rate of 8192 Hz and a total length of 4 s. One is band-limited white noise $\sim \mathcal{N}(0,1)$ and the other is simulated Hanford noise, generated by the Bilby library \cite{Ashton:2018jfp,Romero-Shaw:2020owr}. Next we use the Python package $\textsf{ringdown}$ \cite{ringdown_isi,Isi:2021iql} to estimate their one-sided PSDs with the Welch method \cite{1161901}. On the other hand, we apply a rational filter $\mathcal{F}_{220}$ to both noise series and obtain the time-domain filtered data. Then we repeat the Welch method and obtain the PSDs of the filtered data. The results are plotted in Figs.~\ref{fig:psds_filter_white_noise} and \ref{fig:psds_filter_sim_han} for comparison. We find the filtered PSDs are always identical to the unfiltered PSDs (the black curves completely overlap the red ones), no matter what the values of $M_f$ and $\chi_f$ are used to generate $\mathcal{F}_{220}$.

We then move on to two more realistic cases: LIGO Hanford and Livingston data \cite{TheLIGOScientific:2014jea,Abbott:2016xvh,TheLIGOScientific:2016agk,Harry:2010zz} around GW150914 \cite{Abbott:2016blz,LIGOScientific:2018mvr,GW_open_science_center}. We estimate the corresponding PSDs of 32-s raw time-series data with a sampling rate of 16384 Hz. Again, we apply the rational filter $\mathcal{F}_{220}$ to the frequency-domain data, convert them back to time series, and compute the filtered PSDs. The PSDs of the filtered and unfiltered data are shown as black and red curves in Figs.~\ref{fig:psds_filter_gw150914_han} and \ref{fig:psds_filter_gw150914_liv}, which fully agree with each other. Furthermore, we condition both the Hanford and Livingston raw data by (a) removing frequency components that are below 20 Hz using a high-pass filter and (b) downsampling the data to 2048 Hz. The filtered and unfiltered PSDs of the conditioned data are plotted as dashed orange and blue curves, respectively, in Figs.~\ref{fig:psds_filter_gw150914_han} and \ref{fig:psds_filter_gw150914_liv}. We see that they are also in perfect agreement.



\section{Incorporating the filter into Bayesian inference and searching for the first overtone}
\label{sec:filter_bayes}

After discussing the properties of the rational filter in Sec.~\ref{sec:ration_filter}, we now outline our novel Bayesian framework based on the rational filter. The core step is to construct a new likelihood function that depends only on the detector-frame remnant mass $M_f$ and spin $\chi_f$; and this leads to an alternative approach to computing model evidence and Bayes factor. Below, we elaborate on the details of the framework in Sec.~\ref{subsec:new_likelihood}. Then we use two examples to demonstrate the application of our method. In Sec.~\ref{subsec:injection_likelihood}, we first consider a case where a GW150914-like NR waveform \cite{Lovelace:2016uwp} is injected into Gaussian noise. Next in Sec.~\ref{subsec:GW150914_likelihood}, we apply the method to the real GW150914 event.

\subsection{A new likelihood function}
\label{subsec:new_likelihood}
Suppose we have a BBH GW signal in observational time-series data $d_t$. After applying rational filters with enough QNMs to $d_t$\footnote{Again, $d_t$ needs to be long enough to avoid spectral leakage.} to remove all the QNMs, we expect that the filtered data $d^{F}_t$ should be consistent with pure noise in the ringdown regime, with a likelihood function given by
\begin{align}
    \ln P\,(d_t|M_f,\chi_f,t_{0},\mathcal{F}_{\rm tot})=-\frac{1}{2}\sum_{i,j>I_0}d^{F}_iC_{ij}^{-1}d^F_j, \label{eq:log-p}
\end{align}
where $d^F_i\equiv d^{F}_{t_i}$ are the samples of the filtered data after the ringdown starts $(t_i>t_{0})$, $I_0$ is the index associated with $t_0$, and $C_{ij}$ is the covariance matrix. Note that here we compute the likelihood in the time domain, closely following \cite{Isi:2021iql}. Let us then recall what the formula of $\ln P$ looks like in Ref.~\cite{Isi:2021iql}:  
\begin{align}
    &\ln P\,(d_t|A_{lmn},\phi_{lmn},M_f,\chi_f,t_{0}) \notag \\
    &=-\frac{1}{2}\sum_{i,j>I_0}(d_i-h_i)C_{ij}^{-1}(d_j-h_j), \label{eq:log-p-old}
\end{align}
where $h_t$ is a multiple-QNM ringdown waveform template
\begin{align}
    h_t=\sum_{lmn} A_{lmn}e^{-(t-t_0)/\tau_{l m n}}\cos\left[2\pi f_{l m n}(t-t_0)+\phi_{lmn}\right], \label{eq:qnm_model}
\end{align}
with $A_{lmn}$'s and $\phi_{lmn}$'s being the amplitudes and phases of the QNMs, respectively. We can see that Eq.~\eqref{eq:log-p} is similar to Eq.~\eqref{eq:log-p-old} in the aspect of computing the likelihood of the data being pure noise after removing the GW signals. However, the underlying approaches of the two are quite different. In the usual unfiltered approach, one needs to build a ringdown template in terms of a set of QNMs, based on the amplitude $A_{lmn}$ and phase $\phi_{lmn}$ of each mode, and then subtract the template $h_t$ from the data $d_t$. On the contrary, our approach does not need $A_{lmn}$'s and $\phi_{lmn}$'s ---  the rational filters can completely remove the relevant complex-frequency component corresponding to each QNM from the ringdown no matter what the amplitude and phase are. A direct analogy is that a constant can be eliminated by a derivative no matter what its value is. In the Supplemental Material of our companion paper \cite{Ma_prl}, we explicitly show that the rational filter eliminates the dependency on mode amplitudes and phases through a new maximum likelihood estimator and relates to the usual time-domain approach \cite{Isi:2021iql} via an approximate marginalization. As a result, the new likelihood function implicitly depends on $M_f$ and $\chi_f$ only, given that the rational filter is built using a given set of $M_f$ and $\chi_f$ [see Eq.~\eqref{eq:rational_filter}].

\begin{figure*}
        \includegraphics[width=\columnwidth,clip=true]{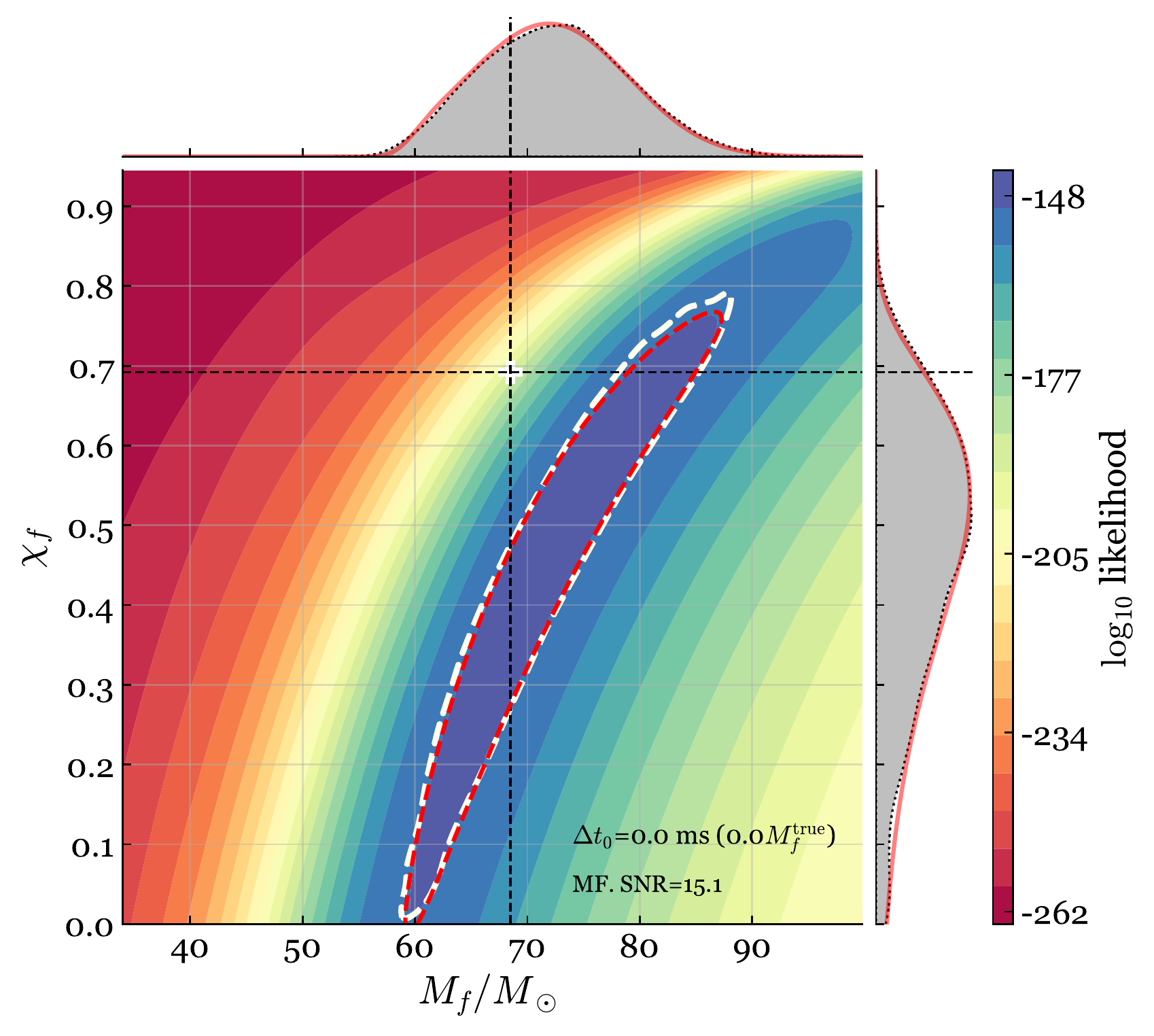}
        \includegraphics[width=\columnwidth,clip=true]{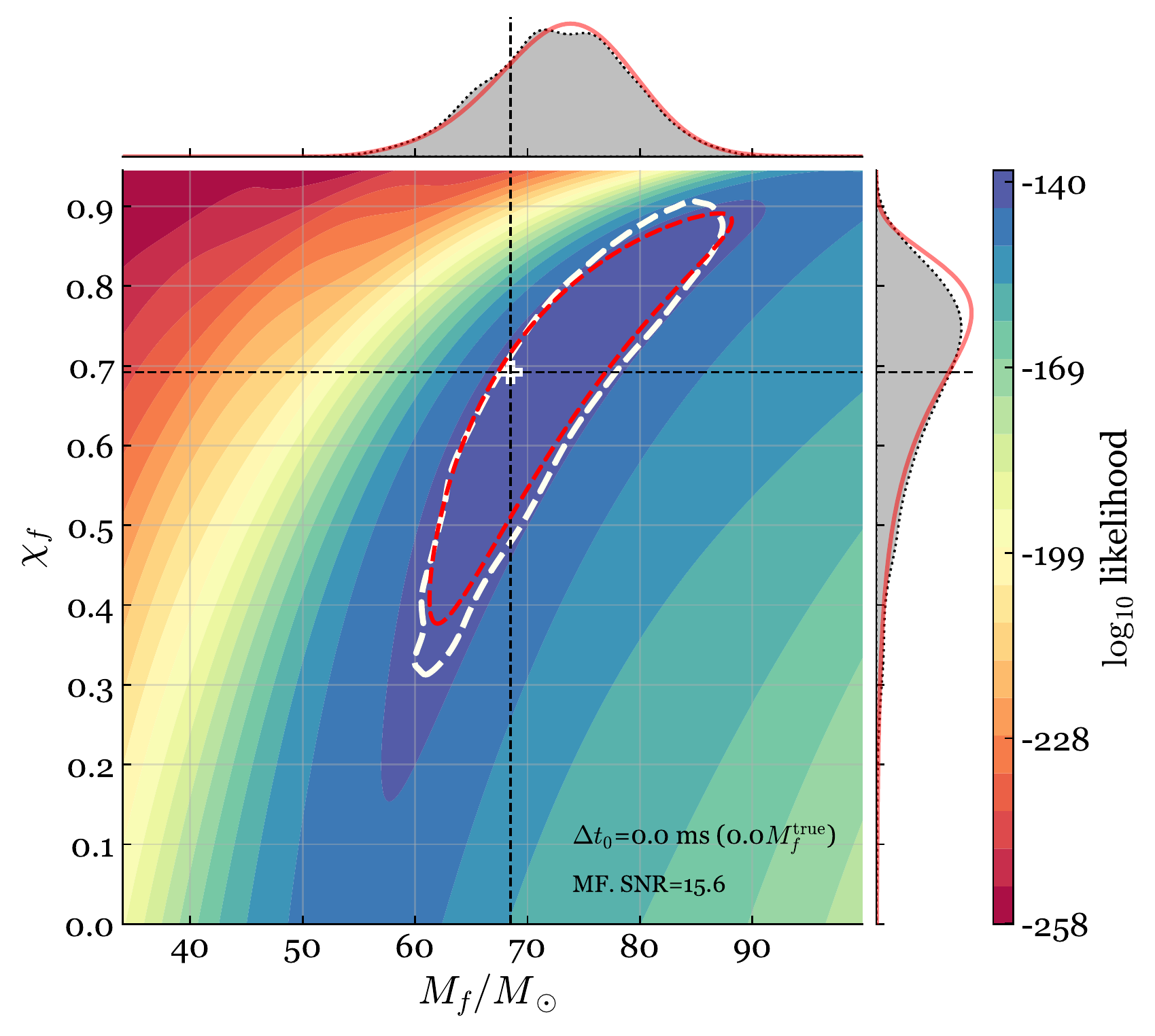} \\
        \includegraphics[width=\columnwidth,clip=true]{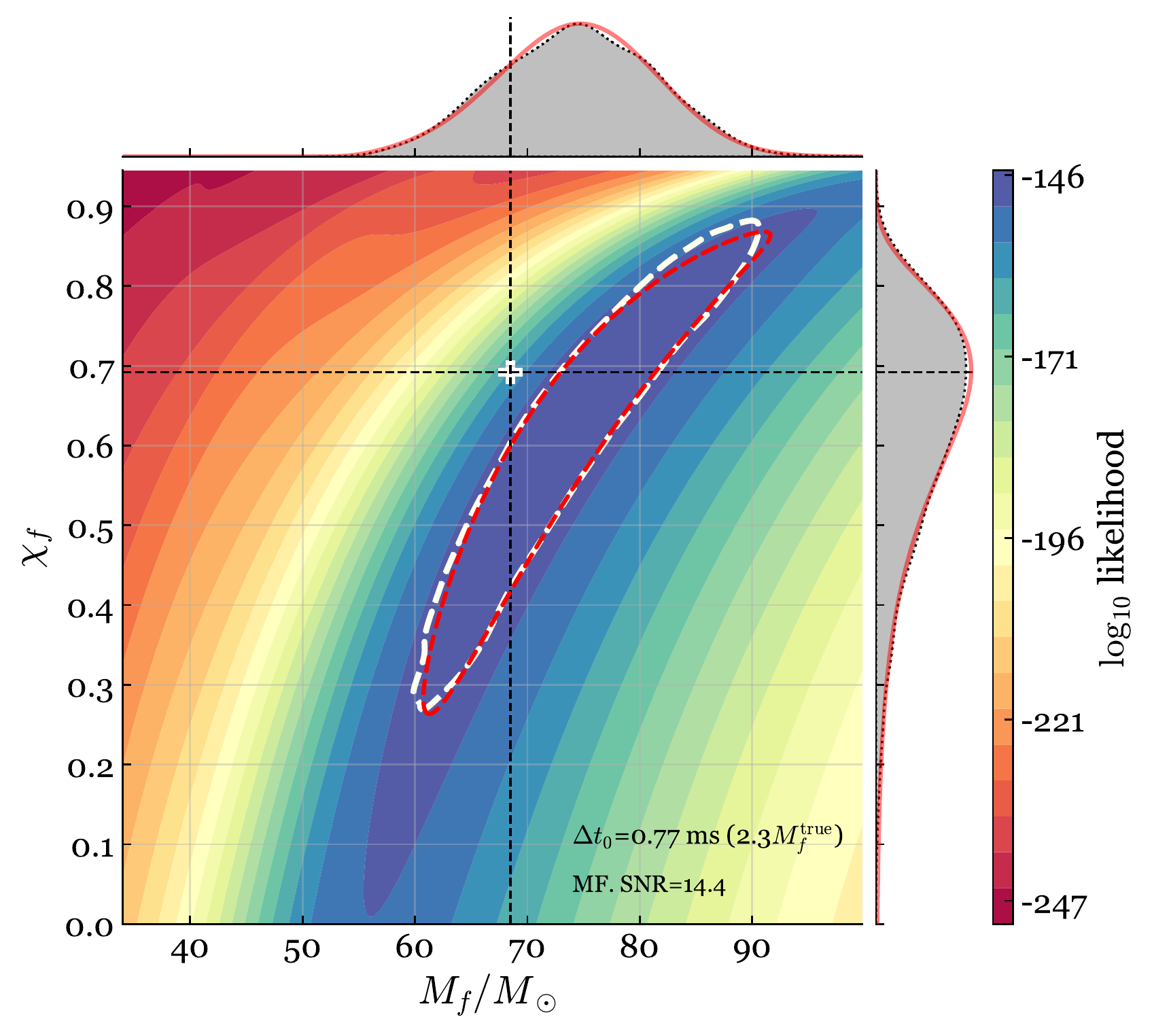}
        \includegraphics[width=\columnwidth,clip=true]{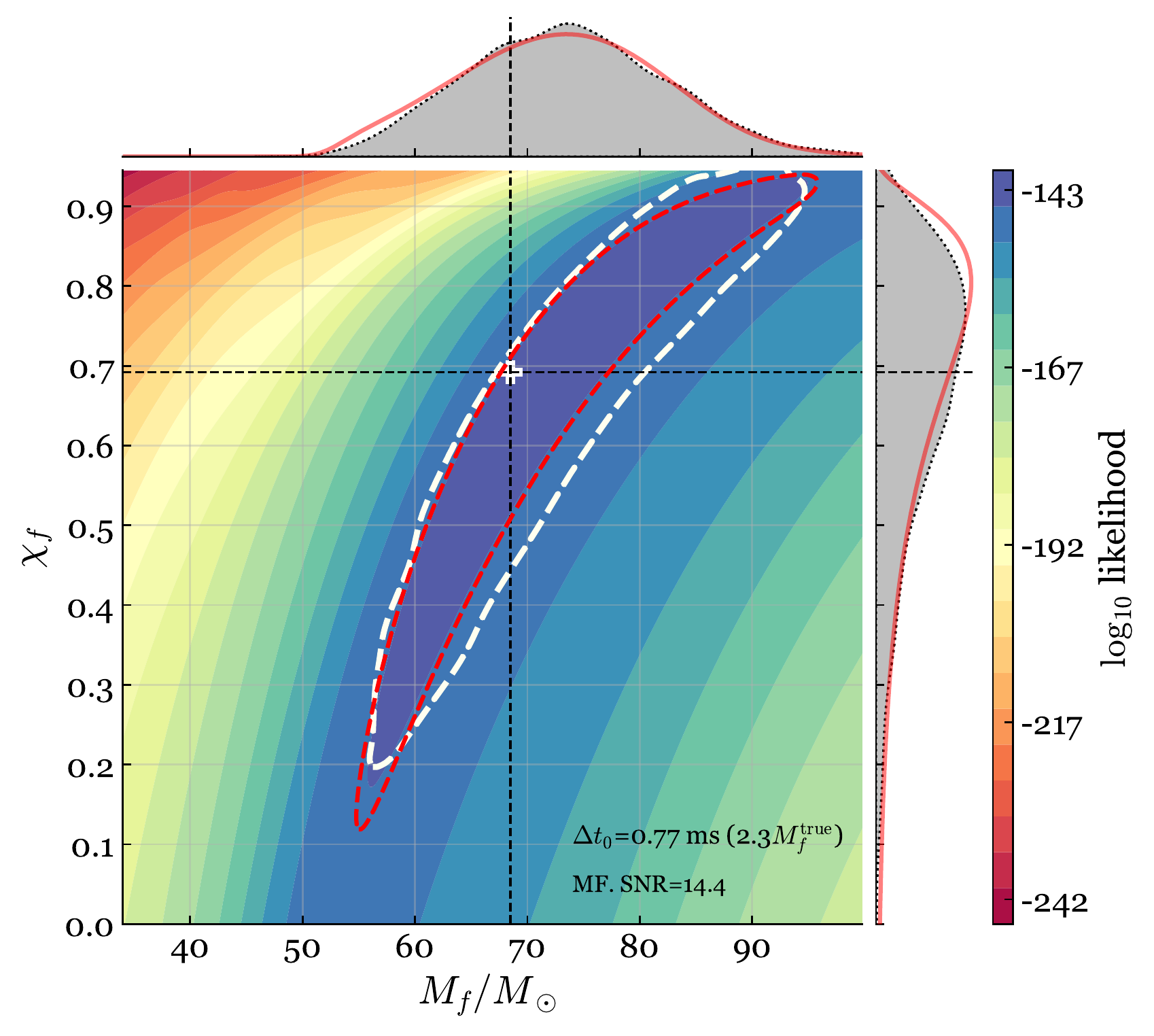}
  \caption{Joint posterior distributions of $M_f$ and $\chi_f$ evaluated with Eq.~\eqref{eq:m_chi_posteriors}. The GW150914-like NR waveform is injected into band-limited white noise. The top and bottom panels represent $\Delta t_0 = 0$ and 0.77 ms, respectively.
  The left and right panels show results from applying the filter for the fundamental mode only, $\mathcal{F}_{220}$, and the filter $\mathcal{F}_{221}\mathcal{F}_{220}$, respectively. The red-dashed contours display the 90\% credible region by integrating our new joint posterior in Eq.~\eqref{eq:m_chi_posteriors}; and the joint distribution is projected to the individual 1D space of $\chi_f$ and $M_f$ (red curves in side panels) using Eq.~\eqref{eq:likelihood_1d_marginal}. The white plus signs stand for the true value of $M_f$ and $\chi_f$ obtained from NR. The white dashed contours show the 90\% credible region from the full-RD MCMC approach. The MCMC results are marginalized to the 1D distributions of $M_f$ and $\chi_f$, shown as the gray-shaded regions in side panels. The value of the matched filter (MF) SNR is also provided.}
 \label{fig:injection1_0305}
\end{figure*}

In this paper, we treat $M_f$ and $\chi_f$ as the parameters of the filtered data, whereas the ringdown start time $t_0$ and the choice of the set of QNMs included in the filter $\mathcal{F}_{\rm tot}$ as hyperparameters. Therefore, we can convert the likelihood to the joint posterior of $M_f$ and $\chi_f$ via
\begin{align}
    &\ln P\,(M_f,\chi_f | d_t,t_{0},\mathcal{F}_{\rm tot}) = \ln P\,(d_t|M_f,\chi_f,t_{0},\mathcal{F}_{\rm tot}) \notag \\
    &+ \ln \Pi\,(M_f,\chi_f)+{\rm constant}, \label{eq:m_chi_posteriors}
\end{align}
where $\ln \Pi\,(M_f,\chi_f)$ is the prior. In our following discussions for the injection study and GW150914, we always place uniform priors on the final mass and spin in the ranges of $M_f\in \left[35M_\odot,140M_\odot\right]$ and $\chi_f\in[0,0.99]$. In addition, $P\,(M_f,\chi_f | d_t,t_{0},\mathcal{F}_{\rm tot})$ can be marginalized by integrating over one dimension (1D) to obtain the distribution of the other dimension:
\begin{subequations}
\label{eq:likelihood_1d_marginal}
\begin{align}
    P\,(\chi_f | d_t,t_{0},\mathcal{F}_{\rm tot})=\int P\,(M_f,\chi_f | d_t,t_{0},\mathcal{F}_{\rm tot})dM_f,\\
    P\,(M_f | d_t,t_{0},\mathcal{F}_{\rm tot})=\int P\,(M_f,\chi_f | d_t,t_{0},\mathcal{F}_{\rm tot})d\chi_f.
\end{align}
\end{subequations}
Since this new $\ln P$ is simply a two-dimensional function, it is computationally cheap enough to directly compute the distribution of $M_f$ and $\chi_f$ without using techniques like MCMC. 
\begin{figure*}
        \includegraphics[width=\columnwidth,clip=true]{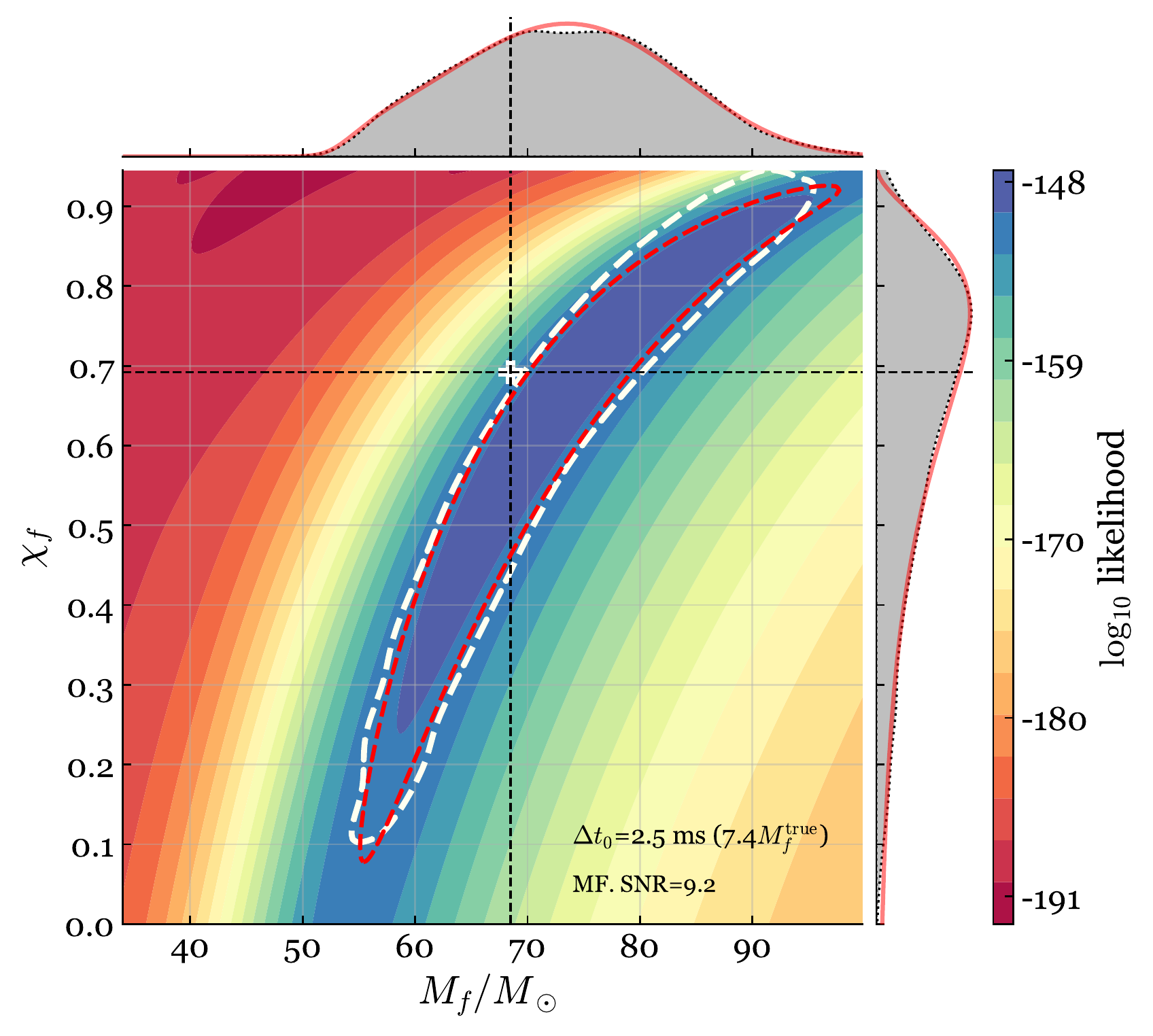}
        \includegraphics[width=\columnwidth,clip=true]{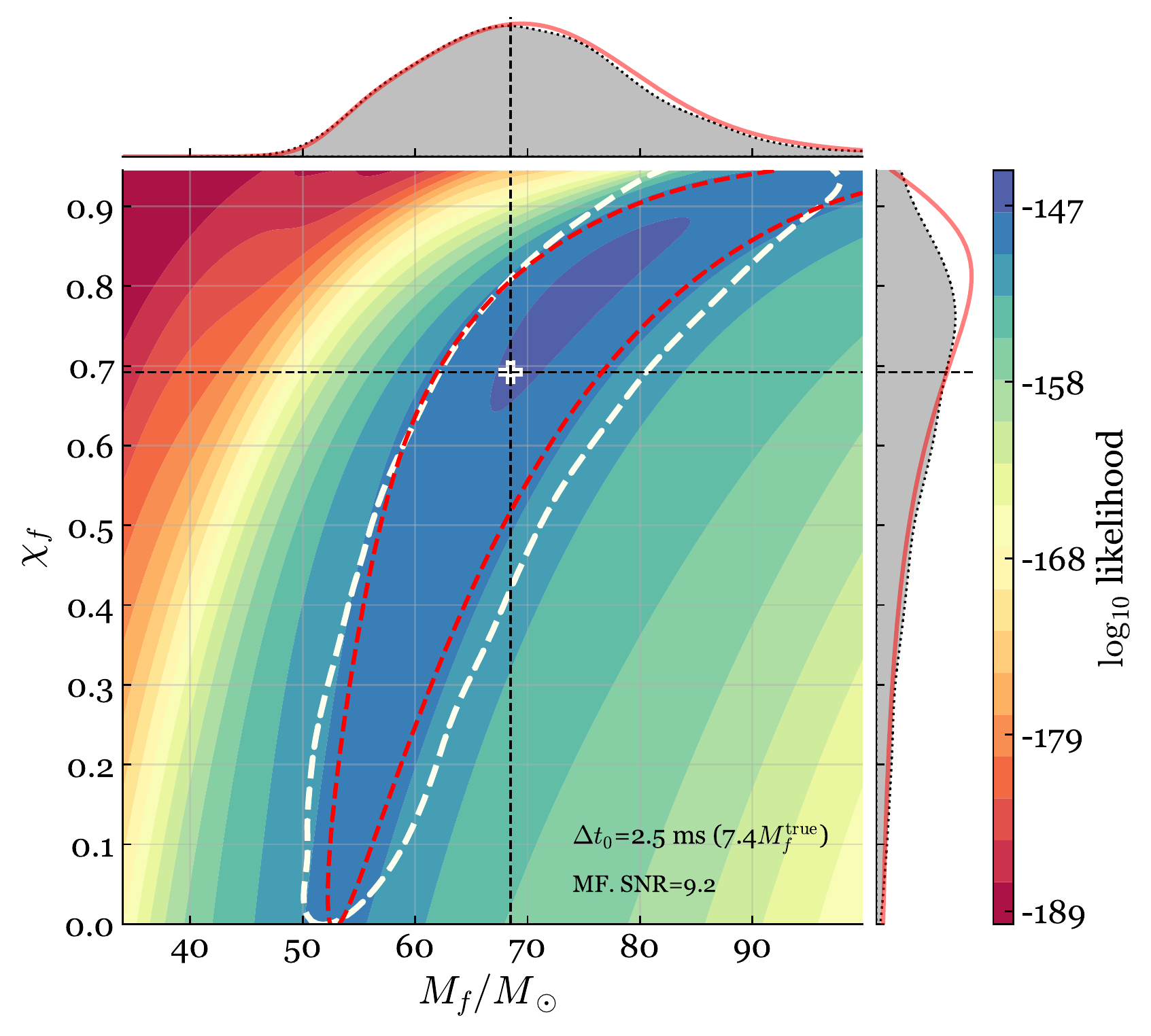} \\
        \includegraphics[width=\columnwidth,clip=true]{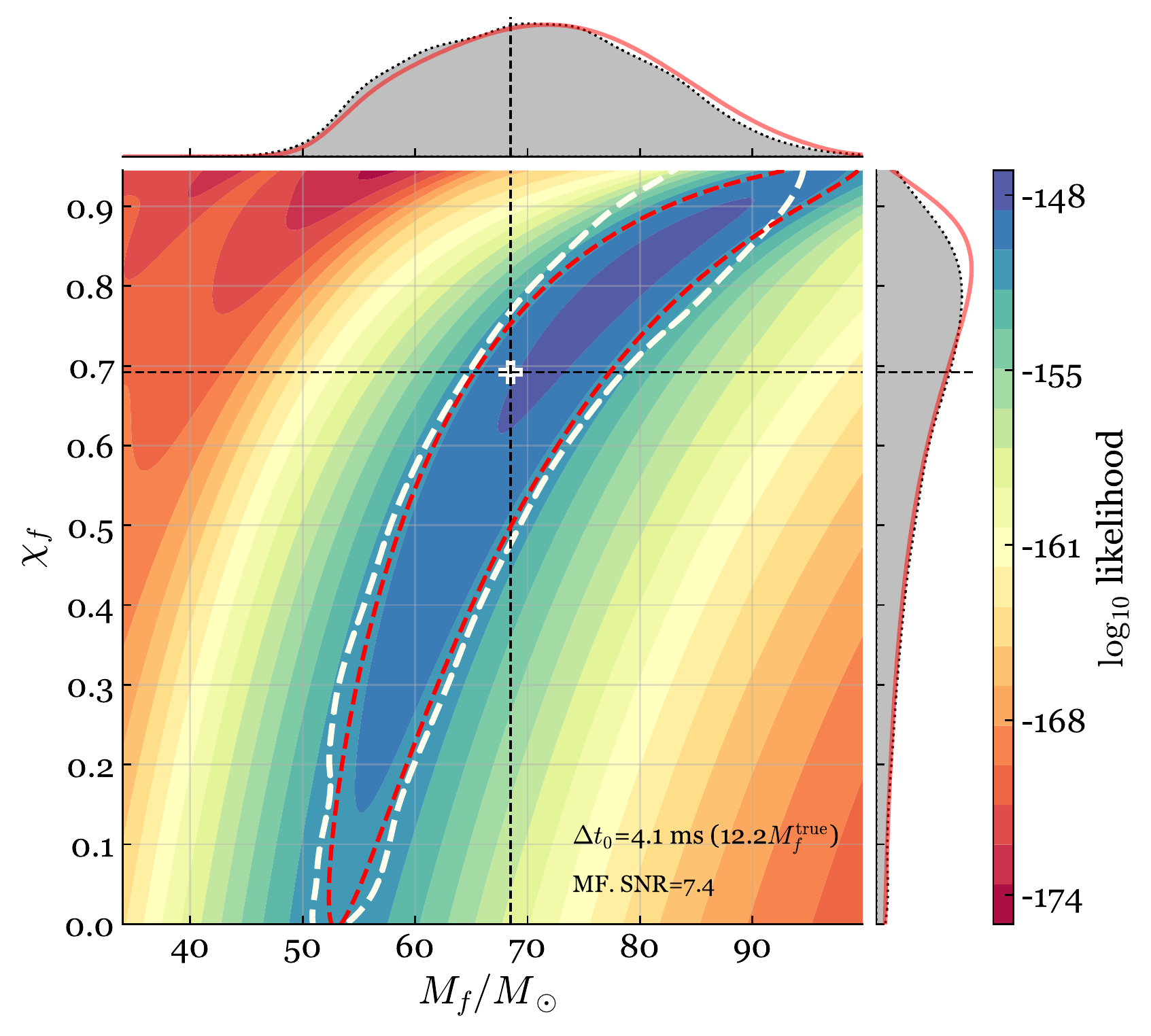}
        \includegraphics[width=\columnwidth,clip=true]{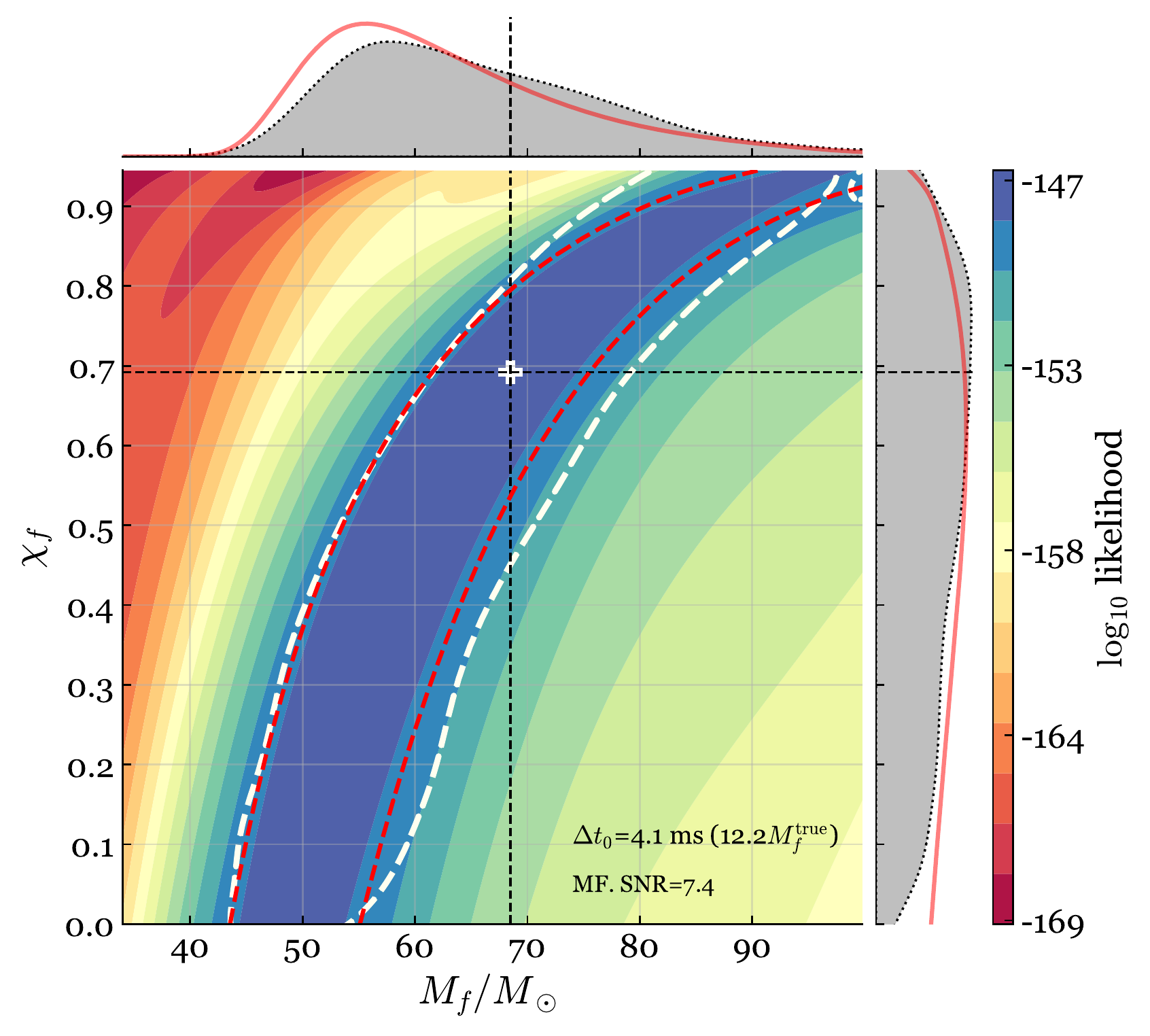} 
  \caption{Joint posterior distributions of $M_f$ and $\chi_f$. Fig.~\ref{fig:injection1_0305} continued; more values of $\Delta t_0$ are tested.}
 \label{fig:injection2_0305}
\end{figure*}

\begin{figure*}
        \includegraphics[width=\columnwidth,clip=true]{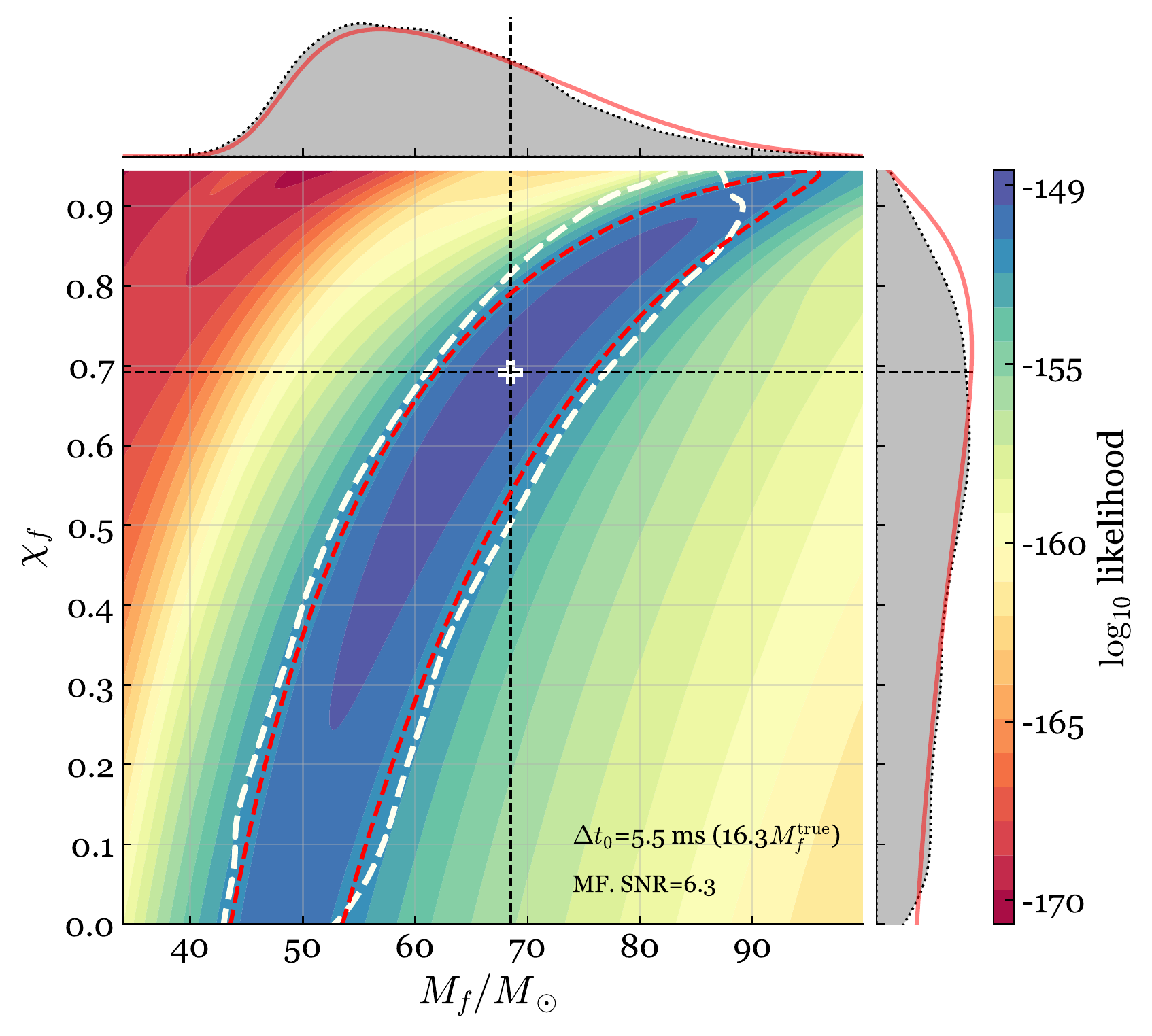}
        \includegraphics[width=\columnwidth,clip=true]{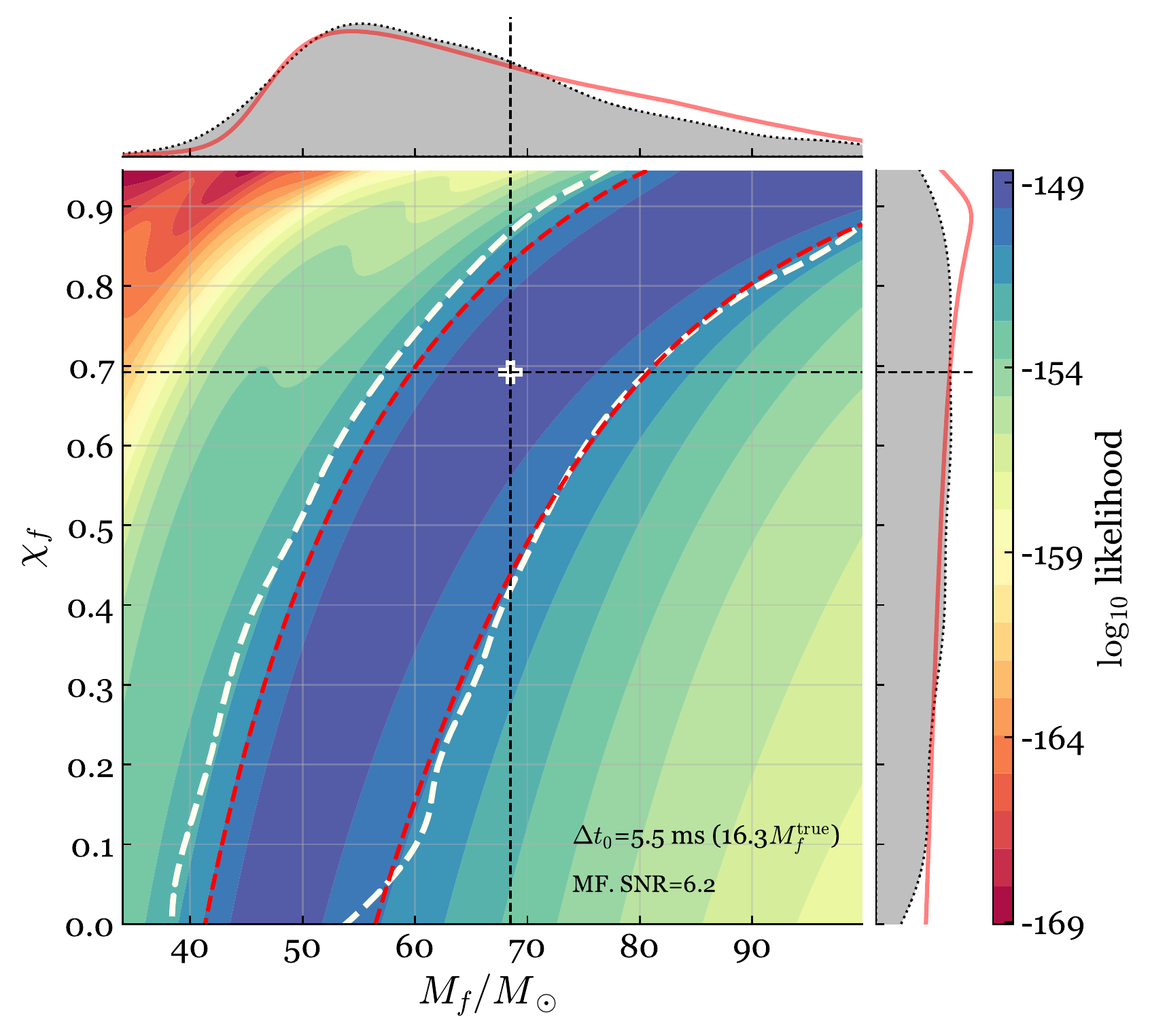} \\
        \includegraphics[width=\columnwidth,clip=true]{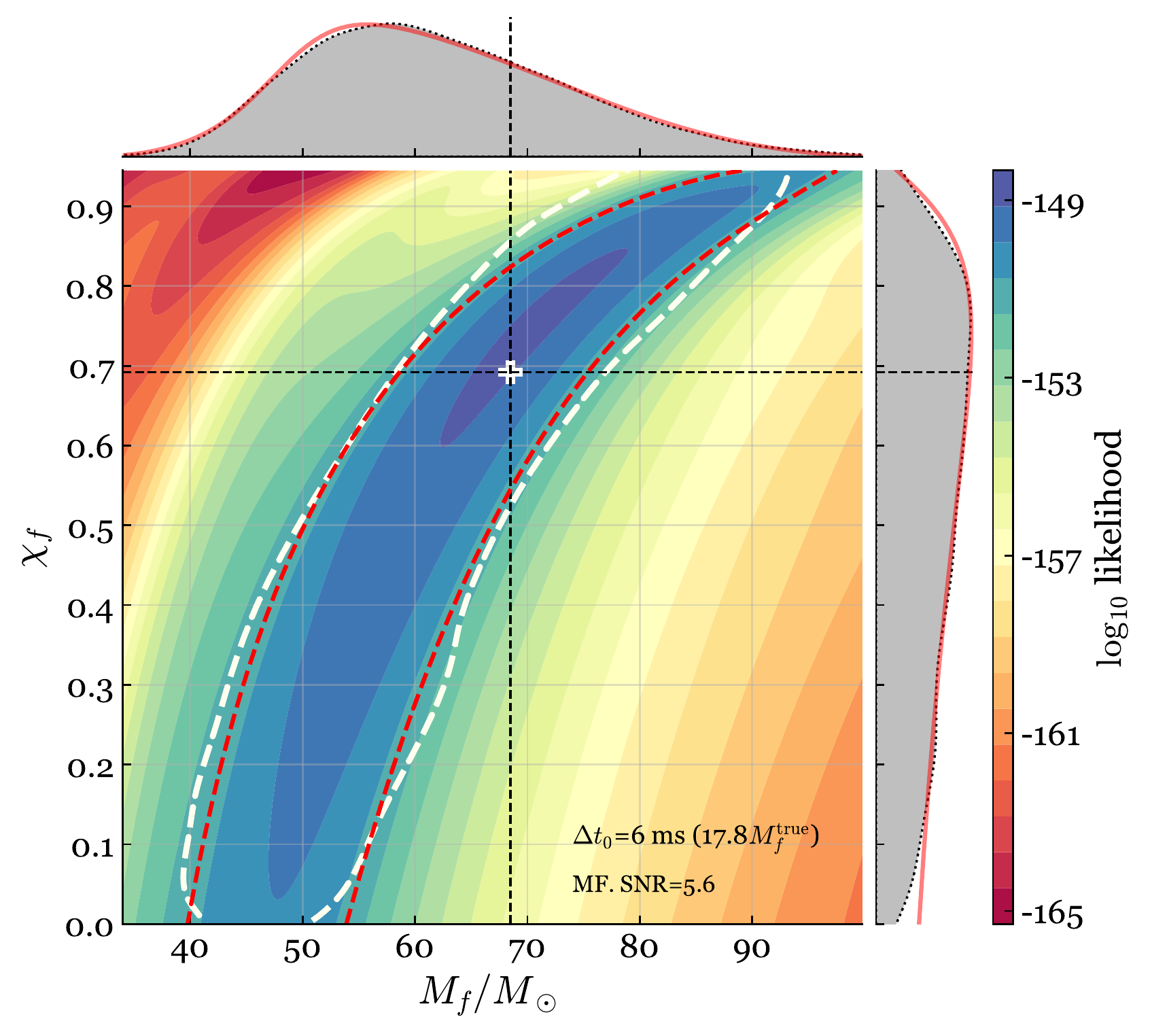}
        \includegraphics[width=\columnwidth,clip=true]{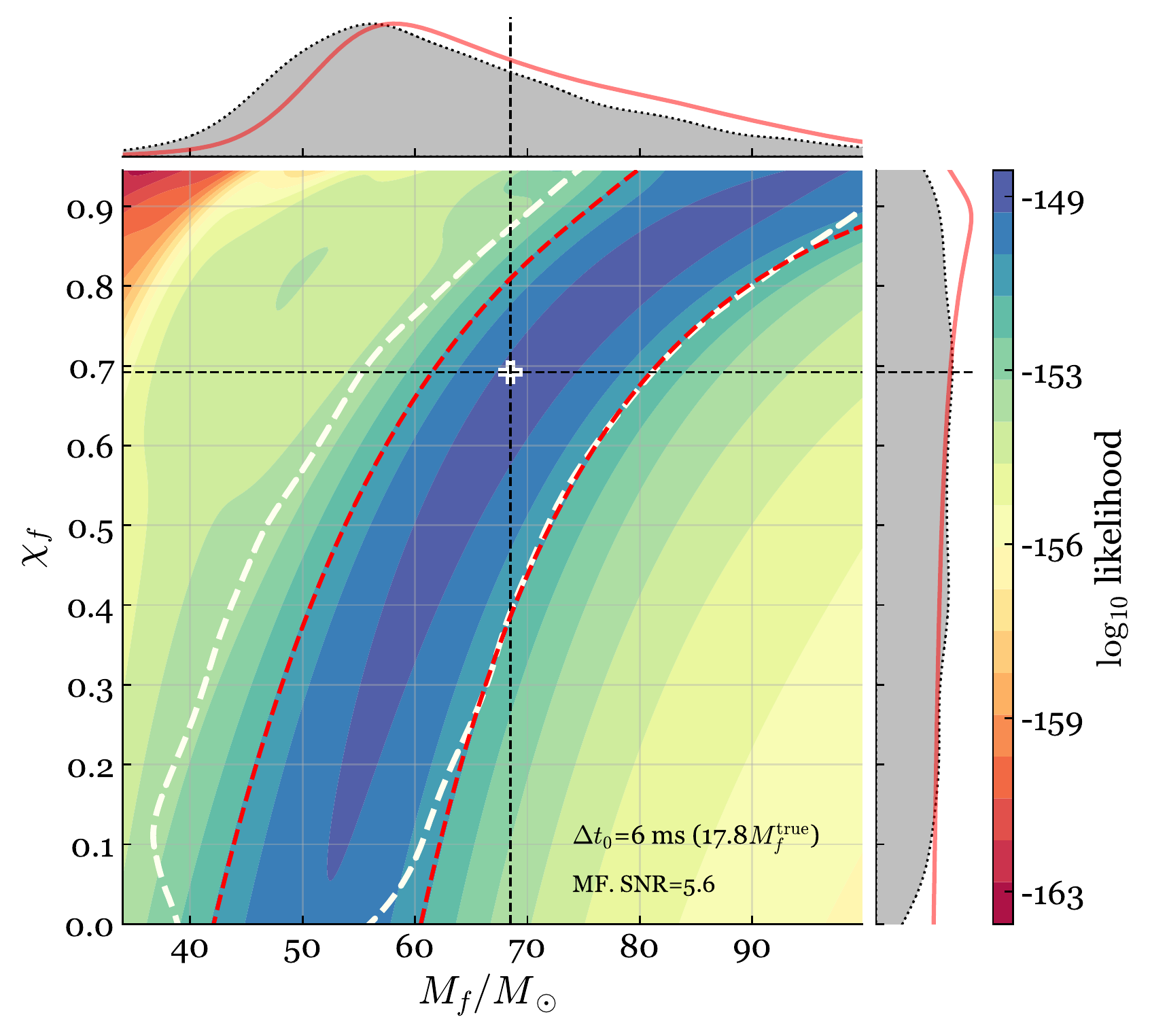}
  \caption{Joint posterior distributions of $M_f$ and $\chi_f$. Fig.~\ref{fig:injection1_0305} continued; more values of $\Delta t_0$ are tested.}
 \label{fig:injection3_0305}
\end{figure*}

\subsection{NR waveform injection}
\label{subsec:injection_likelihood}
We first take the GW150914-like NR waveform \cite{Lovelace:2016uwp} from the SXS catalog \cite{Boyle:2019kee,SXSCatalog} and build a complex strain $h$ from the $(l,m)$ harmonics $h_{lm}$ in the NR waveform, given by
\begin{align}
    h=h_+-ih_\times=\sum_{lm}\tensor[_{-2}]{Y}{_{lm}}(\iota,\beta)h_{lm}, \label{eq:complex_strain}
\end{align}
where $\tensor[_{-2}]{Y}{_{lm}}(\iota,\beta)$ denotes the spin-weighted spherical harmonics, and angles $(\iota,\beta)$ stand for the angular coordinates of an observer within the source frame. Here we choose $(\iota=\pi,\beta=0)$ to simulate the orientation of GW150914 (face-off) \cite{Isi:2019aib}. 
For simplicity, we include only the two most dominant modes $h_{2,\pm2}$ in Eq.~\eqref{eq:complex_strain} and inject the ``$+$'' polarization state to band-limited white noise.
To mimic GW150914, we set the total initial mass of the system (detector frame) to $72 \,M_\odot$ so that the mass of the remnant BH $M_f=68.5\, M_\odot$ agrees with that inferred from the full IMR waveform \cite{Isi:2019aib}. The length of the full NR waveform is 1.38 s. We pad zeros on both ends of the waveform to prolong the length to 4 s.\footnote{Since we can control the simulated noise in this case, a relatively short signal is chosen for efficiency.} The data $d_i\equiv d_{t_i}$ (including white noise and simulated signal) are sampled at 16384~Hz. We also adjust the relative amplitude of the NR waveform so that the ringdown matched filter SNR is $\sim 15$, as measured after the peak of the strain. We then condition $d_i$ to remove contents that are below 20 Hz and downsample the data to 4096 Hz.

To calculate $\ln P$ in Eq.~\eqref{eq:log-p}, we need to estimate the covariance matrix $C_{ij}$ first, which has an exact expression for the given band-limited white noise. We further verify this by estimating the PSD $S(f)$ from 4 s of the off-source white noise with the Welch method, as described in Sec.~\ref{subsec:impact_detector_noise}. We then inverse Eq.~\eqref{eq:PSD_vs_rho} to obtain the value of ACF $\rho(\tau)$
\begin{align}
    \rho_k\equiv\rho(t_k)=\frac{1}{2T}\sum_{j=0}^{N-1}S_j e^{2\pi ijk/N}, \label{eq:rho_from_S}
\end{align}
where $S_j\equiv S(f_j)$ is the PSD value at $f=f_j$, $N$ is the total number of frequency bins in $S(f)$, and $T=4 \,{\rm s}$ is the total length of the noise. Finally, we get the covariance matrix $C_{ij}$ via
\begin{align}
    C_{ij}=\rho(|i-j|). \label{eq:C_ij_injection}
\end{align}

On the other hand, to apply the rational filter to the data $d_i$, we transform the full length of $d_i$ into the frequency domain via FFT
\begin{align}
    \tilde{d}_j=\Delta t \sum_{k=0}^{N-1}d_ke^{-2\pi ijk/N},
\end{align}
with $\Delta t=t_{k+1}-t_k$. Then the filtered data read [see Eq.~\eqref{eq:total_filter} for $\mathcal{F}_{\rm tot}$]
\begin{align}
    \tilde{d}^{F}_j= \mathcal{F}_{\rm tot}(f_j)\tilde{d}_j,
\end{align}
and the corresponding time-domain data are given by
\begin{align}
    d_k^F=\frac{1}{T}\sum_{j=0}^{N-1}\tilde{d}^{F}_je^{-2\pi ijk/N}. \label{eq:injection_dF}
\end{align}
Next, we select the filtered time-series data $d_k^F$ that lie within the time interval of $[t_0,t_0+w]$, which we refer to as a ringdown window with window width $w$. Here we fix $w = 0.08$ s while letting $t_0$ be a free parameter. We discuss the impact of $t_0$ later in this section. Finally, we substitute the value of $C_{ij}$ [Eq.~\eqref{eq:C_ij_injection}] and $d_k^F$ [Eq.~\eqref{eq:injection_dF}] within the ringdown window into Eq.~\eqref{eq:log-p} to compute the likelihood.

For a given start time $t_0$, we build $\mathcal{F}_{\rm tot}$ by choosing a set of remnant black hole parameters, $M_f$ and $\chi_f$ [Eq.~\eqref{eq:rational_filter}], and a set of QNMs [Eq.~\eqref{eq:total_filter}].   We can calculate the posterior distribution of $M_f$ and $\chi_f$ using Eq.~\eqref{eq:m_chi_posteriors} for a given set of QNMs and infer which QNMs are more likely to be present in the signal. On the other hand, for a fixed filter $\mathcal{F}_{\rm tot}$ (built with a given choice of $M_f$, $\chi_f$ and QNMs), we can slide the ringdown window $[t_0,t_0+w]$ by varying $t_0$. The $t_0$ value corresponding to the maximum posterior probability indicates the start time of the QNM(s) included in $\mathcal{F}_{\rm tot}$. In this example, it is found that the $l=m=2$ harmonic can be modeled with multiple overtones $\omega_{22n}$'s right after the time when the strain reaches its peak amplitude \cite{Giesler:2019uxc}, denoted by $t_{\rm peak}$. Therefore, we set the form of the total filter to
\begin{align}
    \mathcal{F}_{\rm tot}=\prod_{n=0}^X\mathcal{F}_{22n}, \label{eq:total_filter_event}
\end{align}
where $X$ is the highest overtone included, and focus on the regime of $\Delta t_0=t_0-t_{\rm peak} \gtrsim 0$.

Figs.~\ref{fig:injection1_0305}, \ref{fig:injection2_0305} and \ref{fig:injection3_0305} show the joint posterior of $M_f$ and $\chi_f$ evaluated at different start times (parameterized by $\Delta t_0$), for the injected signal. Here we display a 2D grid of $M_f\in \left[35M_\odot,100M_\odot\right]$ and $\chi_f\in[0,0.95]$ for better readability. Other regions of the parameter space do not provide extra features. The left and right columns correspond to having $X=0$ (``one QNM'', 220) and $X=1$ (``two QNMs'', 220 and 221), respectively, in $\mathcal{F}_{\rm tot}$. Adding more overtones does not further improve the likelihoods, given the ringdown SNR level at the current stage. The true values of the remnant mass $M_f=68.5\,M_\odot$ and spin $\chi_f=0.692085$ are marked by white plus signs. We compute the 90\% credible region by integrating the joint posterior evaluated with our filter [Eq.~\eqref{eq:m_chi_posteriors}] over the $M_f-\chi_f$ parameter space. The results are shown as red-dashed contours. In the meantime, the marginalized posterior distributions of $M_f$ and $\chi_f$ are plotted as 1D histograms (red curves) in the side panels, calculated by Eq.~\eqref{eq:likelihood_1d_marginal}. For comparison purposes, we also use the Python package $\textsf{ringdown}$ \cite{ringdown_isi,Isi:2021iql} to perform a conventional time-domain full-ringdown Bayesian analysis via MCMC (hereafter ``full-RD MCMC''), in which the likelihoods are evaluated by Eq.~\eqref{eq:log-p-old}. To build the ringdown template $h_t$ in Eq.~\eqref{eq:qnm_model}, we include the same QNM(s) as the one(s) used in the filter $\mathcal{F}_{\rm tot}$. The $90\%$ credible interval joint posteriors evaluated by MCMC are shown as the regions enclosed by white dashed contours. Similarly, we plot the 1D histograms for $M_f$ and $\chi_f$, obtained via MCMC, as gray-shaded regions. For reference, we compute the matched-filter SNRs (SNR$_{\rm MF}$) from the posterior samples via 
\begin{align}
    {\rm SNR_{MF}}=\frac{\braket{h_t|d_t}}{\sqrt{\braket{h_t|h_t}}},
\end{align}
with
\begin{align}
    \braket{h_t|d_t}=\sum_{i,j>I_0} h_{i}C^{-1}_{ij}d_j.
\end{align}

Now let us look at the first row of Fig.~\ref{fig:injection1_0305}, with $\Delta t_0= 0\,{\rm ms}$. In both the left and right panels, the contours obtained from our filters largely agree with the full-RD MCMC results. The true remnant properties lie within the 90\% credible region when we include both the fundamental mode and the first overtone in the filter. On the contrary, there is a strong bias when the first overtone is excluded. Next, in the second row of Fig.~\ref{fig:injection2_0305}, namely $\Delta t_0= 4.1\,{\rm ms}=12.2M_{f}^{\rm true}$, the true remnant mass and spin can be recovered with the fundamental mode alone, whereas the constraints of $M_f$ and $\chi_f$ get worse after adding the filter of the first overtone $\mathcal{F}_{221}$. The reason is as follows: since the first overtone has decayed to a small value at $\Delta t_0= 4.1\,{\rm ms}$, the major impact of the rational filter $\mathcal{F}_{221}$ on the signal is to reduce the amplitude of the fundamental mode by a factor of $B_{220}^{221}=2.053$ [Eq.~\eqref{eq:filter_on_other_qnm}], making it harder to infer the remnant properties from the filtered data. 

\begin{figure}[htb]
        \includegraphics[width=\columnwidth,clip=true]{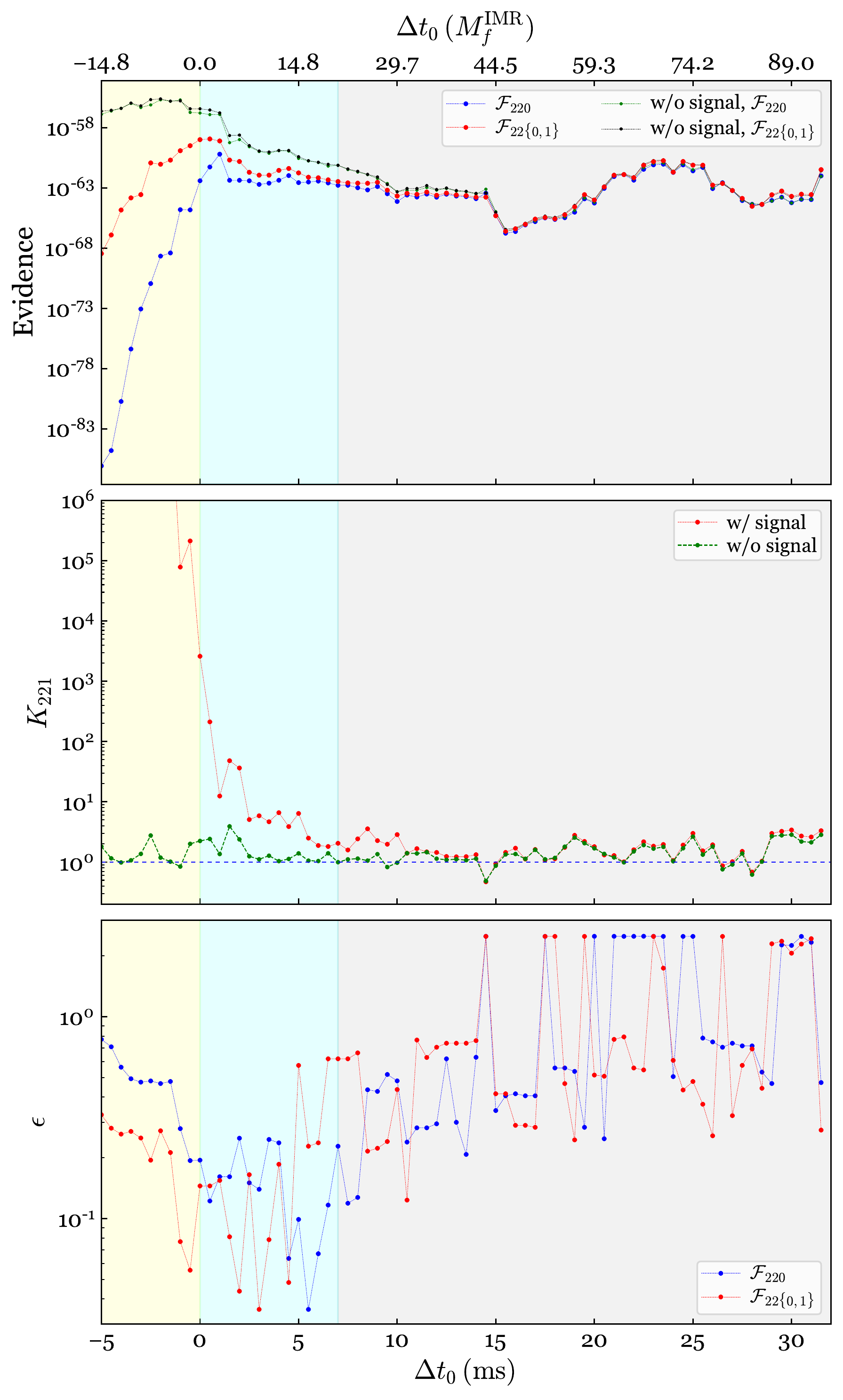}
  \caption{Model comparison at different $\Delta t_0$ for a GW150914-like NR waveform injected in band-limited white noise. Top: Model evidence as a function of $\Delta t_0$, evaluated by Eq.~\eqref{eq:filter_prob}. 
  The blue and red curves indicate the results after applying $\mathcal{F}_{220}$ (clean the fundamental mode only) and $\mathcal{F}_{22\{0,1\}}$ (clean the fundamental mode and the first overtone), respectively. The corresponding results computed with off-source noise are shown in green and black (almost indistinguishable). Middle: Bayes factor ($K_{221}$) of the existence of the first overtone over fundamental mode only (red curve), calculated by Eq.~\eqref{eq:bayes_overtone}. As a comparison, the green curve shows the Bayes factor evaluated with the off-source noise. We take $K_{221}=1$ as a benchmark, indicated by the horizontal dashed line. Bottom: Distance ($\epsilon$) of the MAP values of $M_f$ and $\chi_f$ to the true values, calculated by Eq.~\eqref{eq:MAP_true_distance_eps}.}
 \label{fig:injection_likelihood_time}
\end{figure}

After qualitatively discussing the posteriors obtained using Eq.~\eqref{eq:log-p}, we propose some more quantitative quantities to evaluate the significance of the first overtone. First, given the hyperparameter $\Delta t_0$ and the filter model $\mathcal{F}_{\rm tot}$, we can compute model evidence (marginal likelihood) via
\begin{align}
&P(d_t|\Delta t_0,\mathcal{F}_{\rm tot})=\notag \\
&\iint P\,(d_t|M_f,\chi_f,\Delta t_{0},\mathcal{F}_{\rm tot})\Pi(M_f,\chi_f)dM_fd\chi_f. \label{eq:filter_prob}
\end{align}
where $\Pi(M_f,\chi_f)$ is the prior distribution of $M_f$ and $\chi_f$. Here we simply assume a flat prior in $M_f\in \left[35M_\odot,140M_\odot\right]$ and $\chi_f\in[0,0.99]$. By comparing the evidence across different choices of $\mathcal{F}_{\rm tot}$, one could figure out which QNMs are more likely to be present in the signal. In the top panel of Fig.~\ref{fig:injection_likelihood_time}, we plot $P(d_t|\mathcal{F}_{220},\Delta t_0)$ and $P(d_t|\mathcal{F}_{22\{0,1\}},\Delta t_0)$ as functions of $\Delta t_0$, for the same simulation data set with the injected signal. As a comparison, we also show the results obtained from the off-source noise (green and black curves). The boundary between the yellow and cyan regions stands for the time when the strain reaches its peak, i.e., $\Delta t_0=0$. We can see both the red and blue curves surge rapidly within the regime of $[-5,0]\,{\rm ms}$, implying the onset of the ringdown stage. The evidence of the QNMs starts to grow even before the peak time $\Delta t_0=0$ because the width of our window $w=0.08 \, {\rm s}=237M_{f}^{\rm true}$ is longer than the ringdown duration --- the full ringdown signal already falls in the window when $\Delta t_0 \in [-5, 0]\,{\rm ms}$. Consequently, the QNM model evidence continues to increase as the inspiral-merger part of the signal slides out of the window. Additionally, the evidence of the ``two-QNM'' model $(\mathcal{F}_{22\{0,1\}})$ is greater than the ``one-QNM'' one $(\mathcal{F}_{220})$ near the peak time $(\Delta t_0\sim0)$, indicating that the ``two-QNM'' filter is preferred in the early stage of ringdown. To provide a more quantitative evaluation, we compute the Bayes factor by taking the ratio between the marginal likelihoods of a QNM model with and without the first overtone:
\begin{align}
    K_{221}(\Delta t_0)=\frac{P(d_t|\mathcal{F}_{22\{0,1\}},\Delta t_0)}{P(d_t|\mathcal{F}_{220},\Delta t_0)}. \label{eq:bayes_overtone}
\end{align}
The middle panel of Fig.~\ref{fig:injection_likelihood_time} shows $K_{221}$ as a function of $\Delta t_0$. We also show the off-source  results (green dash-dot curve). As expected, $K_{221}$ drops sharply near the peak time $\Delta t_0\sim0$ and gradually converges to the off-source result at later times. In the absence of the GW signal, the Bayes factor simply oscillates around unity across the entire time tested, which is also expected.

On the other hand, to characterize how well we can recover the remnant properties by applying our filters, we look for the maximum a posteriori (MAP) estimates of $(M_f,\chi_f)$. Following Ref.~\cite{Giesler:2019uxc}, we define a dimensionless quantity $\epsilon$ to describe the distance of the MAP values $(M_f^{\rm MAP},\chi^{\rm MAP}_f)$ to the true values $(M_f^{\rm true},\chi^{\rm true}_f)$, given by
\begin{align}
    &\epsilon(\Delta t_0)= \notag \\
    &\sqrt{\left(\chi_f^{\rm MAP}(\Delta t_0)-\chi^{\rm true}_f\right)^2+\left(\frac{M_f^{\rm MAP}(\Delta t_0)-M^{\rm true}_f}{M^{\rm true}_f}\right)^2}. \label{eq:MAP_true_distance_eps}
\end{align}
The bottom panel of Fig.~\ref{fig:injection_likelihood_time} shows the resulting $\epsilon(\Delta t_0)$. For both QNM models, $\epsilon$ starts to decrease before the peak time $\Delta t_0=0$, for the same reason that the analysis window we use is wider than the span of the ringdown signal. At $\Delta t_0=0$ and right after, $\epsilon$ obtained by using the filter $\mathcal{F}_{22,\{1,2\}}$ is smaller than that using $\mathcal{F}_{220}$, indicating the existence of the first overtone. After $\sim 5$ ms, $\epsilon$ from the filter $\mathcal{F}_{22,\{1,2\}}$ starts to increase as the first overtone decays away. By contrast, the fundamental mode still shows significance at $\Delta t_0\sim 5.5\,{\rm ms}=16.3M_f^{\rm true}$, which leads to a much smaller $\epsilon$. Beyond $\Delta t_0=7 \,{\rm ms}=20.8M_f^{\rm true}$, no precise parameter information can be extracted from the fundamental mode anymore.
Therefore we plot a boundary between the cyan and gray regions in the figure at $\Delta t_0=7 \,{\rm ms}=20.8M_f^{\rm true}$ to indicate the time around which the whole ringdown signal fades away.

\begin{figure*}
        \includegraphics[width=\columnwidth,clip=true]{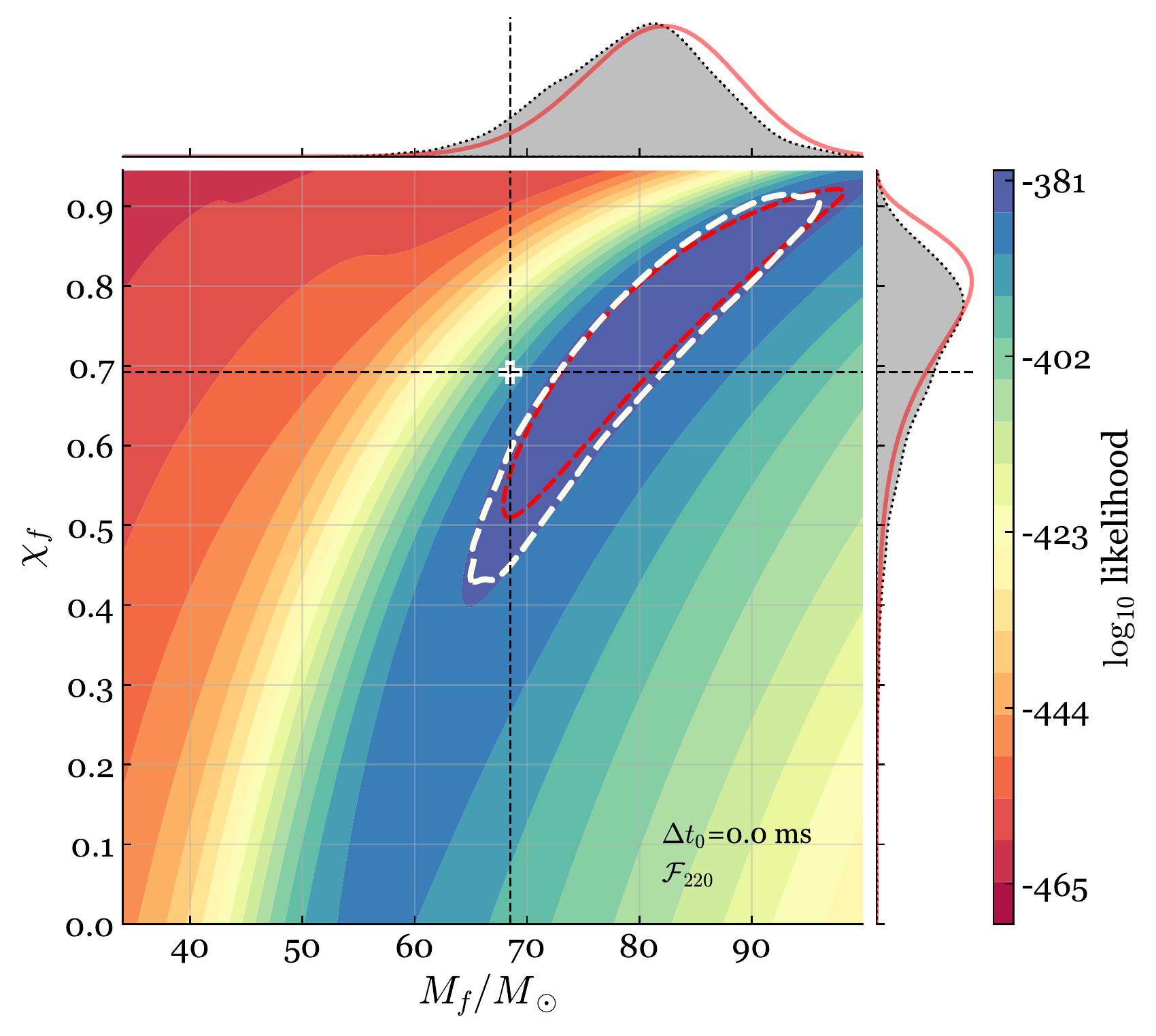}
        \includegraphics[width=\columnwidth,clip=true]{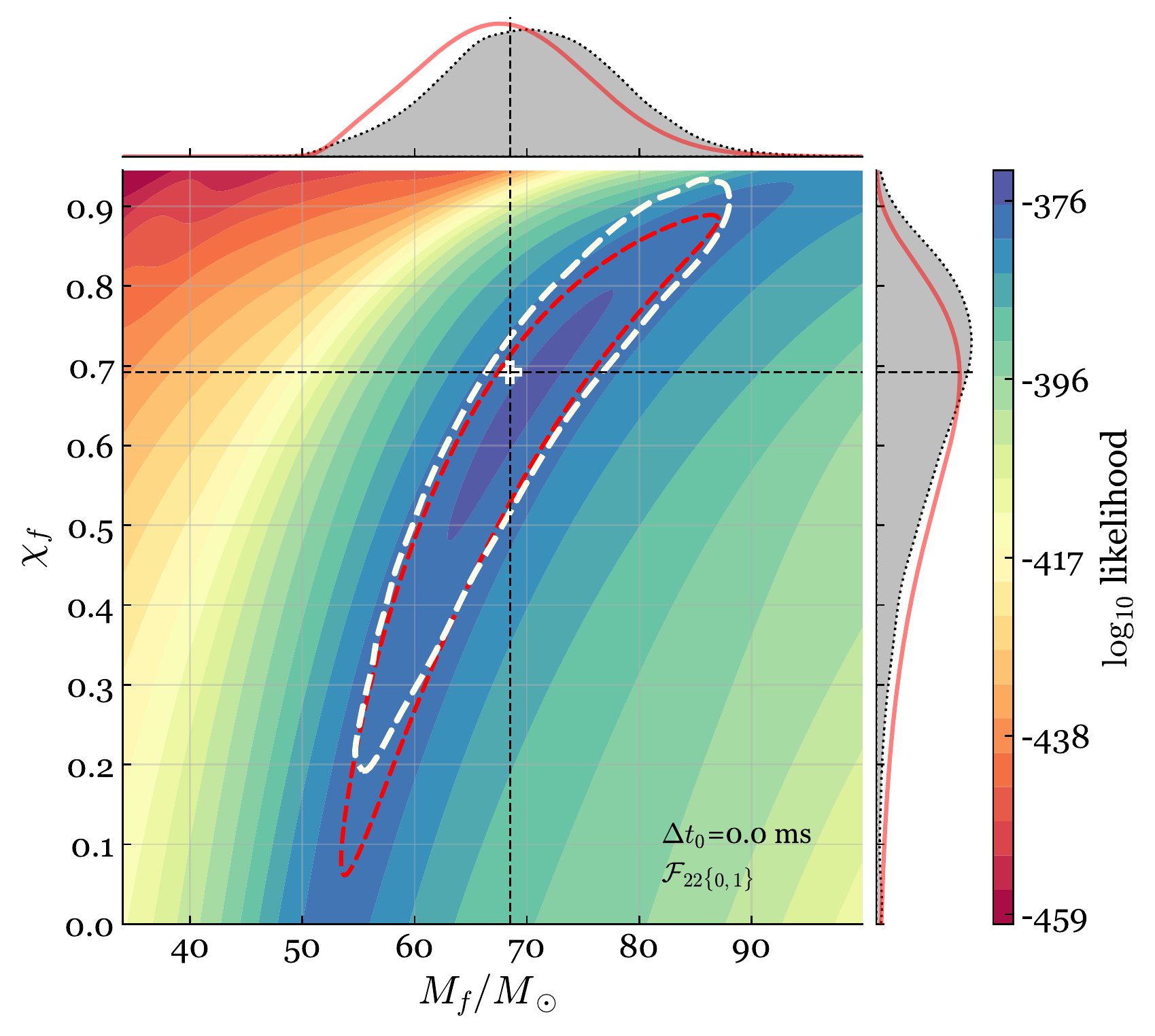} \\
        \includegraphics[width=\columnwidth,clip=true]{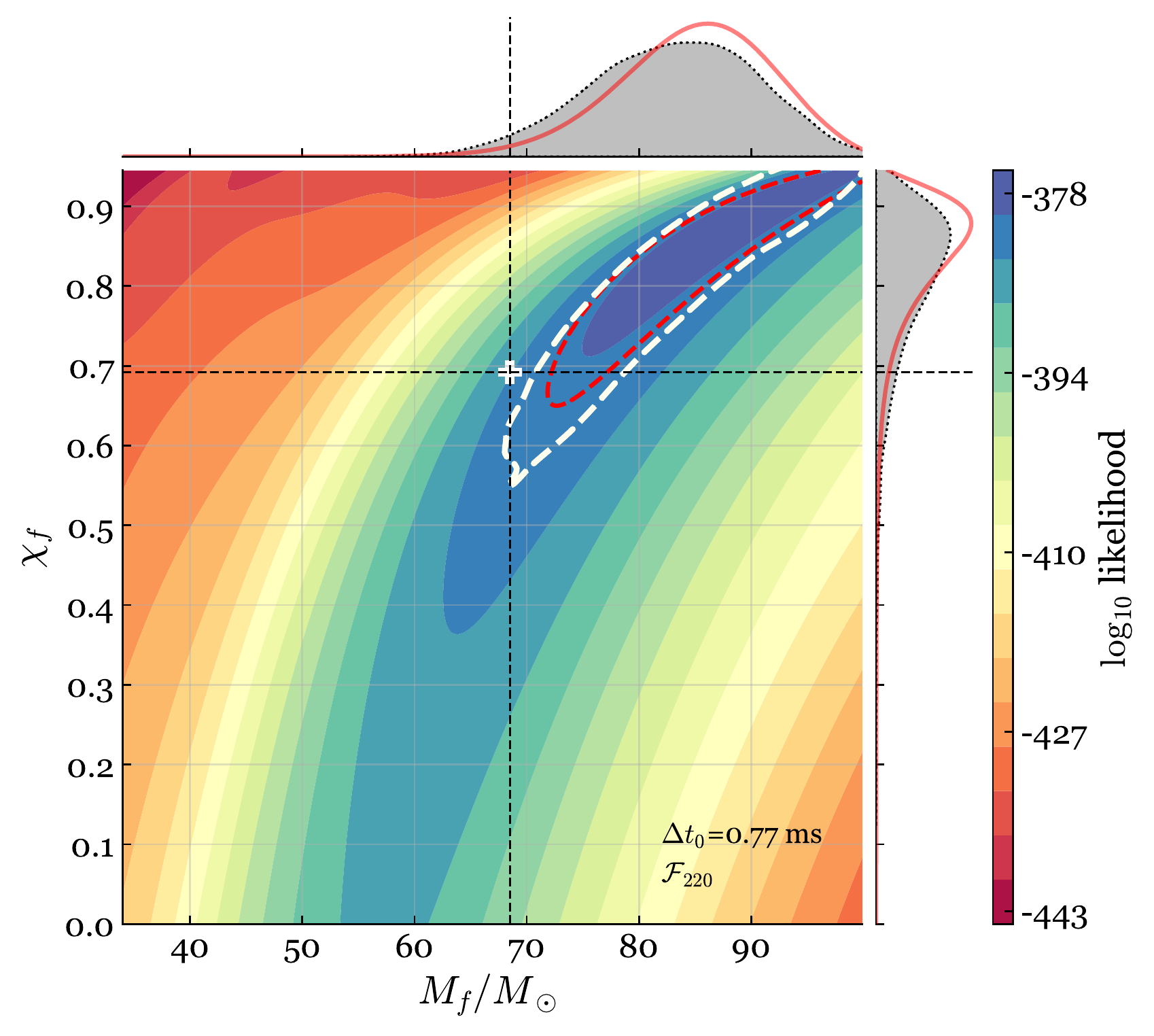}
        \includegraphics[width=\columnwidth,clip=true]{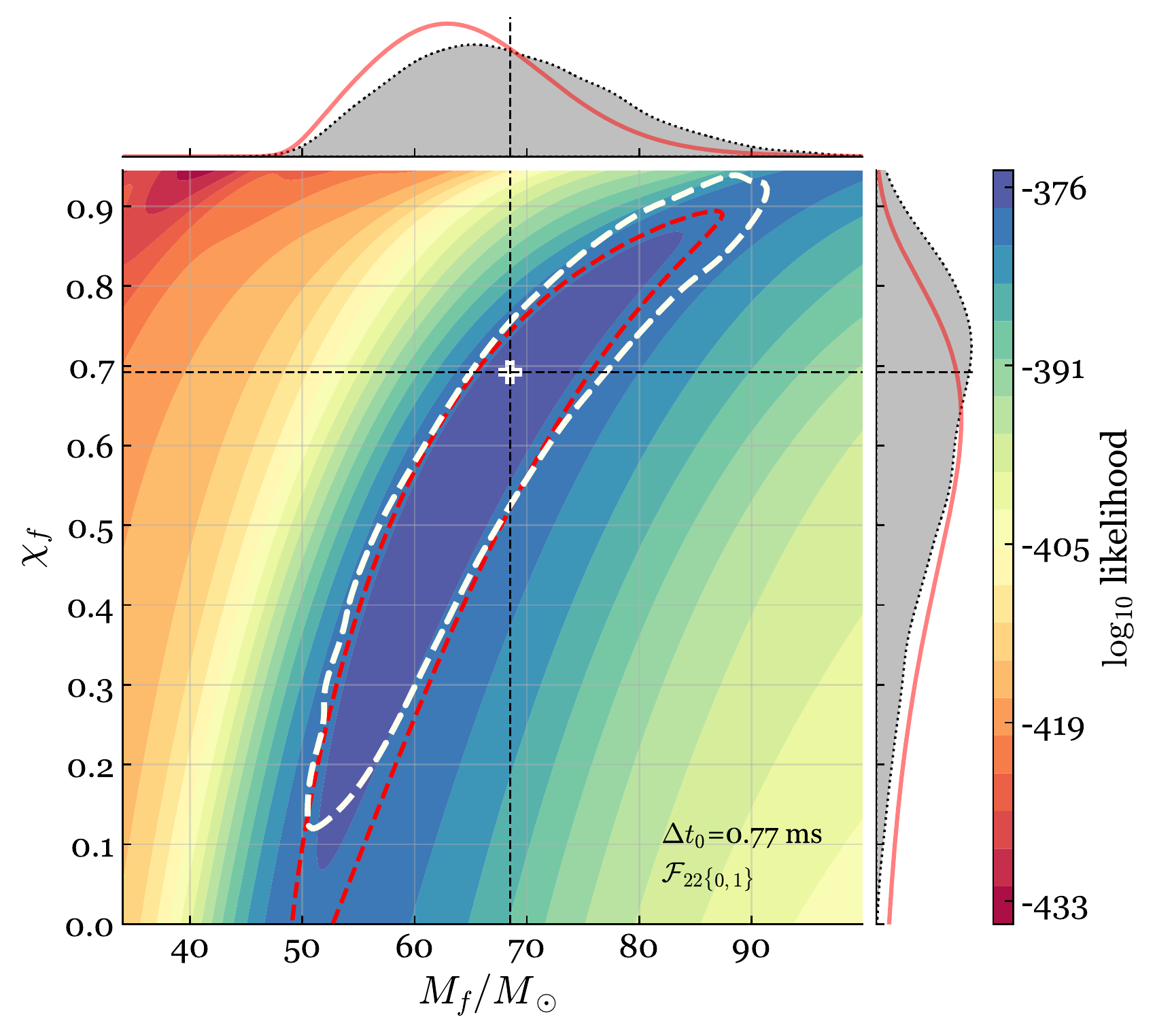} 
  \caption{(Similar to Fig.~\ref{fig:injection1_0305}) Joint posterior distributions of $M_f$ and $\chi_f$ for GW150914 (data collected by the two Advanced LIGO detectors are used). The top and bottom panels represent $\Delta t_0 = 0$ and 0.77 ms, respectively. The left and right panels show results from applying the filter for the fundamental mode only, $\mathcal{F}_{220}$, and the filter $\mathcal{F}_{221}\mathcal{F}_{220}$, respectively. The red-dashed contours display the 90\% credible region by integrating our new joint posterior in Eq.~\eqref{eq:m_chi_posteriors}; and the joint distribution is projected to the individual 1D space of $\chi_f$ and $M_f$ (red curves in side panels), using Eq.~\eqref{eq:likelihood_1d_marginal}. The white plus signs stand for the parameters estimated from the whole IMR waveform. The white dashed contours show the 90\% credible region from the full-RD MCMC approach. The MCMC results are marginalized to the 1D distributions of $M_f$ and $\chi_f$, shown as the gray-shaded regions in side panels.}
 \label{fig:GW150914_contours_1}
\end{figure*}

\begin{figure*}
        \includegraphics[width=\columnwidth,clip=true]{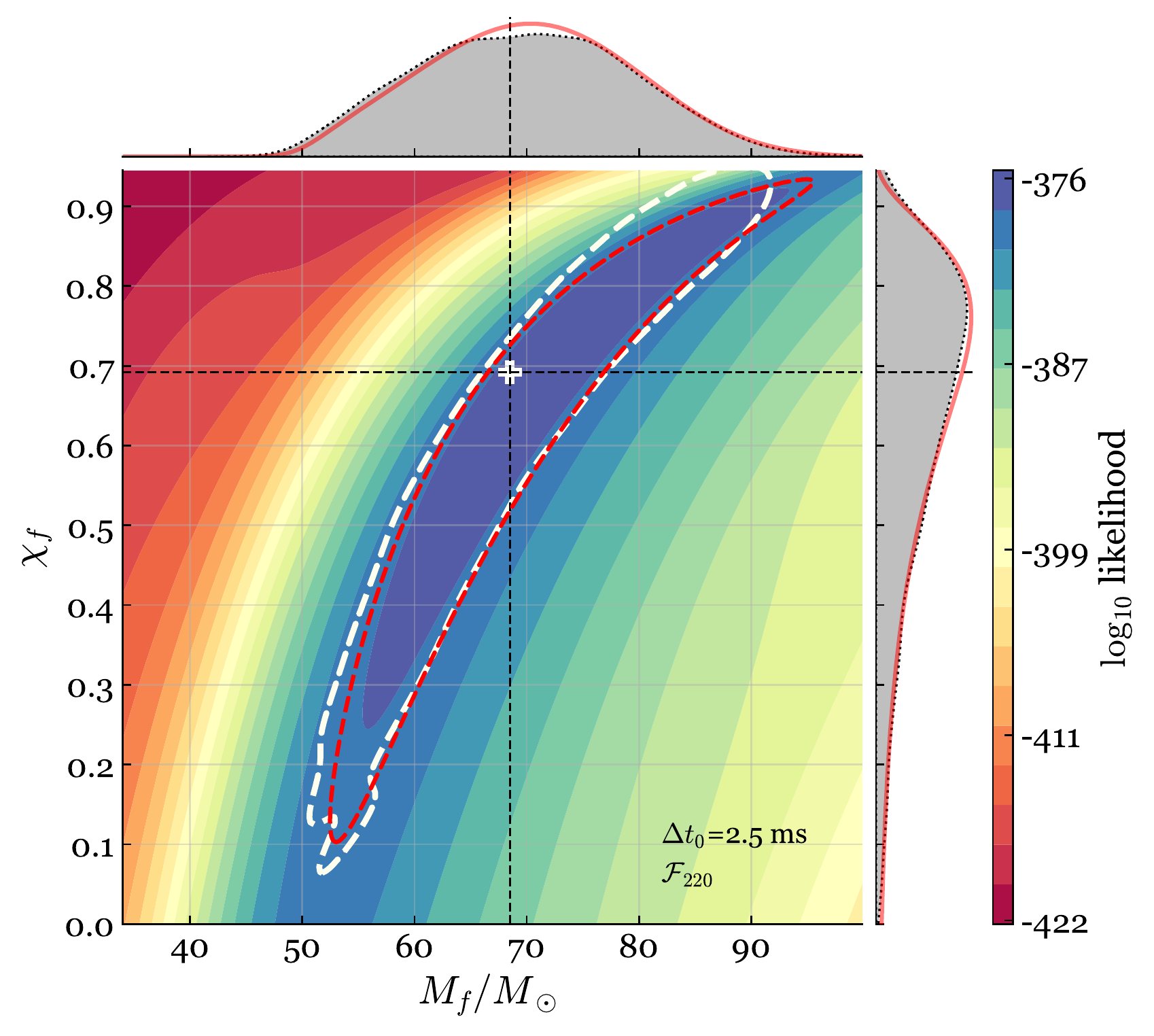}
        \includegraphics[width=\columnwidth,clip=true]{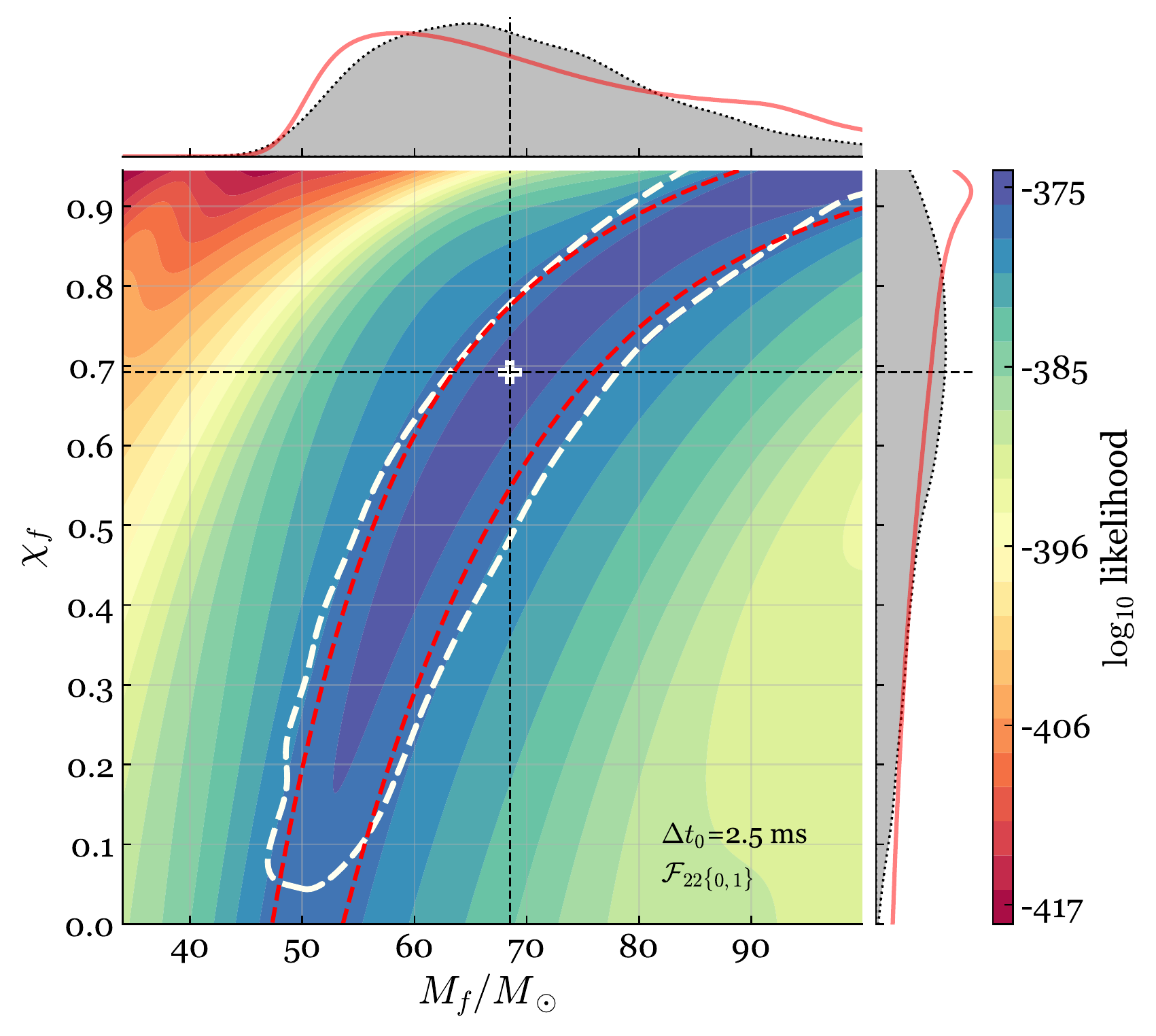}\\
        \includegraphics[width=\columnwidth,clip=true]{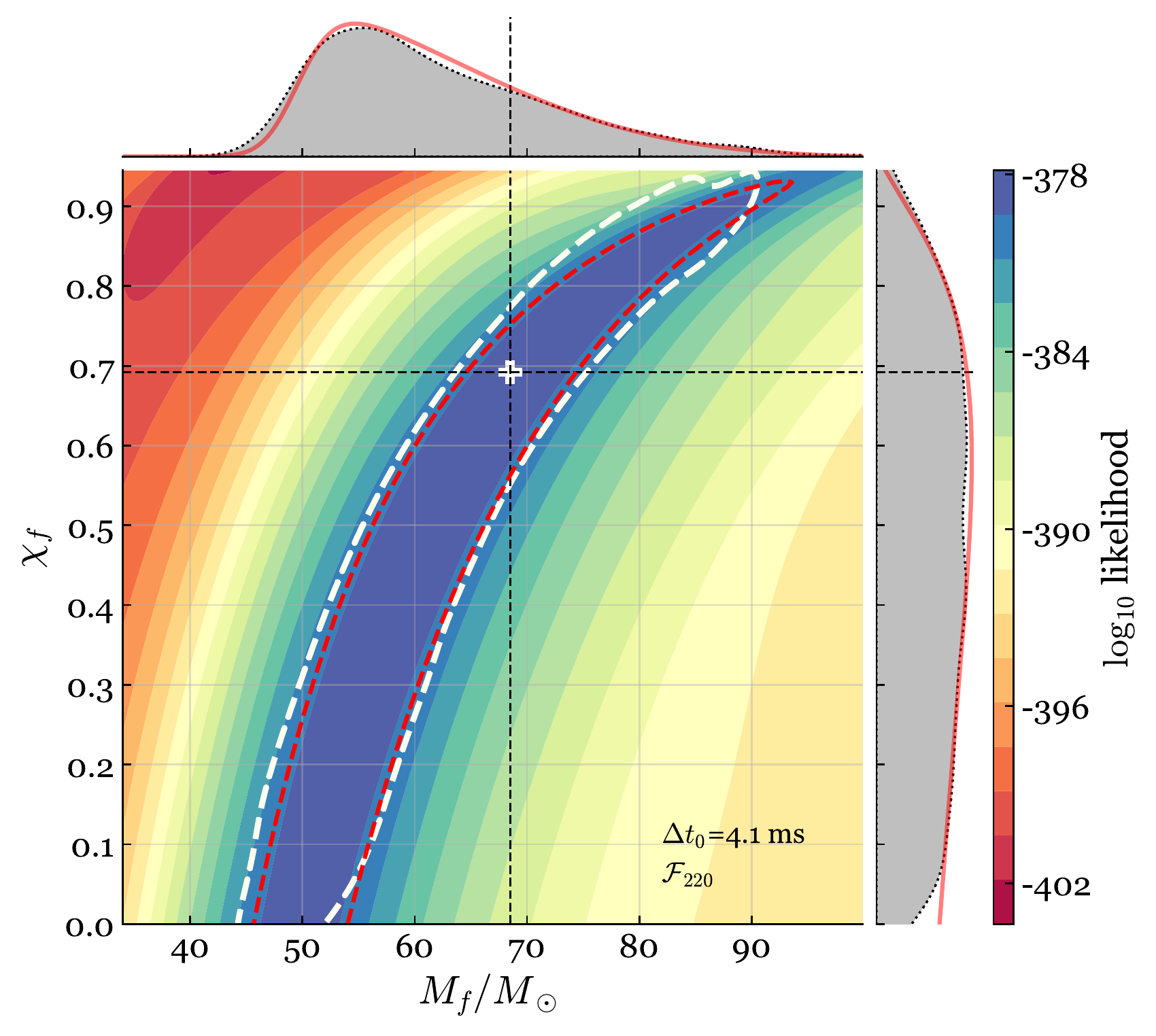}
        \includegraphics[width=\columnwidth,clip=true]{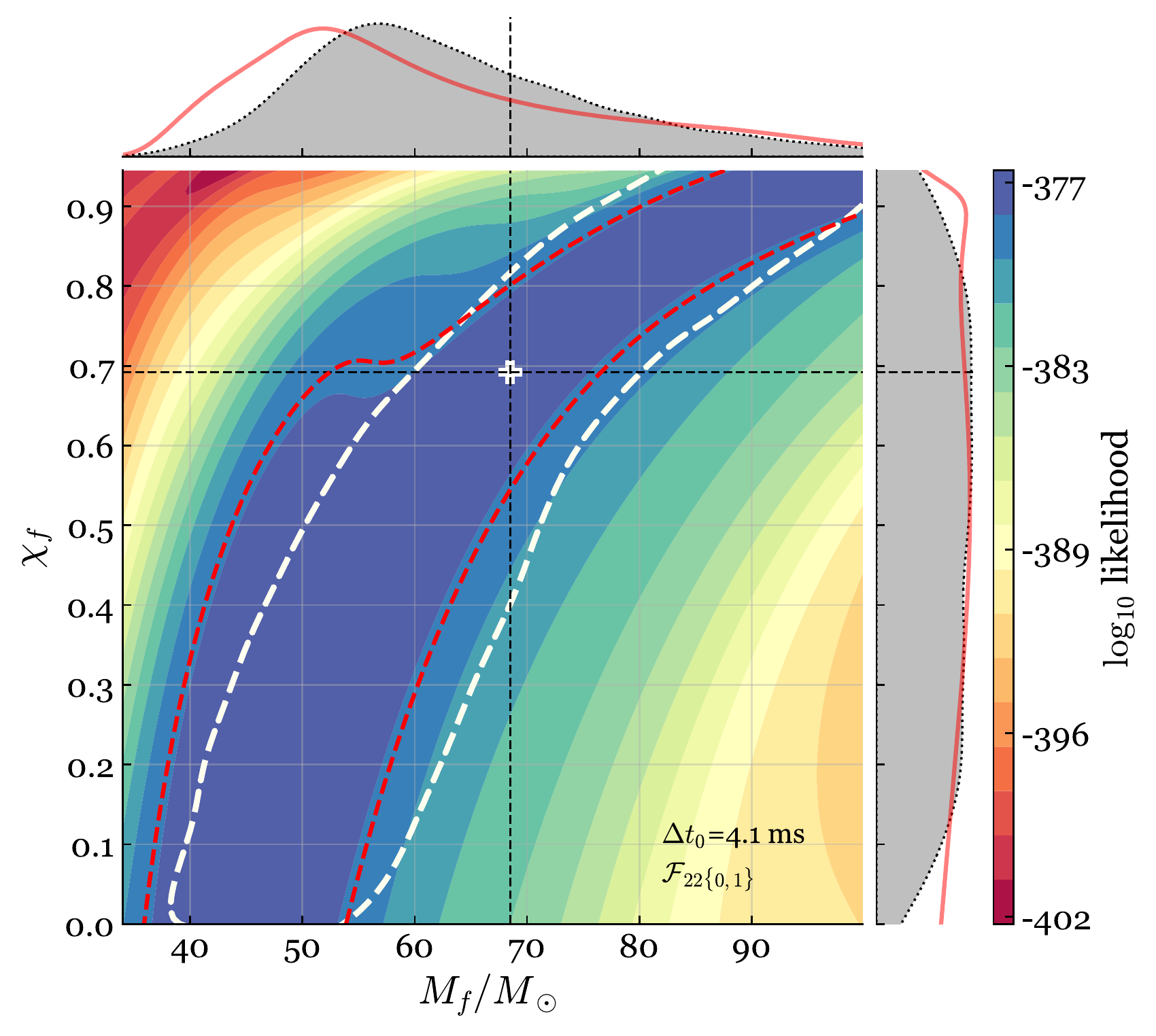}
  \caption{Joint posterior distributions of $M_f$ and $\chi_f$ for GW150914. Fig.~\ref{fig:GW150914_contours_1} continued; more values of $\Delta t_0$ are tested.}
 \label{fig:GW150914_contours_2}
\end{figure*}

\begin{figure*}
        \includegraphics[width=\columnwidth,clip=true]{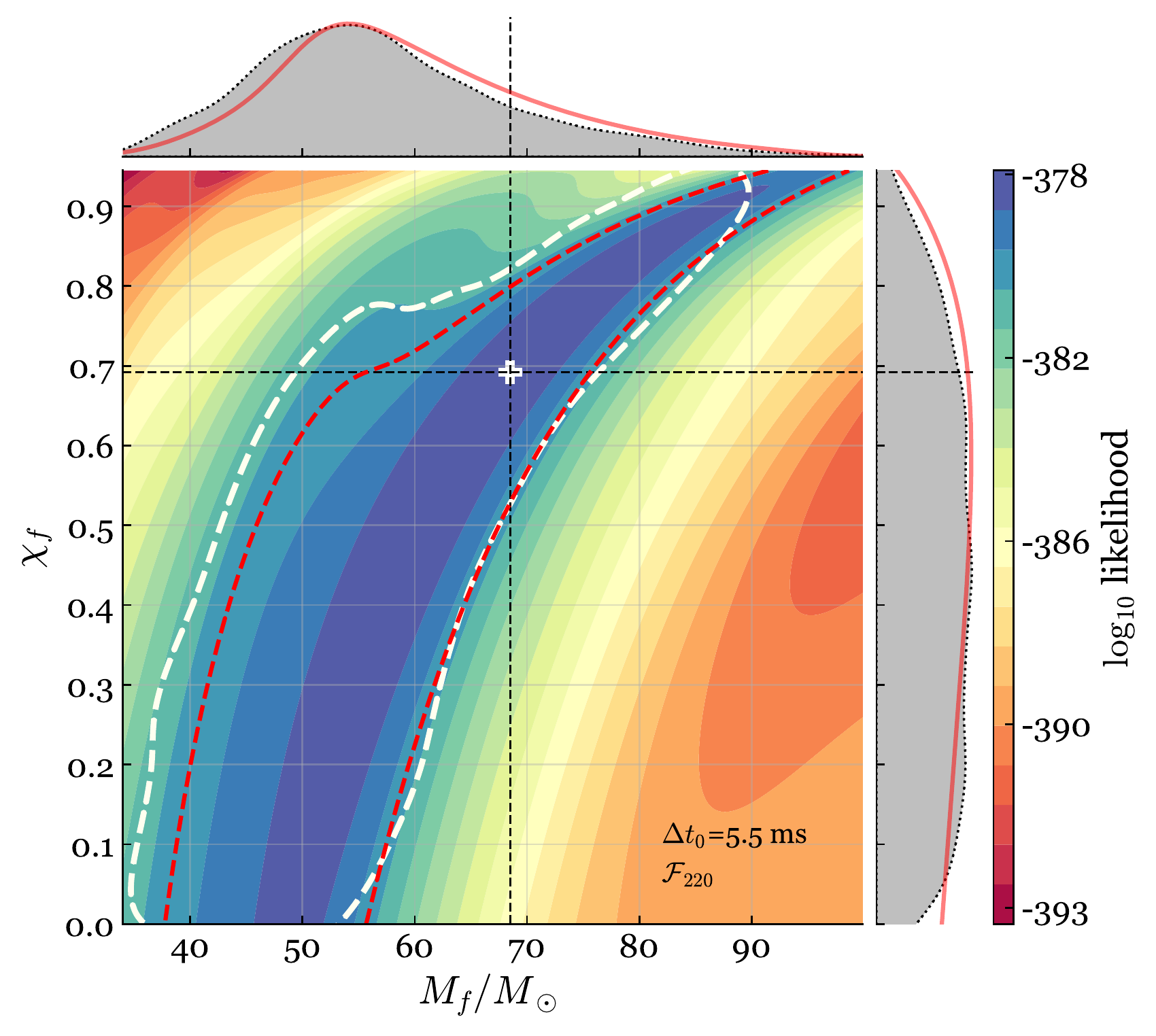}
        \includegraphics[width=\columnwidth,clip=true]{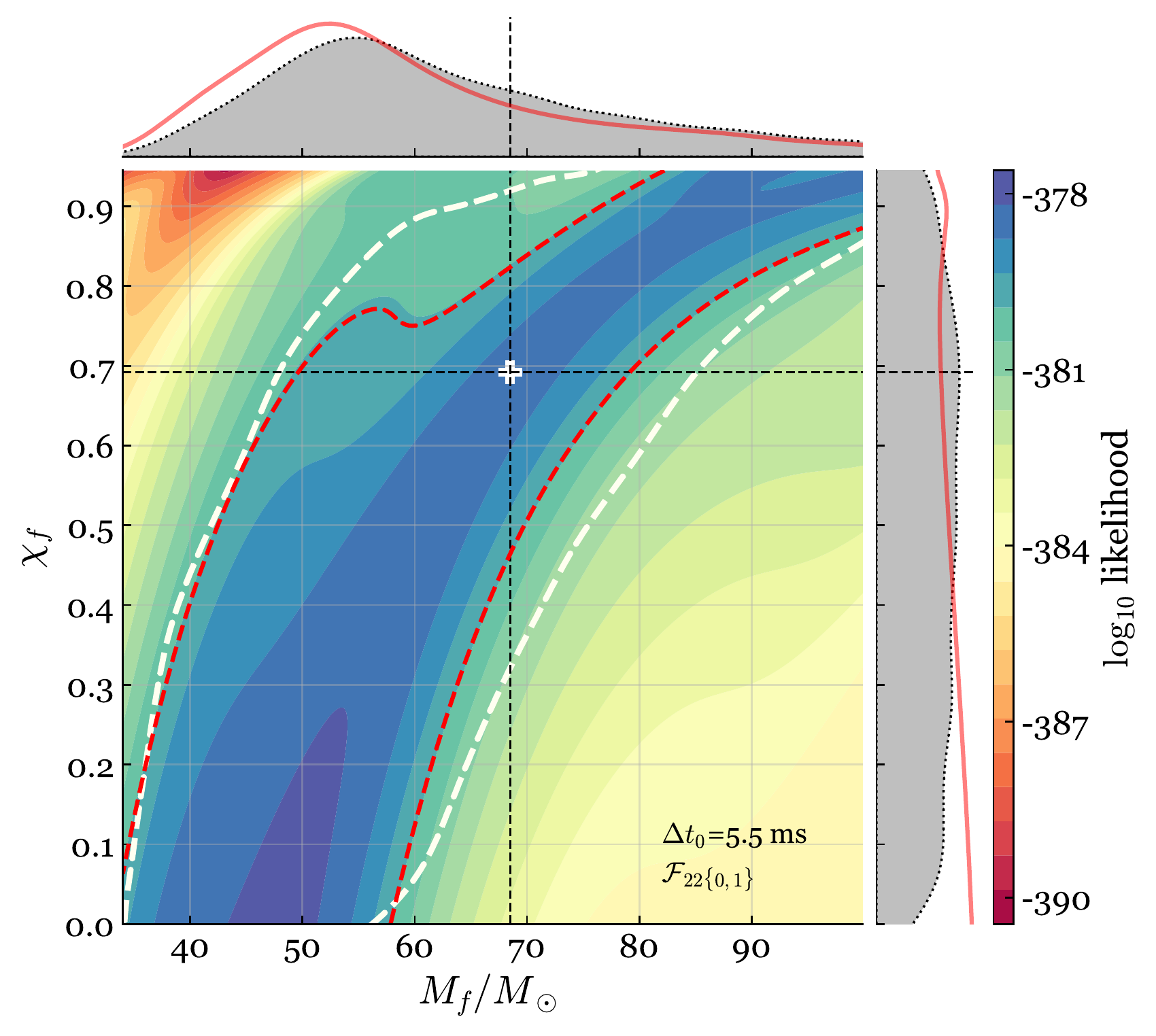}\\
        \includegraphics[width=\columnwidth,clip=true]{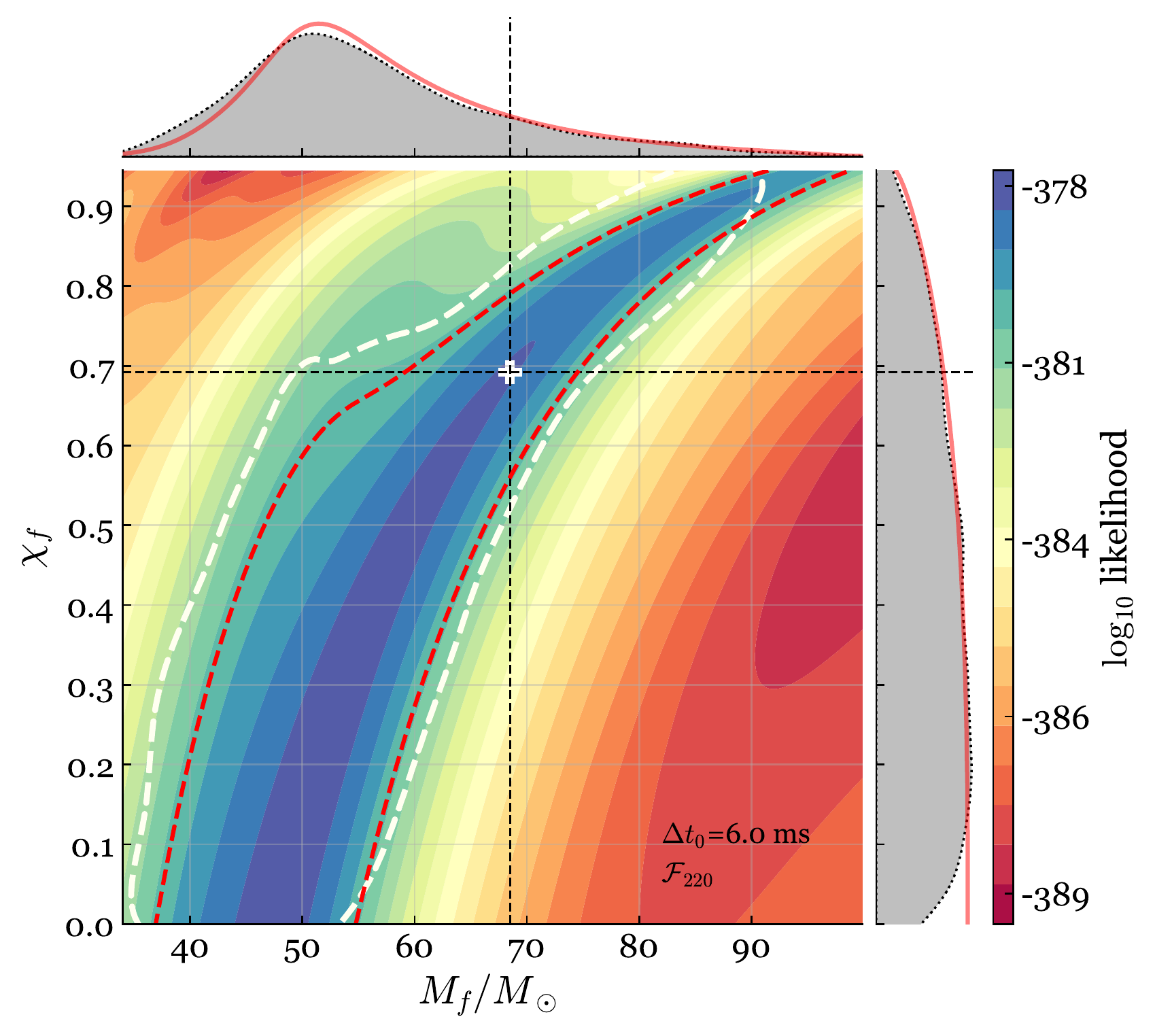}
        \includegraphics[width=\columnwidth,clip=true]{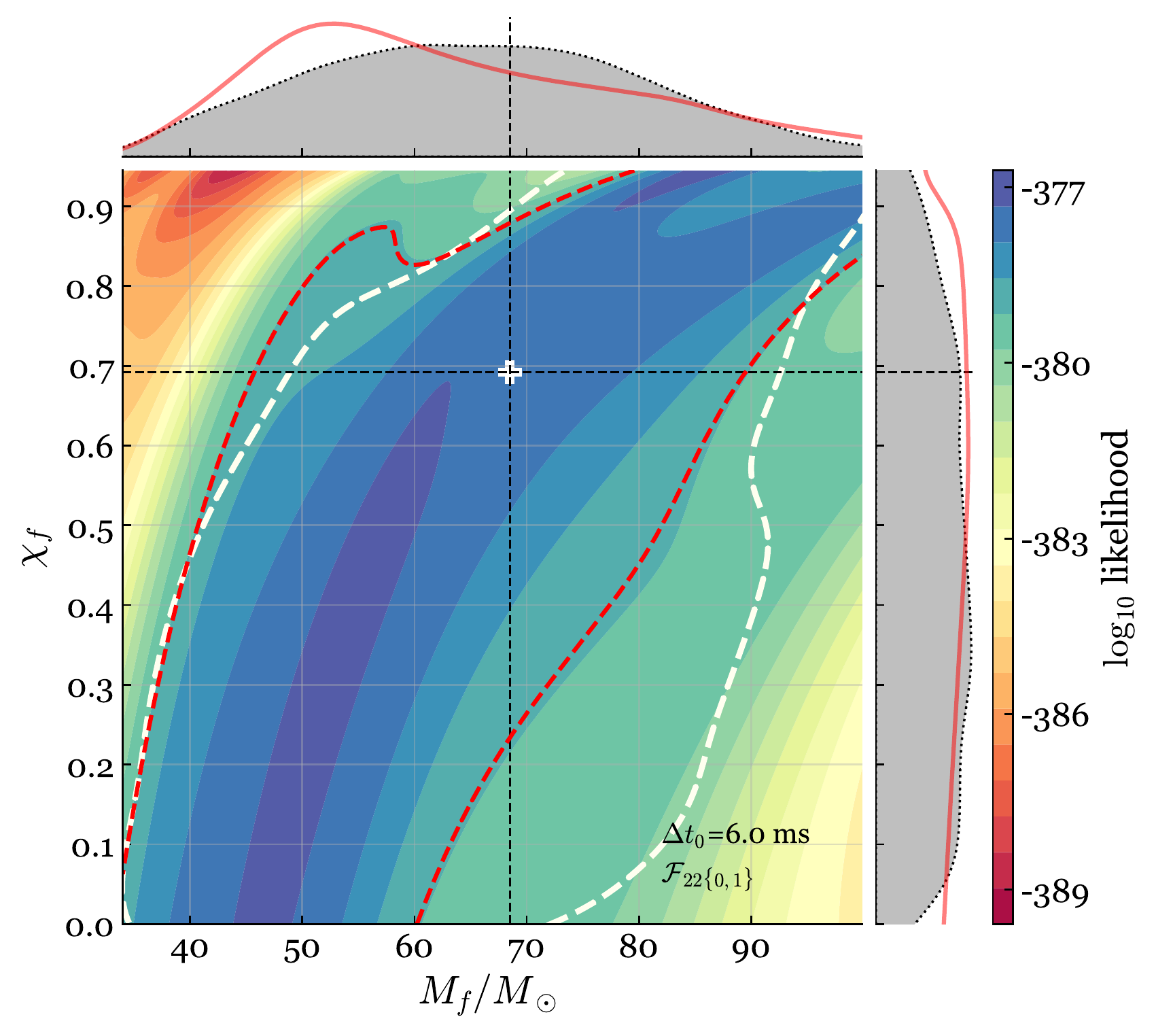}
  \caption{Joint posterior distributions of $M_f$ and $\chi_f$ for GW150914. Fig.~\ref{fig:GW150914_contours_1} continued; more values of $\Delta t_0$ are tested.}
 \label{fig:GW150914_contours_3}
\end{figure*}

\subsection{GW150914}
\label{subsec:GW150914_likelihood}
After studying the injected signal in Sec.~\ref{subsec:injection_likelihood}, we now apply our method to analyzing GW150914 using the data collected in the first observing run of the two Advanced LIGO detectors (Hanford and Livingston) \cite{Abbott:2016blz,LIGOScientific:2018mvr,GW_open_science_center}. We adopt the same procedure described in Sec.~\ref{subsec:impact_detector_noise} to condition the data, and the PSDs are evaluated with 32 s of the conditioned data (see Figs.~\ref{fig:psds_filter_gw150914_han} and \ref{fig:psds_filter_gw150914_liv}). Then, the PSDs are converted to the covariance matrix $C_{ij}$ via Eqs.~\eqref{eq:rho_from_S} and \eqref{eq:C_ij_injection}. Following \cite{Isi:2019aib,Isi:2021iql,ringdown_isi}, we take the inferred GPS time when the signal strain reaches its peak at geocenter, $t_{\rm start}=1126259462.4083$,  and parameterize the analysis start time via $\Delta t_0=t_0-t_{\rm start}$. The information on antenna patterns, polarization, and inclination angles are not needed within our framework, but we do need to time-shift the data to align the signals at the two detectors, based on the sky location of the event, right ascension $\alpha=1.95$ rad and declination $\delta=-1.27$ rad \cite{Isi:2019aib}. To compute the joint posteriors of $(M_f, \chi_f)$ in Eq.~\eqref{eq:m_chi_posteriors}, we fix the width of the ringdown window to $w=0.2$ s, and we consider two types of the rational filter: $\mathcal{F}_{221}\mathcal{F}_{220}$ and $\mathcal{F}_{220}$ (same as the injection study in Sec.~\ref{subsec:injection_likelihood}).

The posterior distributions of $M_f$ and $\chi_f$ at various $\Delta t_0$ are shown in Figs.~\ref{fig:GW150914_contours_1}, \ref{fig:GW150914_contours_2} and \ref{fig:GW150914_contours_3}. The parameter estimation results obtained from the whole IMR signal, $M^{\rm IMR}_{f}=68.5\,M_{\odot}$ and $\chi_{f}^{\rm IMR}=0.69$ \cite{Isi:2019aib}, are marked by the white plus signs. Again, the red-dashed contours show the 90\% credible region evaluated by integrating Eq.~\eqref{eq:m_chi_posteriors}, and the marginalized posterior distributions of $M_f$ and $\chi_f$ are shown as 1D histograms (red curves in side panels). The results are qualitatively similar to the injection study in Sec.~\ref{subsec:injection_likelihood}. At the early stage of ringdown, the ``two-QNM'' results are more consistent with those from the whole IMR analysis, demonstrating the existence of the first overtone. At $\Delta t_0=0.77\,{\rm ms}$, the constraints obtained with $X=1$ [Eq.~\eqref{eq:total_filter_event}] start to degrade because of the first overtone decays. Meanwhile, there still exists a bias in the estimates of $(M_f, \chi_f)$ in the case of $X=0$. This discrepancy between $X=1$ and $X=0$ becomes less significant as we move to $\Delta t_0=2.5\,{\rm ms}$, indicating that the contribution from the first overtone becomes negligible. From this time onward, the constraints of $(M_f, \chi_f)$ are worse when we apply $\mathcal{F}_{221}$ in addition to $\mathcal{F}_{220}$, since there is nearly none first overtone contribution and the extra filter of $\mathcal{F}_{221}$ reduces the amplitude of the fundamental mode, as discussed in Sec.~\ref{subsec:injection_likelihood}. We can see the constraints keep degrading as the ringdown signal decays away.

On the other hand, we also use MCMC to repeat the Bayesian analyses presented in \cite{Isi:2019aib}, in which we assume the system is face-off \cite{Isi:2019aib}, namely $\iota=\pi$. The 90\% credible interval joint posterior distributions are enclosed by the white dashed contours. The corresponding 1D histograms for $M_f$ and $\chi_f$ are represented by the gray-shaded regions. Similar to the injection study in Sec.~\ref{subsec:injection_likelihood}, the MCMC results are consistent with what we obtain from the rational filter.

\begin{figure}[htb]
        \includegraphics[width=\columnwidth,clip=true]{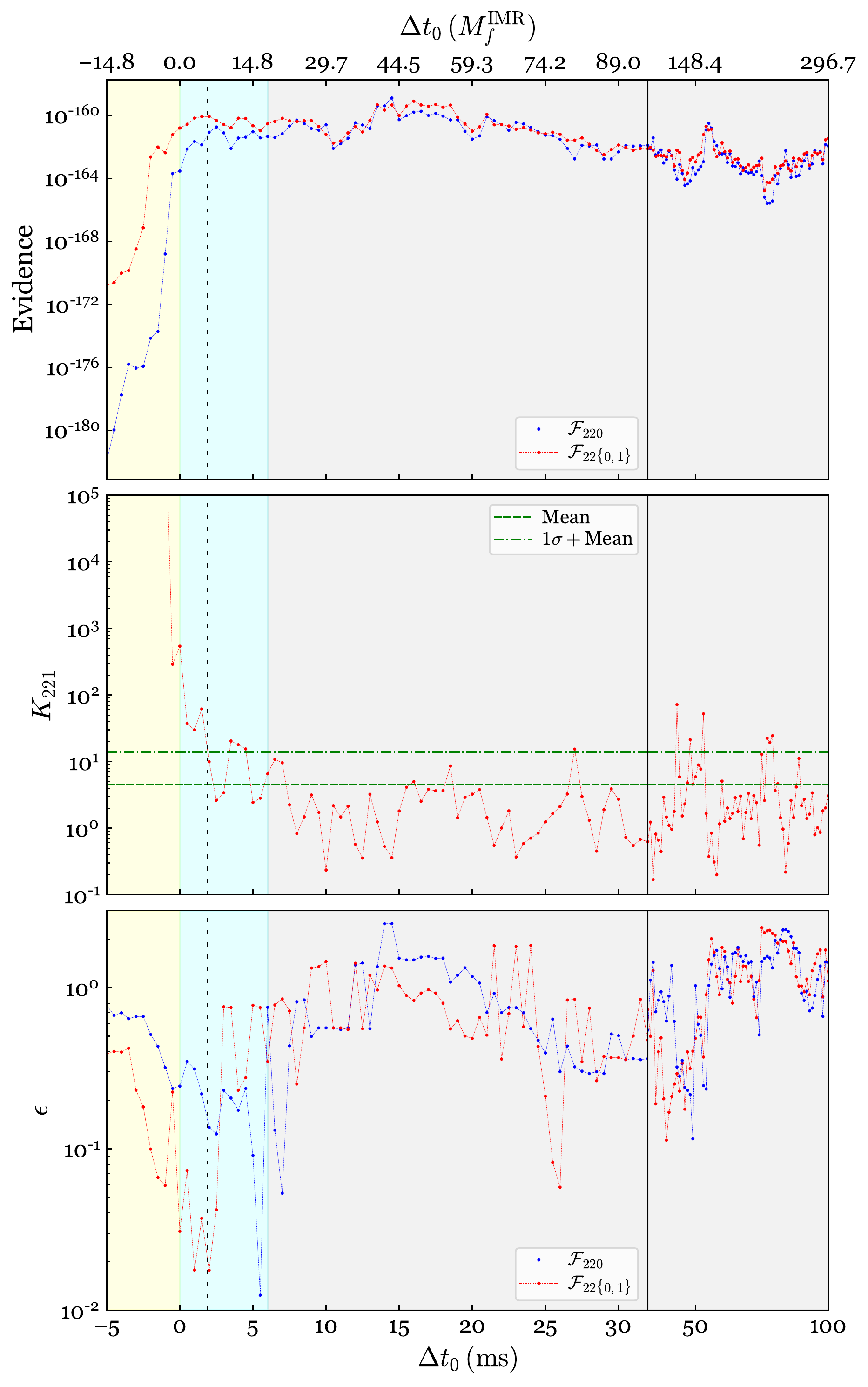}
  \caption{(Similar to Fig.~\ref{fig:injection_likelihood_time}) Model comparison at different $\Delta t_0$ for GW150914. Top: Model evidence as a function of $\Delta t_0$. The blue and red curves indicate the results for applying $\mathcal{F}_{220}$ (clean the fundamental mode only) and $\mathcal{F}_{22\{0,1\}}$ (clean the fundamental mode and the first overtone), respectively. Middle: Bayes factor ($K_{221}$) of the existence of the first overtone over fundamental mode only (red curve). The horizontal dashed and dash-dotted green lines indicate the mean value and the standard deviation within the regime of $\Delta t_0\in [15,100]\, {\rm ms}$, respectively. The red Bayes factor curve intersects the ``$1\sigma+$mean'' line at a time of $\Delta t_0 =1.9$ ms, indicating the time when the first overtone becomes negligible (vertical dashed line). Bottom: Distance ($\epsilon$) of the MAP values of $M_f$ and $\chi_f$ to the values estimated from the whole IMR signal.}
 \label{fig:noval_likelihood_time}
\end{figure}

For more quantitative conclusions, we use Eq.~\eqref{eq:filter_prob} to compute the model evidence $P(d_t|\mathcal{F}_{220},\Delta t_0)$ and $P(d_t|\mathcal{F}_{22\{0,1\}},\Delta t_0)$ as functions of $\Delta t_0$ and show results in the top panel of Fig.~\ref{fig:noval_likelihood_time}. Similar to the injection study, the two evidence curves surge quickly before the time when the signal strain reaches the peak, $\Delta t_0=0$, indicating the onset of the ringdown. The ratio $K_{221}$ [Eq.~\eqref{eq:bayes_overtone}], shown in the middle panel of Fig.~\ref{fig:noval_likelihood_time}, reveals the relative importance of the first overtone. At the peak time $\Delta t_0=0$, the Bayes factor $K_{221}$ is as high as 600. Then its value drops steeply within the first 2 ms. In the case of analyzing this real event, we take the window of $\Delta t_0 \in [15,100]$ ms, a duration when the whole ringdown signal should have decayed, for background estimation. The mean value ($\sim 4.5$) and the mean plus one standard deviation ($\sim 13.8$) of $K_{221}$ computed in the noise-only window are plotted as dashed and dash-dotted horizontal green lines, respectively, in the middle panel of Fig.~\ref{fig:noval_likelihood_time}. The curve of $K_{221}$ intersects with the ``mean $+ 1\sigma$'' and ``mean'' lines at $\Delta t_0=1.9\,{\rm ms}$ and 2.3 ms, respectively. Therefore we can conclude that the first overtone has become negligible around $\Delta t_0 \sim 2\,{\rm ms}$. Indeed, in the first row of Fig.~\ref{fig:GW150914_contours_2}, we see the remnant properties inferred from the fundamental mode alone are consistent with the IMR results at $\Delta t_0=2.5\,{\rm ms}$, a time when the first overtone is deemed vanishing. 

Finally, we use Eq.~\eqref{eq:MAP_true_distance_eps} to evaluate the MAP estimations of $(M_f,\chi_f)$ and the distance to the IMR results $(M_f^{\rm IMR}=68.5,\chi^{\rm IMR}_f=0.69)$. As shown in the bottom panel of Fig.~\ref{fig:noval_likelihood_time}, the results from the ``two-QNM'' filter are better than those from the fundamental-mode-only filter by one order of magnitude right after the peak time $\Delta t_0\sim 0$. Beyond the time when the first overtone mostly decays ($\Delta t_0\sim 2\,{\rm ms}$), the accuracy of parameter estimation using the $\mathcal{F}_{22\{0,1\}}$ filter significantly degrades. Regarding the results from using the ``one-QNM'' filter $\mathcal{F}_{220}$, $\epsilon$ surges after $\Delta t_0 \sim 6$ ms, indicating the time when the fundamental mode also disappears.

\begin{figure*}
        \subfloat[Fundamental mode ($\omega_{220}$), $\Delta t_0=1.5\,{\rm ms}$\label{fig:low_res_injection_A0}]{\includegraphics[width=\columnwidth,clip=true]{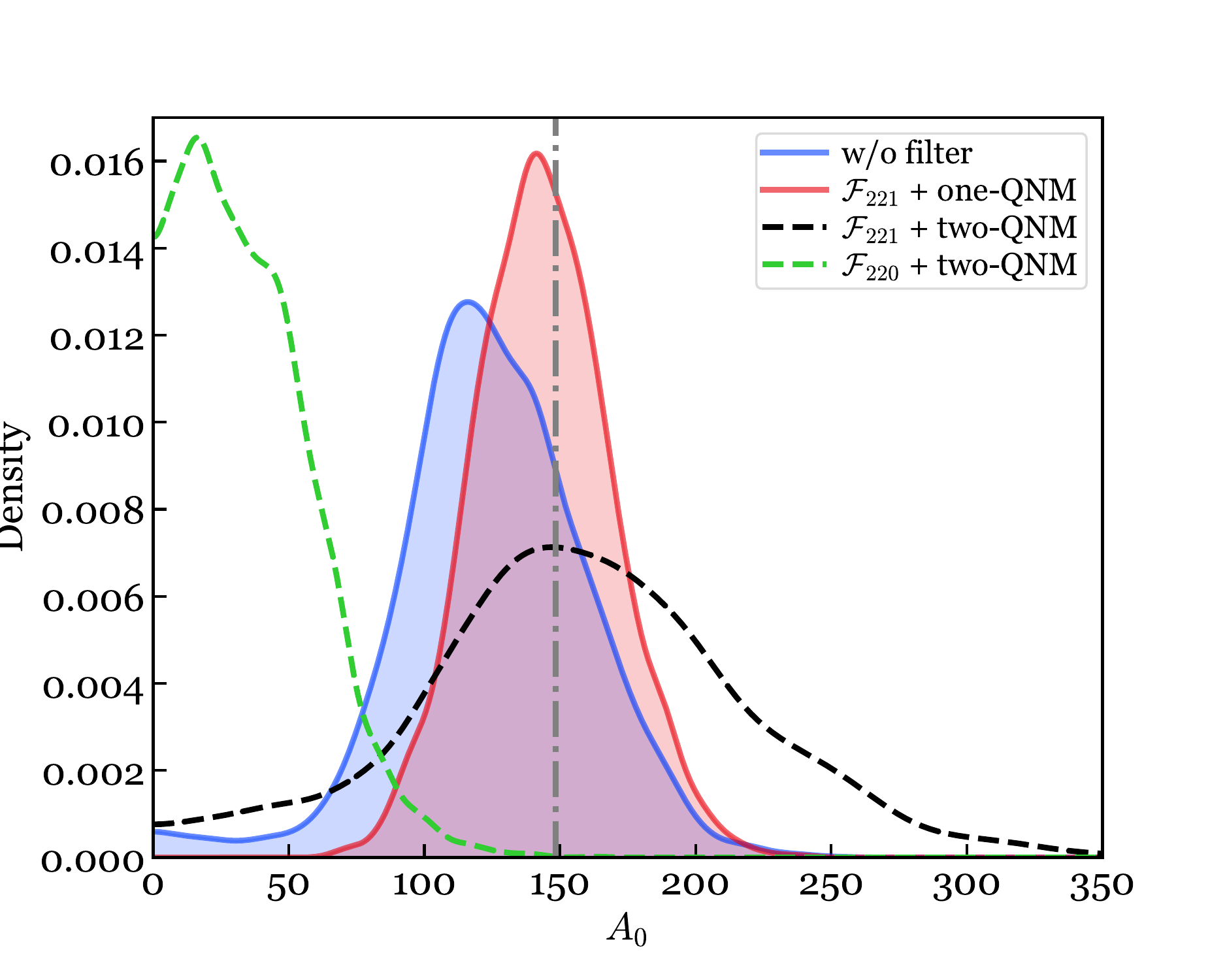}}
        \subfloat[First overtone ($\omega_{221}$), $\Delta t_0=1.5\,{\rm ms}$\label{fig:low_res_injection_A1}]{\includegraphics[width=\columnwidth,clip=true]{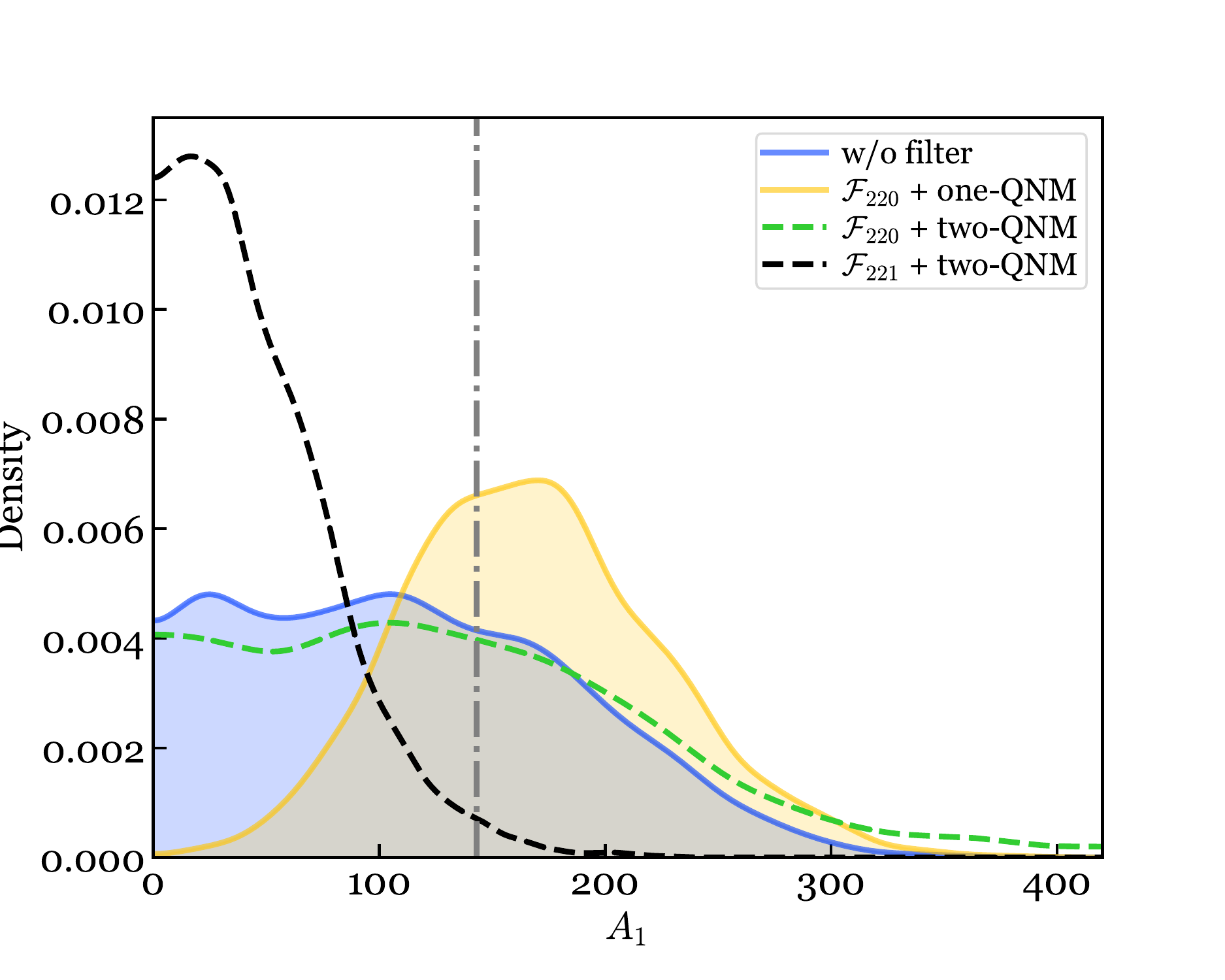}}  \\
        \subfloat[$\Delta t_0=1.5\,{\rm ms}$\label{fig:low_res_1.5_mchi_ligo}]{\includegraphics[width=\columnwidth,clip=true]{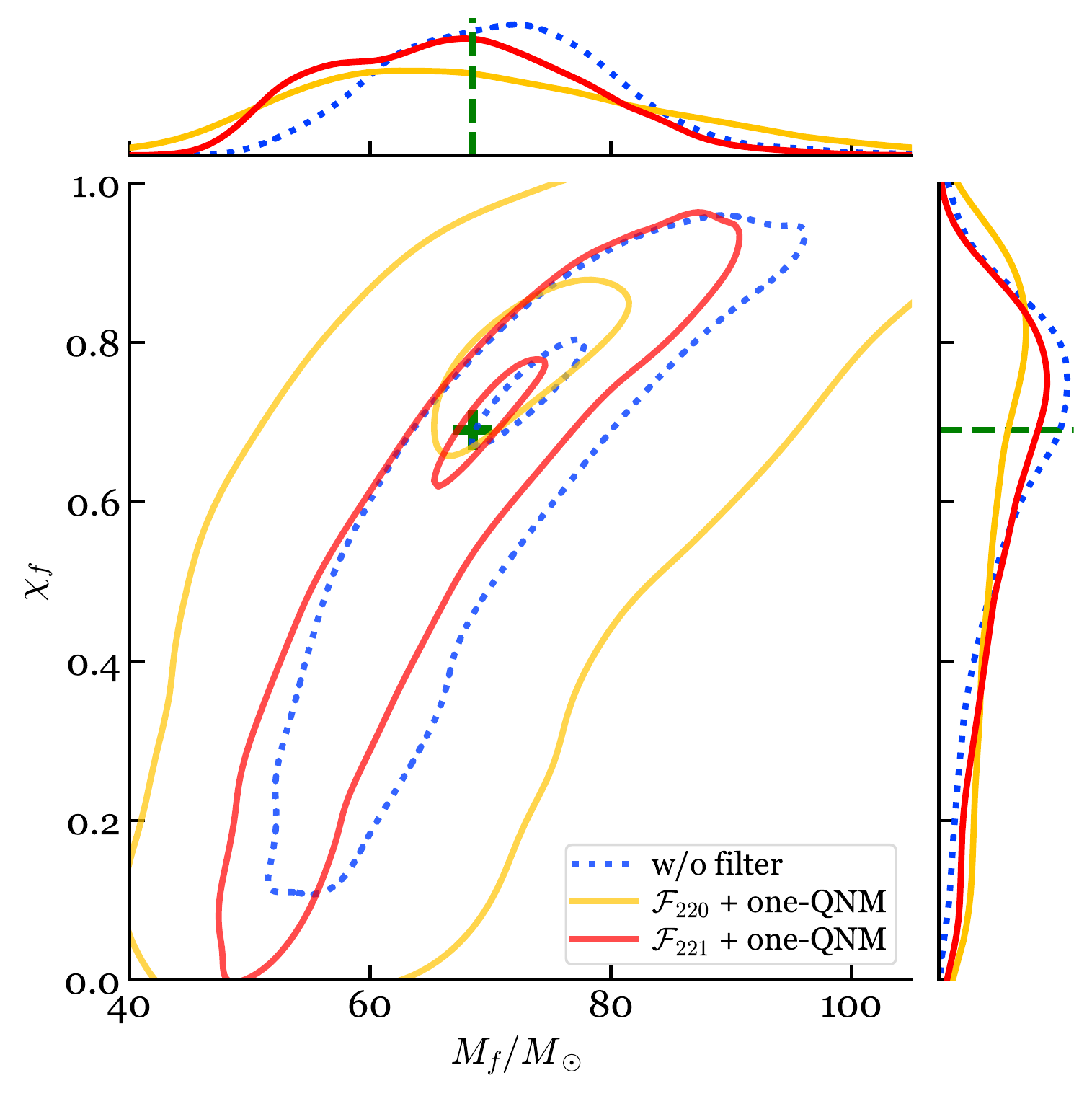}}
        \subfloat[$\Delta t_0=1.0\,{\rm ms}$\label{fig:low_res_1.0_mchi_ligo}]{\includegraphics[width=\columnwidth,clip=true]{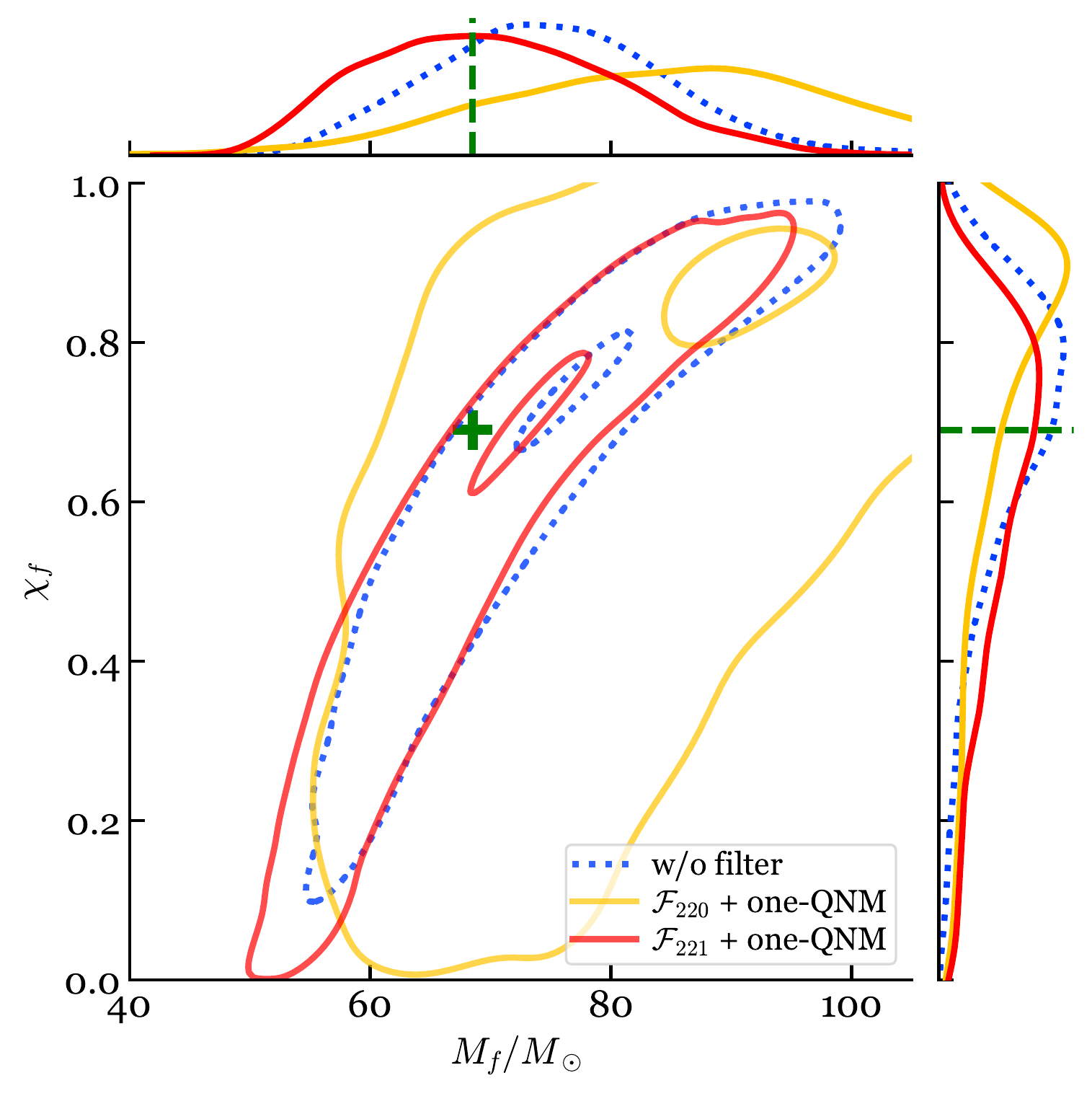}}
  \caption{Estimates of the mode amplitudes and BH properties for the injected signal using the mixed approach. The top panels display the posterior distributions of (a) the fundamental mode amplitude $A_0$, and (b) the first overtone amplitude $A_1$, evaluated at $\Delta t_0 = 1.5$ ms under various filtering conditions. The blue-shaded distributions are obtained via the full-RD MCMC method (without applying the filter). The green dashed curves correspond to removing the $\omega_{220}$ mode using $\mathcal{F}_{220}$ first and fitting the data with a two-QNM ($\omega_{220}$ and $\omega_{221}$) signal template. The same $\mathcal{F}_{220}$-filtered data are also fitted with the one-QNM ($\omega_{221}$) signal template, resulting in the $A_1$ distribution shown in yellow in (b). Similarly, the black dashed curves correspond to removing the $\omega_{221}$ mode using $\mathcal{F}_{221}$ first and fitting the data with a two-QNM ($\omega_{220}$ and $\omega_{221}$) signal template. The $\mathcal{F}_{221}$ filtered data are then fitted with the one-QNM ($\omega_{220}$) signal template, resulting in the $A_0$ distribution shown in red in (a). The two vertical lines indicate the true values of $A_0=148$ and $A_1=143$, computed from the NR waveform. The bottom panels show the posterior distributions of $M_f$ and $\chi_f$ estimated at (c) $\Delta t_0 = 1.5$ ms, and (d) $\Delta t_0 = 1.0$ ms. The yellow curves indicate the results obtained by fitting the $\omega_{220}$-cleaned data with a $\omega_{221}$-only template. The red curves are the results obtained by fitting the $\omega_{221}$-cleaned data with a $\omega_{220}$-only template. The blue dashed curves are the results from the full-RD MCMC analysis without applying any filter. The two contours in each color correspond to the 90\% and 10\% credible intervals.
  }
 \label{fig:injection_amplitude}
\end{figure*}

\section{A Mixed approach to black-hole spectroscopy} 
\label{sec:mixed_appraoch}

We have demonstrated that the rational filter provides a powerful tool for black-hole spectroscopy by cleaning any given QNMs from the data and evaluating a simple and efficient likelihood function for remnant mass and spin [Eq.~\eqref{eq:log-p}]. One limitation of this new method is that we do not obtain information of mode amplitudes and phases in the analysis. By combining our filters with the conventional MCMC method, we could take advantage of both and build another mixed approach for BH spectroscopy: after removing a subset of the QNMs using the rational filters, we analyze the filtered data with a ringdown model that consists of the uncleaned QNMs, using the MCMC approach. The advantage of this hybrid approach is as follows. The conventional full-RD MCMC analysis (without filter) for subdominant QNMs is likely to be biased when the strongest mode is present, especially at low-SNR regime.
The mixed approach allows us to clean the most dominant mode and thus eliminate potential impacts from it when carrying out parameter estimations using the MCMC method. 
In addition, we can carry out a consistency check by comparing the mode amplitudes and remnant properties inferred from the subdominant QNMs (after the strongest mode is cleaned by the filter) to those obtained from the unfiltered data (when the strongest mode is present). 
When we observe events with high ringdown SNRs, we can even clean a set of stronger modes and study the remaining weak ones. In this way, we are able to test the no-hair theorem from a new perspective.

Below, we detail the analysis procedure and results using this mixed approach. In Sec.~\ref{sec:mixed_injection}, we study the NR signal injected into Gaussian noise (same as the one in Sec.~\ref{subsec:injection_likelihood}). In Sec.~\ref{sec:mixed_gw150914}, we analyze GW150914.

\subsection{NR waveform injection}
\label{sec:mixed_injection}
We first analyze the injection of the NR waveform. We describe the estimates of mode amplitudes in Sec.~\ref{subsubsec:mode_amplitude} and the remnant properties in Sec.~\ref{subsubsec:remnant_properties}. Further discussion about the features seen in the mixed approach is provided in Sec.~\ref{subsubsec:further_discussions}.
\subsubsection{Estimate mode amplitudes}
\label{subsubsec:mode_amplitude}
With the NR signal injected into Gaussian noise, we first choose a start time for the analysis, $\Delta t_0=1.5 \,{\rm ms}=4.2M_f$, and use the conventional full-RD MCMC method to fit the unfiltered simulation data with a two-QNM ($\omega_{220}$ and $\omega_{221}$) ringdown template. The analysis is performed with the Python package $\textsf{ringdown}$ \cite{ringdown_isi,Isi:2021iql}. The posterior distributions of the amplitudes of the fundamental mode, $A_0$, and the first overtone, $A_1$, are plotted as the blue shaded regions in Figs.~\ref{fig:low_res_injection_A0} and \ref{fig:low_res_injection_A1}. Meanwhile, we compute what the values of $A_0$ and $A_1$ should be in the injected signal by decomposing the NR waveform (the $l=m=2$ harmonic) 
into a superposition of the fundamental mode and the first overtone with a least-square fit. Here we include up to the first overtone for the least-square fit, to be consistent with the templates used in the MCMC analysis, even though Giesler \etal \cite{Giesler:2019uxc} points out more overtones are needed to model ringdown at such an early stage ($\Delta t_0=1.5 \,{\rm ms}=4.2M_f$). The lack of higher overtones in the least-square fit leads to a bias in the estimates of the mode amplitudes. Nevertheless, it is a fair comparison between the MCMC results and the ``should-be'' values (vertical dash-dotted lines in Figs.~\ref{fig:low_res_injection_A0} and \ref{fig:low_res_injection_A1}) obtained from the least-square fit. We find the MCMC posteriors are consistent with the values indicated by the vertical lines, $A_0=148$ and $A_1=143$. In fact, the same feature has been pointed out by Finch and Moore (see Fig.~7 and discussions in Sec.~III B in \cite{Finch:2022ynt}). We provide more detailed discussions in Sec.~\ref{subsubsec:further_discussions}.

\begin{table}[tbh]
    \centering
    \setlength{\tabcolsep}{15pt}
	\renewcommand\arraystretch{1.2}
    \caption{Combinations of filters and fitting templates for the mixed approach. We have two choices of the filter: $\mathcal{F}_{220}$ and $\mathcal{F}_{221}$, and two choices of the fitting template: two-QNM $(\omega_{220}\&\omega_{221})$ template, ignorant of mode cleaning, and one-QNM template for the remaining mode.}
    \begin{tabular}{c c c} \hline\hline
\diagbox[width=10em]{Filter}{Template} & two-QNM & one-QNM \\ \hline
$\mathcal{F}_{220}$ & $\omega_{220}\, \&\, \omega_{221}$ & $\omega_{221}$  \\ \hline
$\mathcal{F}_{221}$ & $\omega_{220}\, \&\, \omega_{221}$ & $\omega_{220}$   \\ \hline\hline
     \end{tabular}
     \label{table:strategy}
\end{table}

We then use the mixed approach. There are four options from the combinations of the two choices of the filters and two choices of the fitting templates (see Table~\ref{table:strategy}). We can choose to clean the fundamental mode (the first overtone) by applying the filter $\mathcal{F}_{220}$ ($\mathcal{F}_{221}$). After the filtering, we also have two choices of the ringdown template to fit the data and run MCMC: we can (a) continue to use the two-QNM model, assuming both modes exist in the data and we have no knowledge of the mode cleaning (b) use a single-mode template for the remaining QNM. We first apply the filter $\mathcal{F}_{220}$, built from the true remnant mass and spin, to remove the fundamental mode. Then we use the two-QNM template to run MCMC against the filtered data. The posteriors of $A_0$ and $A_1$ are plotted as the green dashed curves in Figs.~\ref{fig:low_res_injection_A0} and \ref{fig:low_res_injection_A1}, respectively. After applying $\mathcal{F}_{220}$, it is expected that there is no $\omega_{220}$ component left in the filtered data. Indeed, we see the distribution of $A_0$ is pushed close to $0$, demonstrating that the fundamental mode no longer exists in the data. By contrast, the posterior distribution of $A_1$ is only slightly impacted by the filtering (compare the green dashed curve to the blue-shaded region in Fig.~\ref{fig:low_res_injection_A1}). We emphasize that, as discussed in Sec.~\ref{subsec:impact_irre_qnm}, the amplitude of the first overtone is reduced by the filter $\mathcal{F}_{220}$ by a factor of $B_{220}^{221}=2.053$ [Eq.~\eqref{eq:filter_on_other_qnm}]. In Fig.~\ref{fig:low_res_injection_A1}, the green dashed curve is obtained by multiplying the original $A_1$ distribution from MCMC by a factor of $B_{221}^{220}=2.053$, so that we can make a fair comparison to the blue distribution. On the other hand, we also fit the filtered data ($\omega_{220}$ component is cleaned) with the single-QNM model, composed of the first overtone alone, which gives the distribution shown in yellow in Fig.~\ref{fig:low_res_injection_A1}. The estimate is more constrained than that from the two-QNM model, and the MAP value is closer to the ``injected'', although biased, mode amplitude (the vertical dash-dotted line).

Similarly, we apply $\mathcal{F}_{221}$ to the original simulation data to clean the first overtone. Fitting with the two-QNM template, the posteriors of $A_0$ and $A_1$ are plotted as black dashed curves in Figs.~\ref{fig:low_res_injection_A0} and \ref{fig:low_res_injection_A1}. This time, $A_1$ is consistent with 0, as expected; whereas the MAP value of $A_0$ is mildly impacted [again, after multiplying the reduction factor $B^{221}_{220}=2.053$ to the original distribution output by MCMC; see Eq.~\eqref{eq:filter_on_other_qnm}]. If we use the one-QNM template of the fundamental mode, the estimated $A_0$ (the red shaded region in Fig.~\ref{fig:low_res_injection_A0}) is more constrained, and the MAP value is closer to the injected mode amplitude (the vertical dash-dotted line).


\subsubsection{Estimate remnant properties}
\label{subsubsec:remnant_properties}
We now estimate the remnant properties ($M_f$ and $\chi_f$) after a certain mode is cleaned. Here we use the one-QNM template to fit the filtered data and show the parameter estimation results obtained from MCMC in Fig.~\ref{fig:low_res_1.5_mchi_ligo} (at $\Delta t_0=1.5\,{\rm ms}=4.2M_f$, to be consistent with Figs.~\ref{fig:low_res_injection_A0} and \ref{fig:low_res_injection_A1}), i.e., we fit for the first overtone after applying $\mathcal{F}_{220}$ (yellow) and fit for the fundamental mode after applying $\mathcal{F}_{221}$ (red). 
The two contours for each case correspond to the 90\% and 10\% credible intervals. For comparison, the estimates obtained by the full-RD MCMC method without applying the filters are plotted as blue dashed contours. The green plus sign stands for the true values. The posterior distributions obtained solely from the first overtone (yellow) are still informative, consistent with the results from the fundamental mode (red) and the full-RD MCMC approach (blue), albeit less constrained. 

As we show in Sec.~\ref{subsubsec:mode_amplitude}, there is inevitably a bias in the estimates of the mode amplitudes at an early time of ringdown $(\Delta t_0=1.5\,{\rm ms}=4.2M_f)$ due to the lack of higher overtones in the model.
In Fig.~\ref{fig:low_res_1.5_mchi_ligo}, however, we see the remnant mass and spin inferred from the fundamental mode (red), the first overtone (yellow), and both modes (blue) are all consistent with the true NR values, indicating that the constraints of $M_f$ and $\chi_f$ are less impacted by the residuals contributed by higher overtones and even the merger signal (if there is any). More details are discussed in Sec.~\ref{subsubsec:further_discussions}. On the other hand, as shown in Fig.~\ref{fig:low_res_1.0_mchi_ligo}, we do see the remnant properties are less consistent with the true value at an earlier time $\Delta t_0=1.0\,{\rm ms}=2.8M_f$. In particular, the constraints from the first overtone (yellow) deviate more significantly from the true value than the other two estimates.

Finally, we note that in terms of estimating the remnant properties ($M_f$ and $\chi_f$) using the filtered waveform when certain modes are cleaned, e.g., Figs.~\ref{fig:low_res_1.5_mchi_ligo} and \ref{fig:low_res_1.0_mchi_ligo}, the results can also be obtained by purely using the filters, instead of running the MCMC analysis. We provide the details in Appendices \ref{app:sec:varying_filter} and \ref{app:sec:devitation_from_kerr}.

\subsubsection{Further discussions}
\label{subsubsec:further_discussions}
In Secs.~\ref{subsubsec:mode_amplitude} and \ref{subsubsec:remnant_properties}, we demonstrate the statistical significance of the first overtone in the injected ringdown signal. 
Given that our two-QNM fitting is carried out at an early time ($\Delta t_0=1.5\,{\rm ms}=4.2M_f$), criticisms might be raised since we do not include higher overtones at such an early time close to the signal strain peak \cite{Giesler:2019uxc}. How do we know the results are not biased by other residual effects? We discuss this from two aspects: (a) estimates of the remnant properties, and (b) estimates of the mode amplitudes.

\begin{figure}[htb]
        \includegraphics[width=\columnwidth,clip=true]{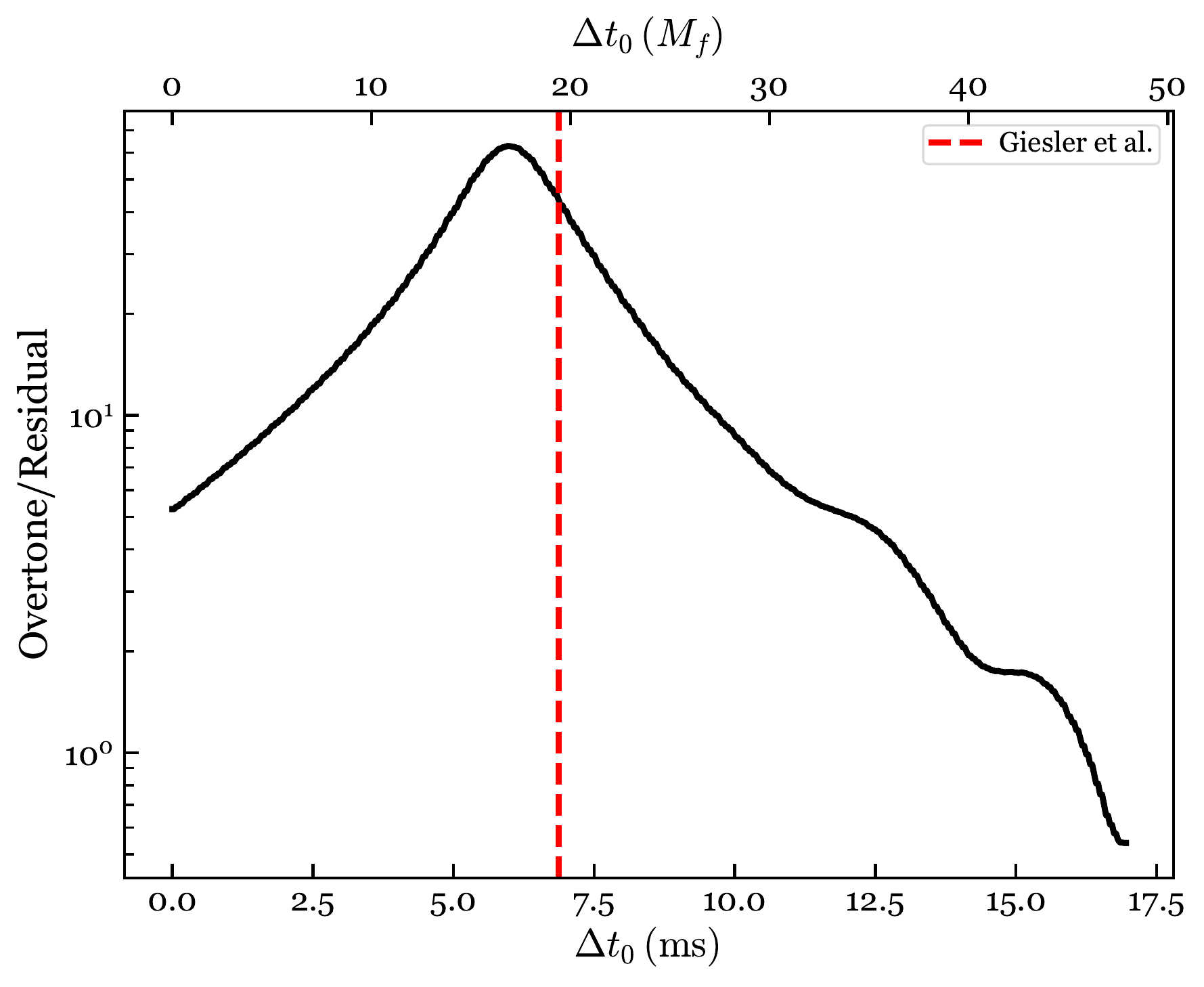}
  \caption{Ratio between the cumulative SNRs of the first overtone and the fitting residual as a function of $\Delta t_0$. The $l=m=2$ harmonic of the GW150914-like NR waveform is fitted with a two-QNM model $(\omega_{220}+\omega_{221})$ at different starting times. The residual is the difference between the $l=m=2$ harmonic in the NR waveform and the fitted two-QNM model template. At early times, the residual corresponds to the systematic bias due to the missing higher overtones in the model template. The cumulative SNRs are computed via Eq.~\eqref{eq:cumulative_SNR_residual}.}
 \label{fig:energy_ratio}
\end{figure}

First, the measurement of a mode frequency (or equivalently, the estimates of $M_f$ and $\chi_f$) needs sufficient mode cycles (duration) to accumulate a high-enough SNR. A missing mode does not bias the measurement when its cumulative SNR is small. To quantify the impact of the residual, we carry out a least-square fit to the $l=m=2$ harmonic of the GW150914-like NR waveform using a two-QNM model ($\omega_{220}+\omega_{221}$). At an early starting time, the fitting residual comes from the modes that are not included in the template (higher overtones). We compute the cumulative SNRs of the constructed first overtone and the fitting residual via
\begin{align}
    {\rm SNR} \,(\Delta t_0)=\sqrt{\int_{\Delta t_0}^{90M_f}|h(t)|^2dt}. \label{eq:cumulative_SNR_residual}
\end{align}
The ratio between the SNR of the first overtone and that of the residual as a function of $\Delta t_0$ is shown in Fig.~\ref{fig:energy_ratio}. We see the cumulative SNR of the first overtone is $\sim 5$ times larger than that of the residual even starting from the peak of the waveform $(\Delta t_0=0)$. The ratio continues to grow when $\Delta t_0<17M_f$, because the residual modes decay faster and the waveform becomes more consistent with the two-QNM model. After $17M_f$, the residual hits the error floor of the NR simulation and remains at that level. Thus, starting from $17M_f$, the ratio decreases exponentially as the first overtone decays away. Note that the maximum at $\Delta t_0=17M_f$ is close to the starting time of the two-QNM regime at $\Delta t_0=19M_f$ estimated by Giesler \etal (see the second row of Table I in \cite{Giesler:2019uxc}). Here we convert the mass unit from the total binary mass $M_{\rm tot}$ in \cite{Giesler:2019uxc} to the remnant mass $M_f$ using $M_f=0.95M_{\rm tot}$. 

In Bayesian analysis, the fact that the cumulative SNR of the first overtone is a few factors stronger than the residual modes allows us to perform a two-QNM MCMC analysis and infer the remnant properties from the first overtone right after the signal peak, when the full ringdown waveform has a low SNR and the systematic error caused by the residual modes is smaller than the uncertainties of the inferred parameters. In such a low-SNR regime, we might want to push our analysis as early as possible to increase the cumulative SNR of the first overtone, as long as the SNR contribution from the residual modes stays low, e.g., Fig.~\ref{fig:low_res_1.5_mchi_ligo} ($\Delta t_0=1.5\,{\rm ms}=4.2M_f$). However, as Fig.~\ref{fig:energy_ratio} suggests, the residual modes play a stronger role at earlier times, and thus we should be careful and avoid conducting the analysis too close to the signal peak, otherwise the residual modes can lead to systematic bias, such as the results shown in Fig.~\ref{fig:low_res_1.0_mchi_ligo}. On the other hand, if we detect a high-SNR ringdown signal, the contribution of the residual modes becomes more significant, and leads to a bias non-negligible compared to the parameter uncertainty range. In that case, following Fig.~\ref{fig:energy_ratio}, the analysis should be moved to later times (although still earlier than what the pure NR waveform suggests) to reduce the systematic bias caused by the residual modes. In an extreme case, when the ringdown signal becomes strong enough, the analysis reduces to the least-square fitting of the NR waveform. Then we can choose the maximum point in Fig.~\ref{fig:energy_ratio} as the starting time to perform the two-QNM fit, just as what Giesler \etal \cite{Giesler:2019uxc} did. In summary, the starting time of the analysis using a two-QNM model should be chosen based on the signal SNR; the higher the SNR, the later the starting time (in the range from $\Delta t_0 = 0$ to the maximum point in Fig.~\ref{fig:energy_ratio}).

That said, in Figs.~\ref{fig:low_res_injection_A0} and \ref{fig:low_res_injection_A1}, the mode amplitudes inferred from the Bayesian analysis are still biased, as mentioned by Finch and Moore \cite{Finch:2022ynt}. This is because the measurement of a mode amplitude depends more heavily on the first mode cycle than the whole cumulative SNR. Therefore, the estimate of the mode amplitude is more sensitive to the existence of the residual modes at an early stage. Nevertheless, the consistency among the ``$\mathcal{F}_{220}$ + one-QNM'' test, the conventional full-RD analysis, and the NR least-square fit (Figs.~\ref{fig:low_res_injection_A1}) implies that after cleaning the fundamental mode, the remaining signal is stably consistent with the first overtone\footnote{This is not to be confused with the noise fluctuations raised by Cotesta \etal \cite{Cotesta:2022pci}.}. 
In other words, at a relatively early time, e.g., $\Delta t_0=1.5\,{\rm ms}=4.2M_f$ in Fig.~\ref{fig:low_res_injection_A1}, the data-driven analysis leads to a measurement of an effective first overtone, with a correct mode frequency and decay rate albeit a biased mode amplitude \cite{Finch:2022ynt}.

\begin{figure*}
        \subfloat[Fundamental mode ($\omega_{220}$), $\Delta t_0=0.77\,{\rm ms}$\label{fig:mode_amplitudes_fun}]{\includegraphics[width=\columnwidth,clip=true]{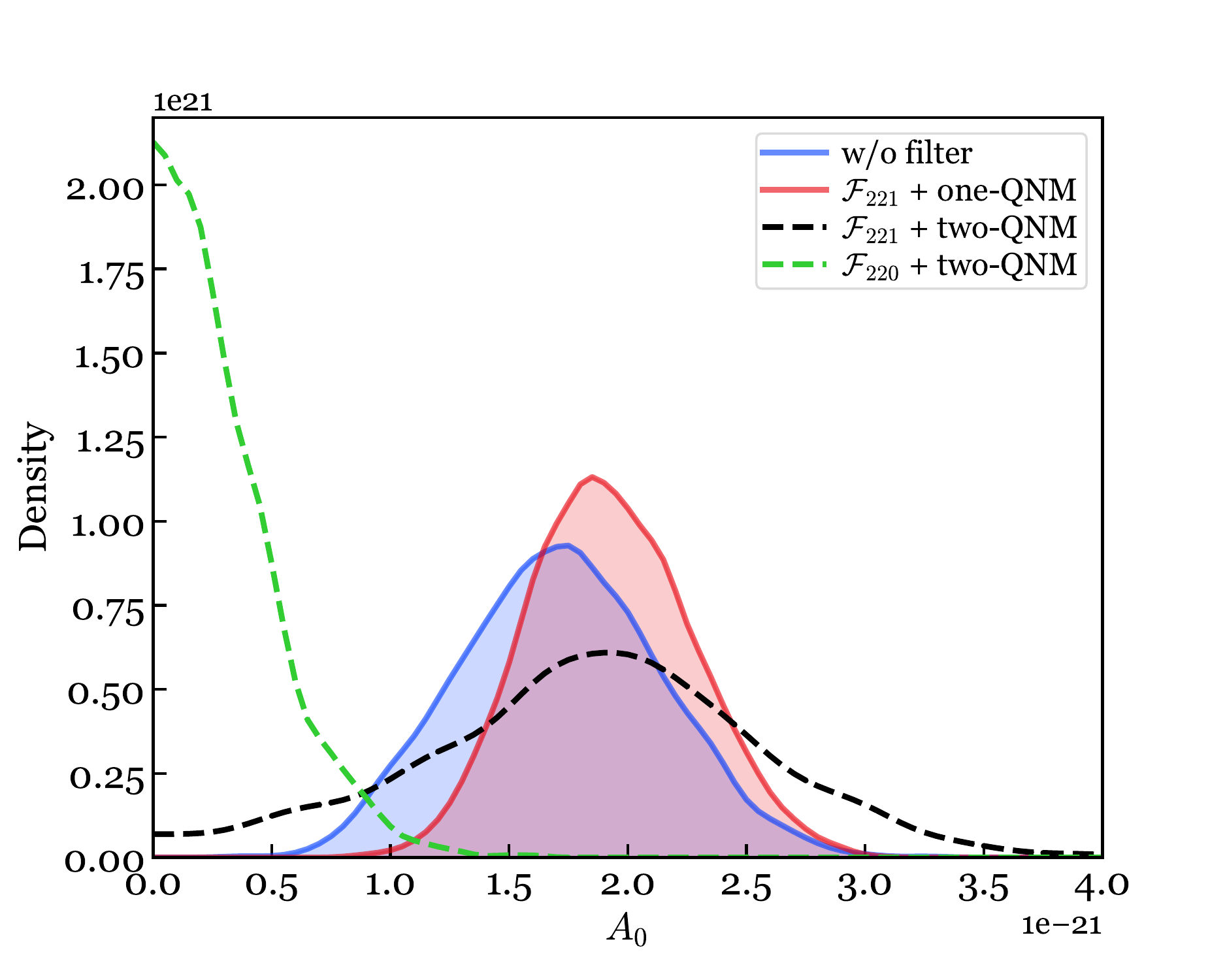}}
        \subfloat[First overtone ($\omega_{221}$), $\Delta t_0=0.77\,{\rm ms}$\label{fig:mode_amplitudes_overtone}]{\includegraphics[width=\columnwidth,clip=true]{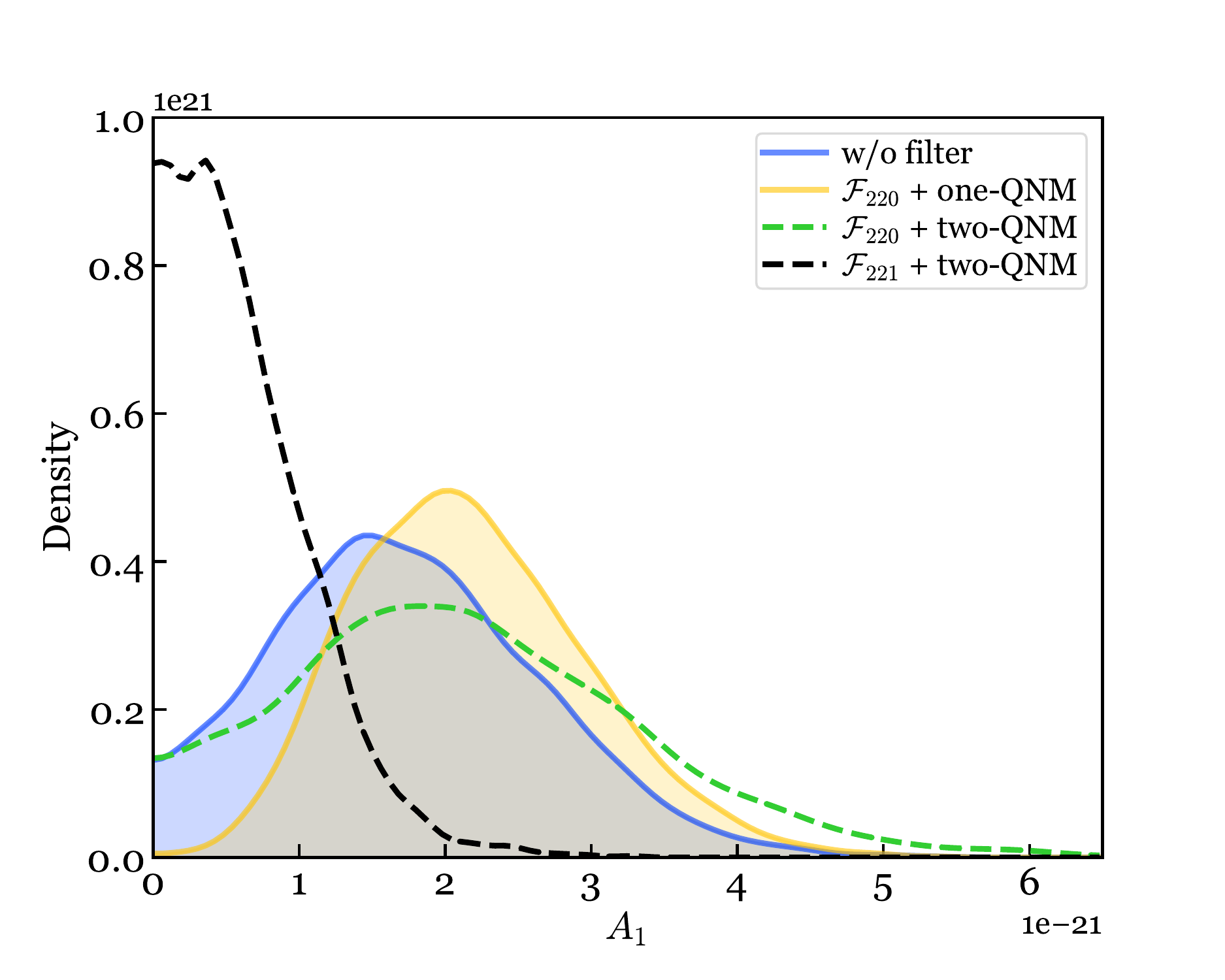}} \\
        \subfloat[$\Delta t_0=0.77\,{\rm ms}$\label{fig:mchi_ligo}]{\includegraphics[width=\columnwidth,clip=true]{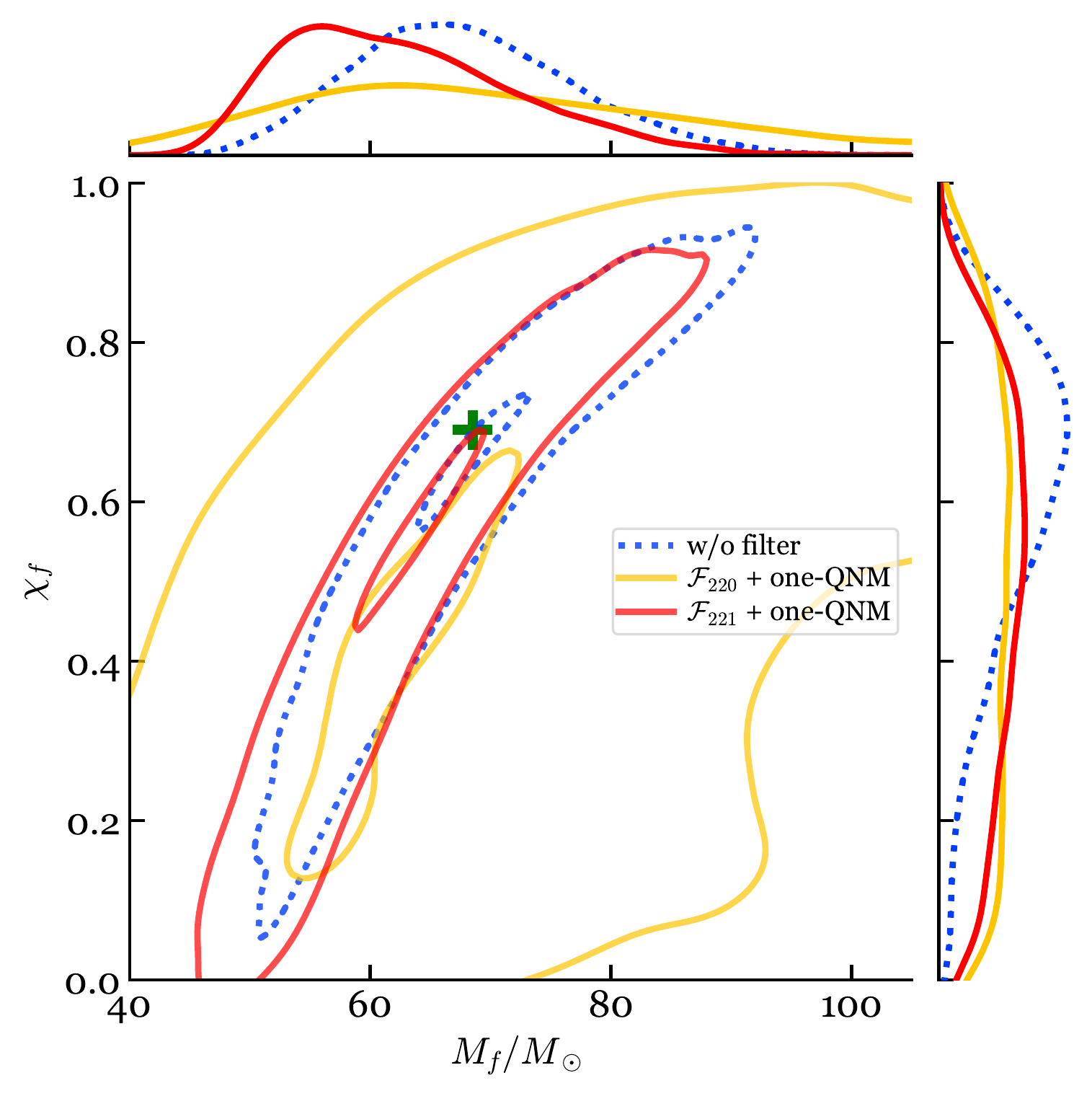}}
        \subfloat[$\Delta t_0=0.1\,{\rm ms}$\label{fig:mchi_ligo_1em1}]{\includegraphics[width=\columnwidth,clip=true]{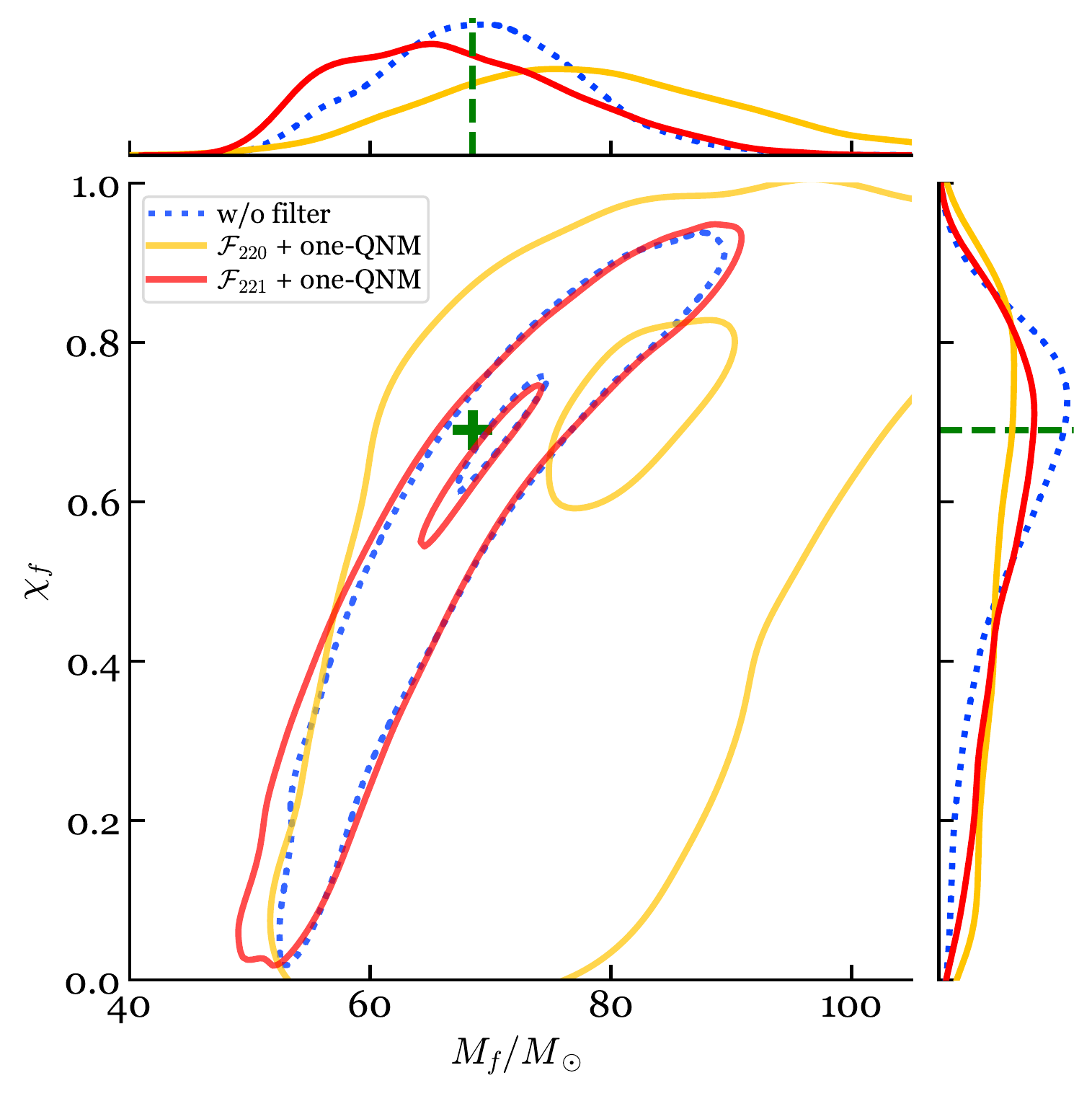}}
  \caption{(Similar to Fig.~\ref{fig:injection_amplitude}) Estimates of the mode amplitudes and BH properties for GW150914 using the mixed approach. See Fig.~\ref{fig:injection_amplitude} caption for detailed descriptions. Note that different start times are used here: $\Delta t_0 = 0.77$ ms in Fig.~\ref{fig:mode_amplitudes_fun}, \ref{fig:mode_amplitudes_overtone} and \ref{fig:mchi_ligo}, and $\Delta t_0 = 0.1$ ms in Fig.~\ref{fig:mchi_ligo_1em1}. 
  } 
 \label{fig:mode_amplitudes}
\end{figure*}

\begin{figure}[htb]
        \includegraphics[width=\columnwidth,clip=true]{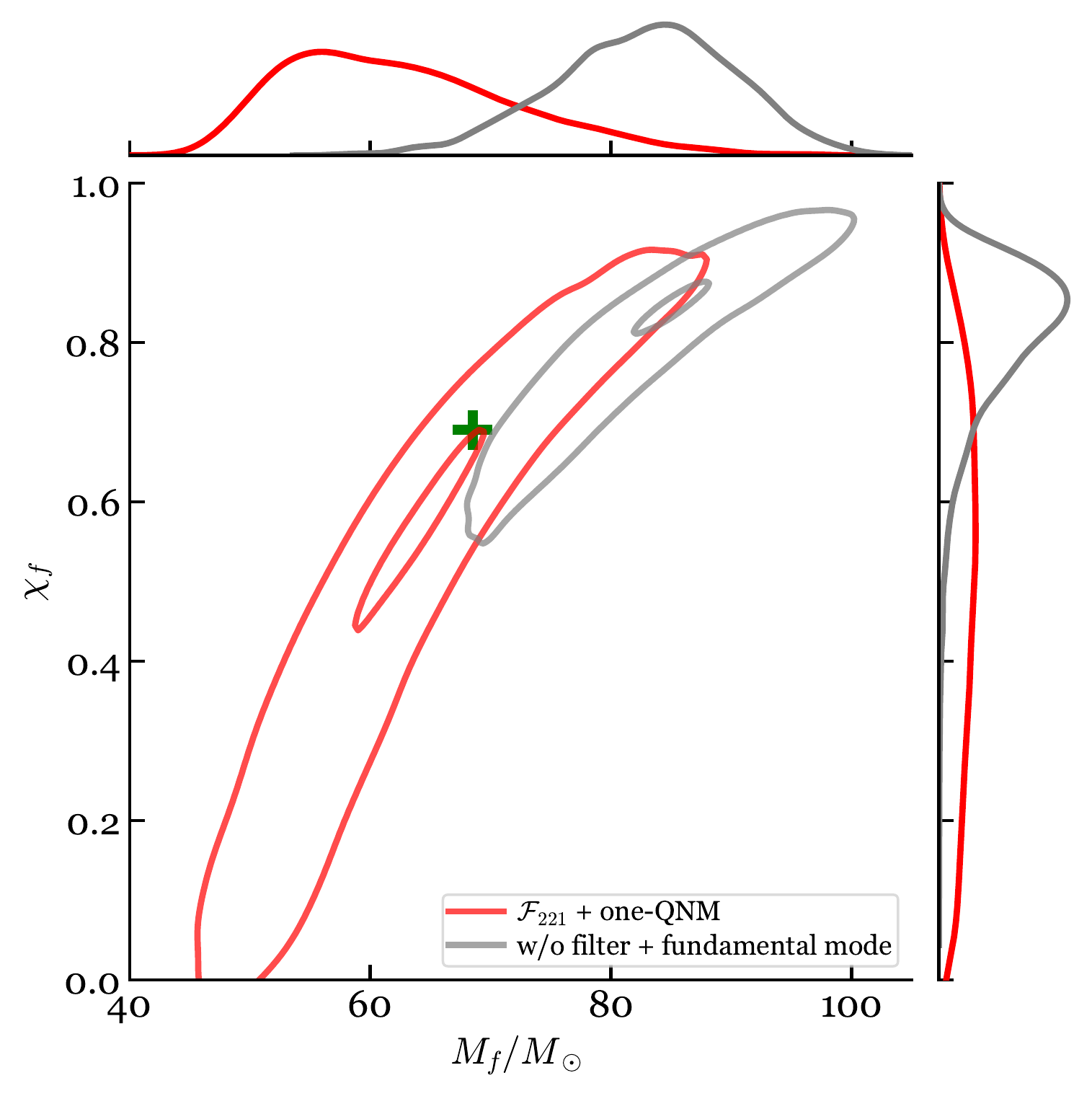}
  \caption{Posterior distributions of $M_f$ and $\chi_f$ for GW150914 estimated at $\Delta t_0=0.77\,{\rm ms}$. The gray contours are obtained from the conventional full-RD MCMC analysis, where the unfiltered data (without mode cleaning) are fitted with the fundamental-mode-only template. The red contours are the same as the ones in Fig.~\ref{fig:mchi_ligo}, where we first apply the filter $\mathcal{F}_{221}$ to remove the first overtone, and then fit the filtered data with the template of the fundamental mode. }
 \label{fig:mchi_ligo_no_filter_one_mode}
\end{figure}

\begin{figure}
        \includegraphics[width=\columnwidth,clip=true]{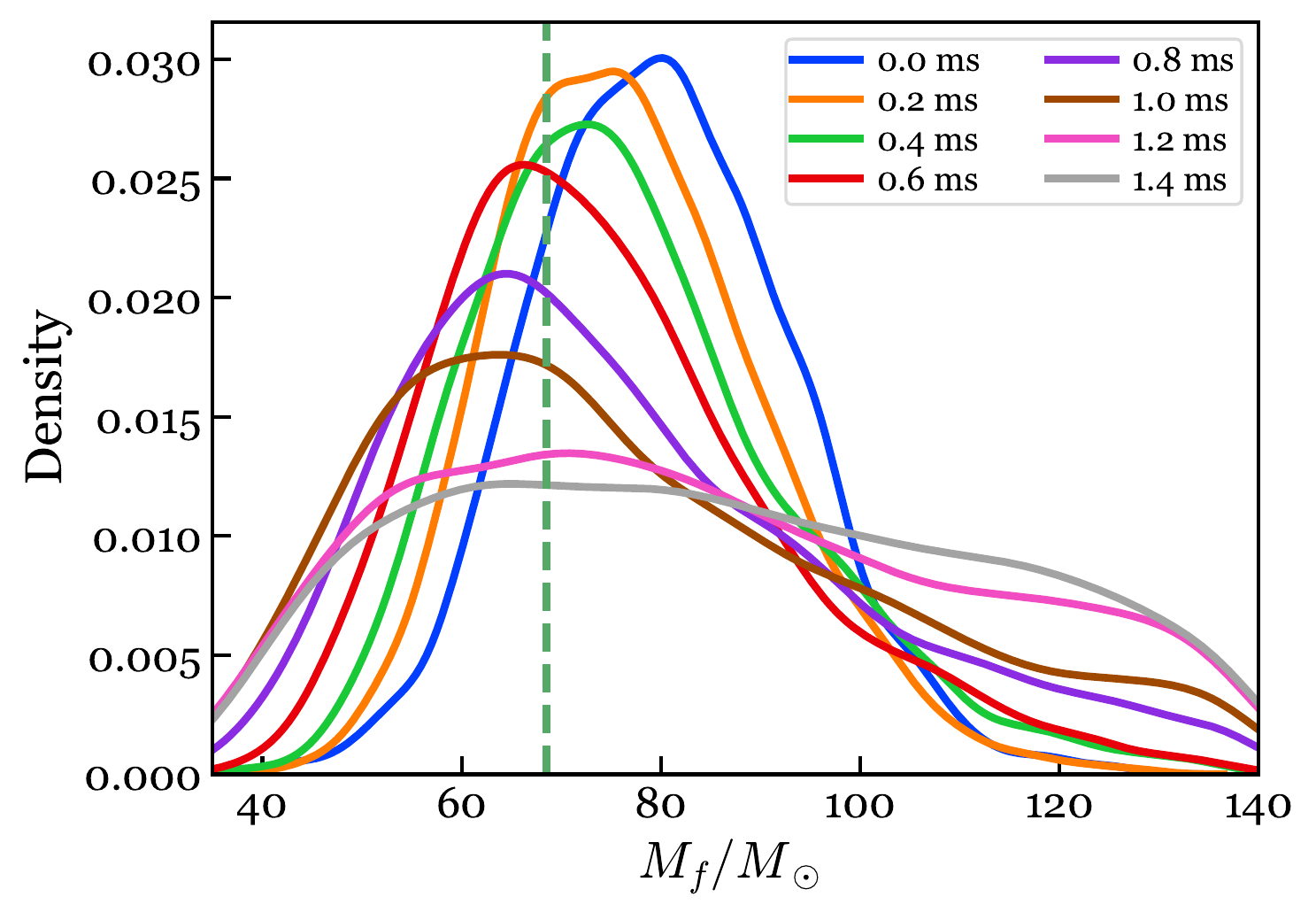} \\
        \includegraphics[width=\columnwidth,clip=true]{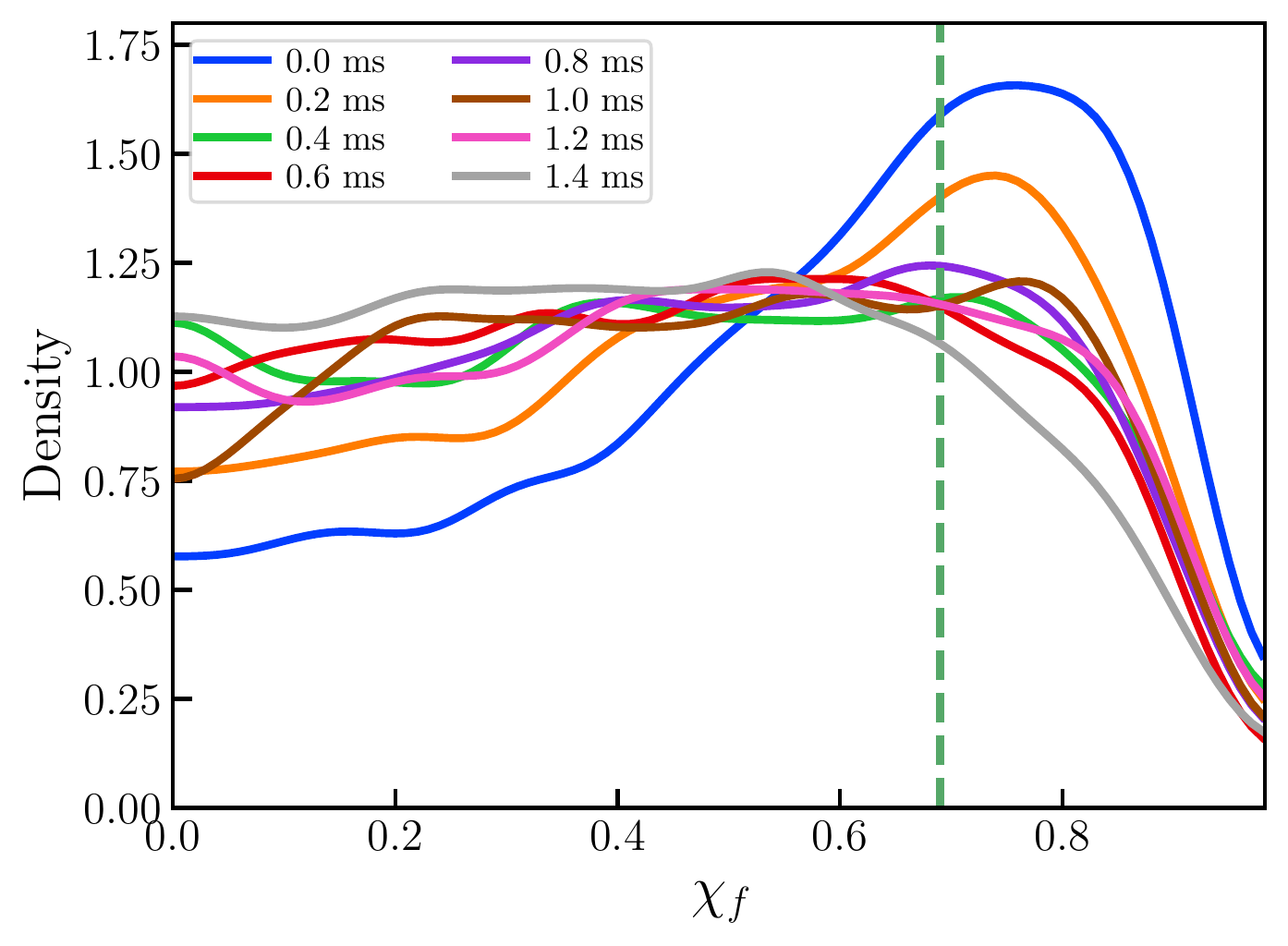}
  \caption{Posterior distributions of $M_f$ (top) and $\chi_f$ (bottom) solely inferred from the first overtone in the ringdown of GW150914, at different $\Delta t_0$ times. We first apply the filter $\mathcal{F}_{220}$ to remove the fundamental mode, and then fit the filtered data with the template of the first overtone. We set uniform priors in the ranges of $M_f\in[35M_\odot, 140M_\odot]$ and $\chi_f\in[0,0.99]$ (as shown in the horizontal axes in the plot). The vertical dashed lines indicate the estimates obtained from the full IMR signal.} 
 \label{fig:GW150914_mix_marginal}
\end{figure}

\subsection{GW150914}
\label{sec:mixed_gw150914}
We now apply the mixed approach to GW150914. 
Similar to the NR simulation, the full-RD MCMC fitting and the four filtering scenarios listed in Table \ref{table:strategy} are tested; the results are shown in Fig.~\ref{fig:mode_amplitudes}.
For mode cleaning, we use the BH properties estimated from the IMR signal \cite{Isi:2019aib}, $M_f^{\rm IMR}=68.5,\chi_f^{\rm IMR}=0.69$, to build the filter.
The MCMC fitting is conducted at a start time of $\Delta t_0=0.77\,{\rm ms}$ and the window length of $w=0.2\, {\rm s}$.
In Figs.~\ref{fig:mode_amplitudes_fun} and \ref{fig:mode_amplitudes_overtone},
the estimates of $A_0$ and $A_1$ under all scenarios qualitatively agree with the injection study in Sec.~\ref{sec:mixed_injection} (Fig.~\ref{fig:injection_amplitude}).
These results demonstrate: (1) the fundamental mode or the first overtone can be successfully cleaned from the ringdown of GW150914 by the filters, and (2) the $M_f$ and $\chi_f$ values obtained from the IMR signal are consistent with the evolution of the QNMs in ringdown (so that the modes can be correctly cleaned).


In Fig.~\ref{fig:mchi_ligo}, we show the estimates of $M_f$ and $\chi_f$ at the start time $\Delta t_0=0.77\,{\rm ms}$ under the three scenarios, similar to the injection case shown in Fig.~\ref{fig:low_res_1.5_mchi_ligo}. The constraints obtained from the unfiltered and the filtered data generally agree, and are consistent with the estimates obtained from the IMR signal. In particular, we note that for the chosen start time,  one needs to include both modes $\omega_{220}$ and $\omega_{221}$ to perform the usual full-RD MCMC analysis (blue-dashed contours), fitting with only the fundamental mode $\omega_{220}$ leads to a strong bias in the inferred $M_f$ and $\chi_f$. The gray contours in Fig.~\ref{fig:mchi_ligo_no_filter_one_mode} display the corresponding joint posterior of $M_f$ and $\chi_f$ when the first overtone is omitted from the MCMC analysis (also see Fig.~3 of Ref.~\cite{Isi:2019aib}). But the discrepancy with the IMR results is reduced if we first apply the filter $\mathcal{F}_{221}$ to clean the first overtone $\omega_{221}$ (red contours in Fig.~\ref{fig:mchi_ligo_no_filter_one_mode}). 

On the other hand, the yellow contours in Fig.~\ref{fig:mchi_ligo} represent the posterior inferred from the first overtone alone. We see the estimates are still informative, although less constrained than the ``$\mathcal{F}_{221}$+one-QNM'' case.
If we start the analysis at an earlier time, e.g., $\Delta t_0=0.1 \,{\rm ms}$ (Fig.~\ref{fig:mchi_ligo_1em1}), the constraints obtained from the first overtone alone deviate more from the IMR results, despite the fact that the contour of the 90\% credible intervals is still consistent with the IMR result. 
Presumably, the shift is caused by the existence of other signal features in addition to the first overtone (e.g., higher overtones, similar to Fig.~\ref{fig:low_res_1.0_mchi_ligo}), although no evidence for the existence of higher overtones is found in the ringdown of GW150914. For more choices of $\Delta t_0$, we focus on the marginalized posteriors of $M_f$ and $\chi_f$ obtained from the ``$\mathcal{F}_{220}+$one-QNM'' scenario. As shown in Fig.~\ref{fig:GW150914_mix_marginal}, the posterior distribution of $M_f$ gradually shifts toward smaller values and moves closer to the IMR result (the vertical dashed line) when $\Delta t_0 \in [0,1]$ ms. For later times $(\Delta t_0 \gtrsim 1\,{\rm ms})$, the distribution widens and becomes less informative. On the other hand, the posterior distribution of $\chi_f$ flattens quickly as $\Delta t_0$ increases and becomes consistent with the prior.

\section{Conclusion} 
\label{sec:conclusion}
In this paper, we incorporate the novel rational filter \cite{Ma:2022wpv} into a Bayesian framework (outlined in our companion paper \cite{Ma_prl}), and obtain several pieces of evidence for the existence of the first overtone in the ringdown of GW150914. We first demonstrate that the rational filter has no impact on the statistical properties of the noise (Gaussianity, stationarity, and PSDs). We then construct a 2D likelihood function that depends only on the mass and spin of the remnant BH and implement an efficient algorithm to obtain the posteriors of mass and spin without running MCMC. We use an NR injection and the GW150914 event to demonstrate that the posteriors obtained using our method are consistent with those from the full-RD MCMC approach. By applying our method to the data of GW150914 near the peak of the strain, we confirm the conclusion of Ref.~\cite{Isi:2019aib}: The inferred remnant BH properties are more consistent with the IMR results when the first overtone is included; the contribution from the first overtone gradually fades away at later times.

Next, we compute the model evidence for filters built with different sets of QNM(s) by integrating the new likelihood function over the 2D parameter space. The evidence depends on the fitting start time, showing a sharp rise around the onset of a ringdown signal, which in turn agnostically reflects the starting time of the corresponding QNM(s). The ratio between two evidence values from two filters with different sets of QNM(s) indicates the Bayes factor for a QNM model. For GW150914, we find a Bayes factor of 600 for the model with the first overtone over the fundamental-mode-only model at the inferred strain peak. This Bayes factor decreases and levels off at later times. 

Finally, we combine the mode-cleaning procedure using the rational filter with the usual MCMC method to build a mixed approach for BH spectroscopy. After cleaning the fundamental mode in the GW150914 data and fitting the filtered data using MCMC, we find the posterior of the fundamental mode amplitude gets close to zero, confirming the successful subtraction of the fundamental mode. On the other hand, the amplitude of the first overtone is barely impacted. We also use the mixed approach to infer the remnant BH mass and spin from only the fundamental mode and only the first overtone. The results from the first overtone alone are still informative, showing consistent constraints on $M_f$ with the full-IMR and fundamental-mode-only analyses. The recovery of the remnant BH properties (mass and mode amplitudes) from the first overtone alone serves as a more direct piece of evidence supporting the existence of the first overtone, in addition to other indicators (e.g., the Bayes factor).

This novel framework is not only powerful at revealing subdominant QNMs; it also has superior computational efficiency compared to the existing MCMC approach. In the GW150914 analysis, it takes $\sim 8$ seconds on a general laptop to produce a low-resolution 2D $M_f$--$\chi_f$ posterior distribution for the remnant BH that is good enough to reveal the key features (the evaluation of the likelihood function in Eq.~\eqref{eq:log-p} for each pair of $M_f$ and $\chi_f$ takes milliseconds). For a production run with a high resolution, e.g., a panel in Fig.~\ref{fig:GW150914_contours_1}, it takes $\sim 3$ minutes on a cluster (a single node with 24 cores). The performance can be further improved by fully parallelizing the calculation, since the likelihood evaluation for each pair of mass and spin is completely independent. In addition, the code's efficiency is not impacted by including more QNMs, given that multiple filters can be applied simultaneously [Eq.~\eqref{eq:total_filter}]. On the contrary, including more QNMs significantly increases the computing cost of the full-RD MCMC calculation, because the dimension of the parameter space is largely increased.

With this framework, the ringdown analysis can be easily extended to investigate more subdominant modes in addition to overtones.
Future work is being planned to investigate another controversial event, GW190521 \cite{Capano:2021etf,Capano:2022zqm}, and new detections in the upcoming O4 run. In addition, here we combine and align the data at two LIGO detectors based on the constraint of the event sky location. It is worth studying whether we could analyze the data from each detector individually and  use the common features in the results, e.g., the time when the evidence rises sharply (see Fig.~\ref{fig:noval_likelihood_time}), to help constrain the sky position of an event.


\begin{acknowledgments}
We thank Maximiliano Isi and Will M. Farr for their helpful suggestions and comments. We also thank Eric Thrane, Paul Lasky, Emanuele Berti, Mark Ho-Yeuk Cheung, Roberto Cotesta, and all the attendees of the ringdown workshop held at CCA, Flatiron Institute for useful discussions. Computations for this work were performed with the Wheeler cluster at Caltech.
YC and SM acknowledge support from the Brinson Foundation, the  Simons Foundation (Award Number 568762), and by NSF Grants PHY-2011961, PHY-2011968, PHY--1836809.
LS acknowledges the support of the Australian Research Council Centre of Excellence for Gravitational Wave Discovery (OzGrav), Project No. CE170100004. This material is based upon work supported by NSF's LIGO Laboratory which is a major facility fully funded by the National Science Foundation.
\end{acknowledgments}

\appendix 


\begin{figure*}
        \subfloat[Reproduce the red contours in Fig.~\ref{fig:low_res_1.5_mchi_ligo}.\label{fig:mixed_injection_contour_t1.5} ]{\includegraphics[width=\columnwidth,clip=true]{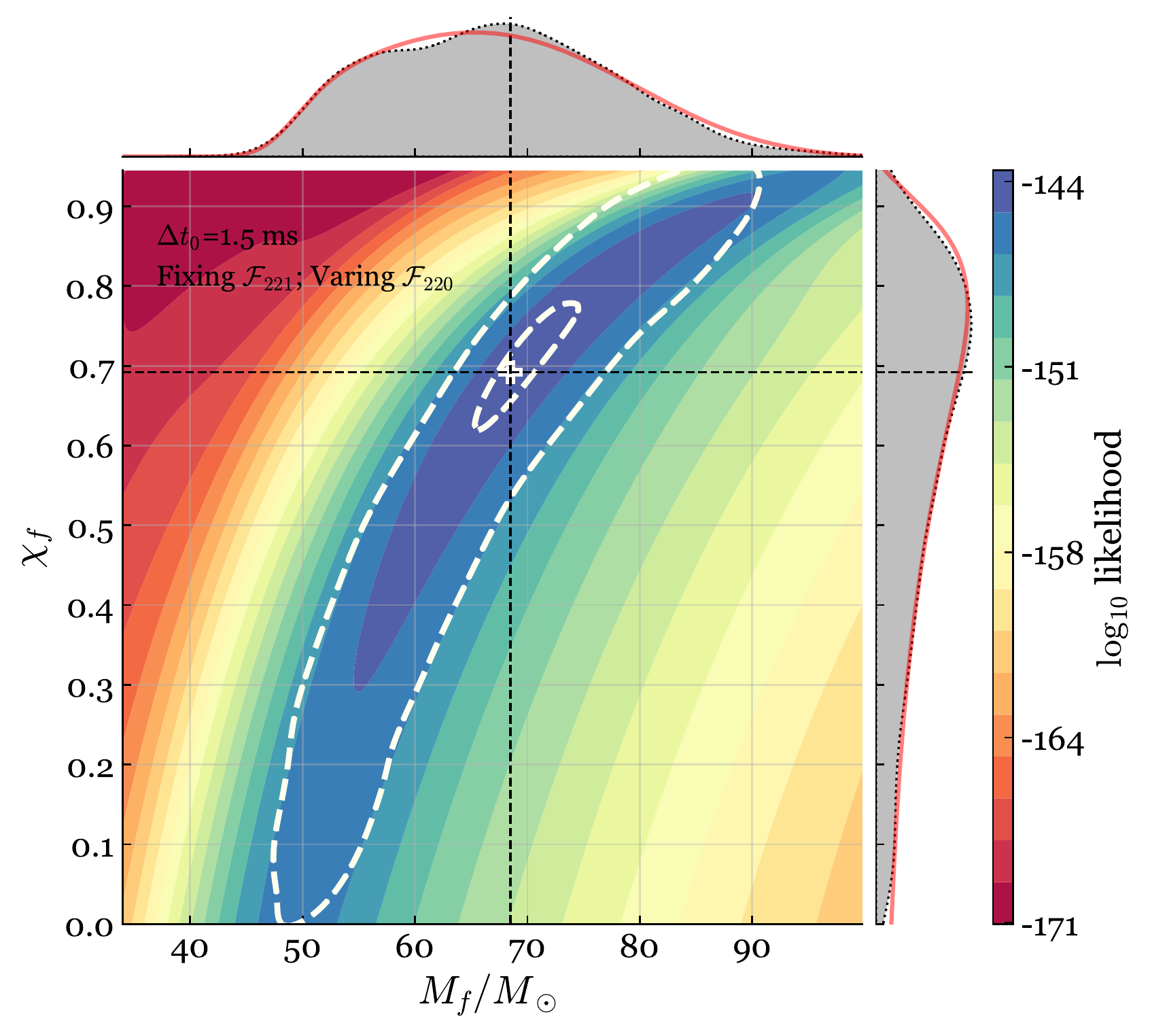}}
        \subfloat[Reproduce the yellow contours in Fig.~\ref{fig:low_res_1.5_mchi_ligo}.\label{fig:mixed_injection_contour_overtone_t1.5}]{\includegraphics[width=\columnwidth,clip=true]{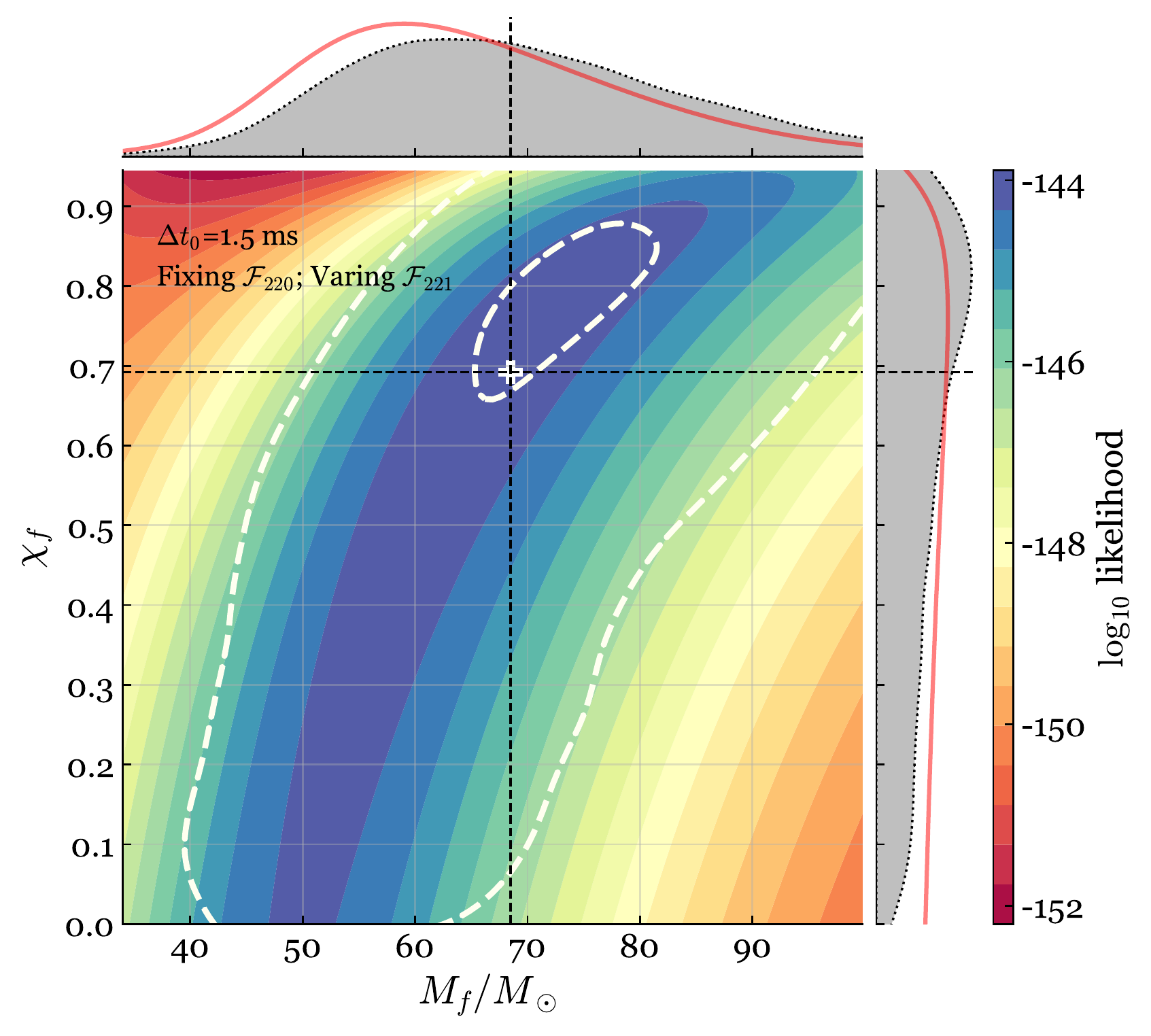}} \\
        \subfloat[Reproduce the red contours in Fig.~\ref{fig:low_res_1.0_mchi_ligo}.\label{fig:mixed_injection_contour_t1.0}]{\includegraphics[width=\columnwidth,clip=true]{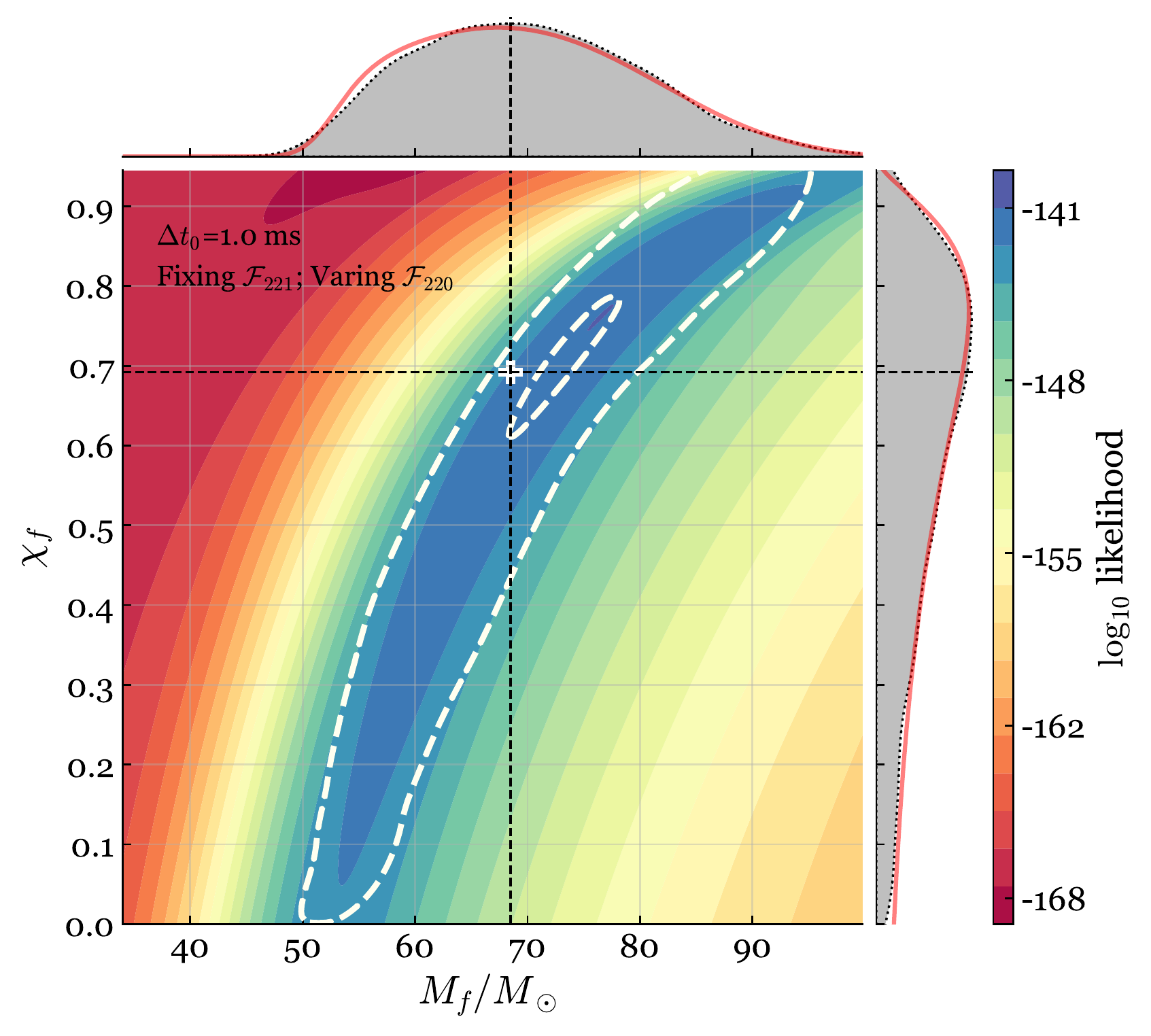}}
        \subfloat[Reproduce the yellow contours in Fig.~\ref{fig:low_res_1.0_mchi_ligo}.\label{fig:mixed_injection_contour_overtone_t1.0}]{\includegraphics[width=\columnwidth,clip=true]{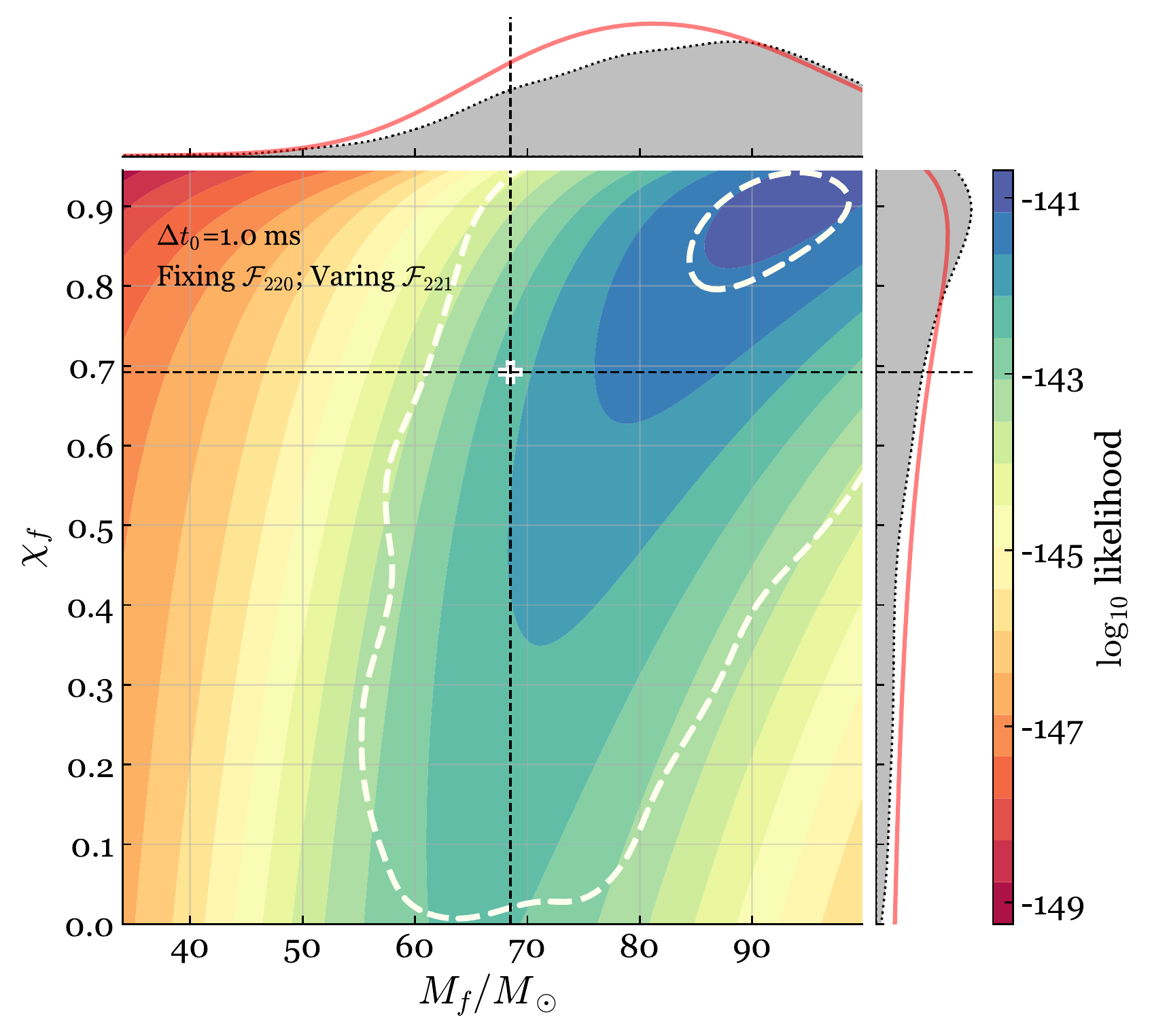}}
  \caption{Reproduce the estimates of $M_f$ and $\chi_f$ in Figs.~\ref{fig:low_res_1.5_mchi_ligo} and \ref{fig:low_res_1.0_mchi_ligo}, using the filters. The top and bottom panels are for $\Delta t_0 = 1.5$ ms (cf. Fig.~\ref{fig:low_res_1.5_mchi_ligo}) and 1.0 ms (cf. Fig.~\ref{fig:low_res_1.0_mchi_ligo}), respectively. The left and right columns correspond to analyzing the fundamental mode only (cf. red contours in Fig.~\ref{fig:injection_amplitude}) and analyzing the first overtone only (cf. yellow contours in Fig.~\ref{fig:injection_amplitude}), respectively. The filters used to clean either the first overtone in the left panels or the fundamental mode in the right panels are built with the true values of $M_f$ and $\chi_f$ for the injected system. The red and yellow contours in Figs.~\ref{fig:low_res_1.5_mchi_ligo} and \ref{fig:low_res_1.0_mchi_ligo} are shown as white dashed contours in this figure.}
 \label{fig:mixed_injection_contour}
\end{figure*}

\begin{figure*}
        \subfloat[Reproduce the red contours in Fig.~\ref{fig:mchi_ligo}.\label{fig:mixed_contour_t0.77}]{\includegraphics[width=\columnwidth,clip=true]{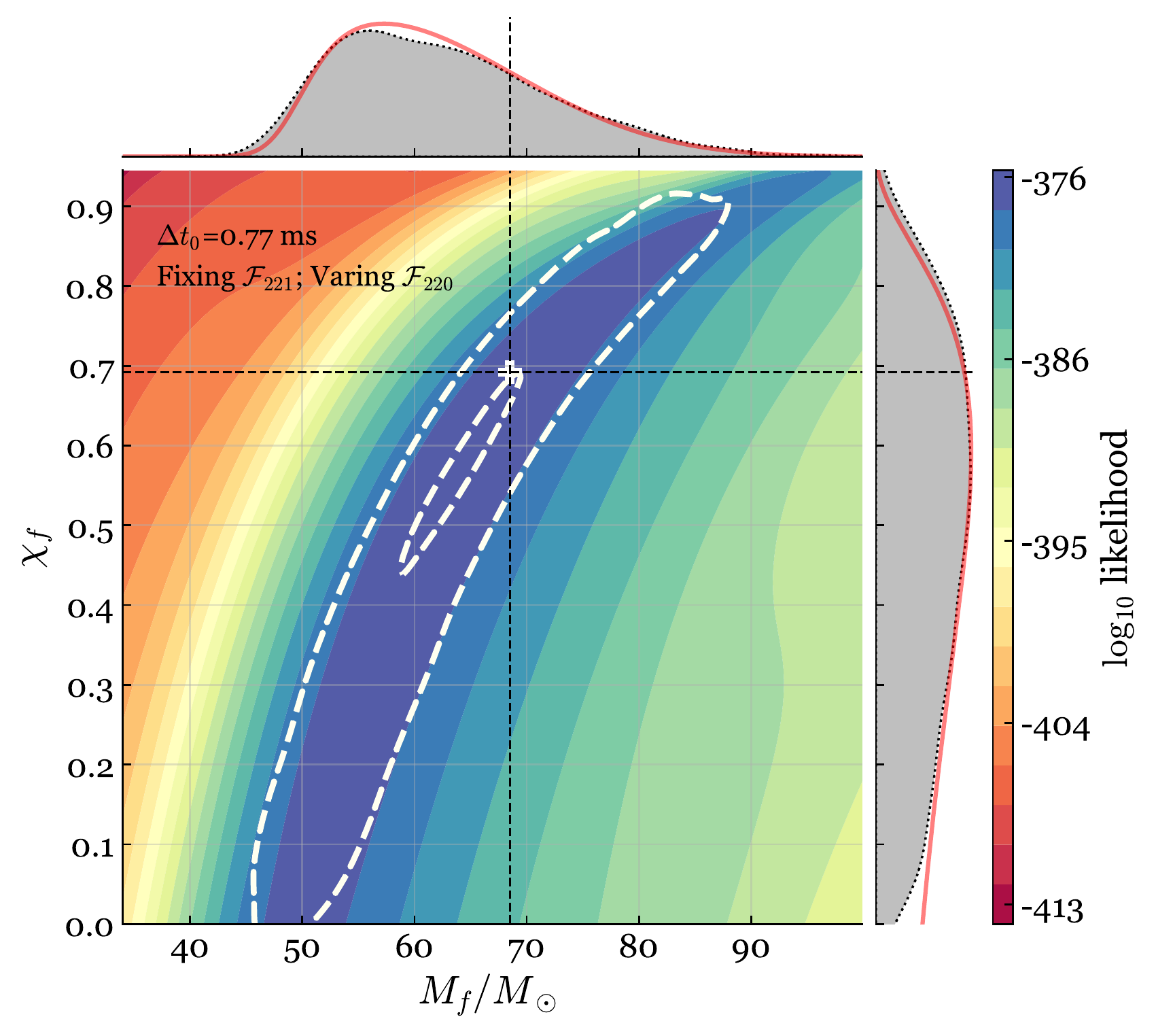}}
        \subfloat[Reproduce the yellow contours in Fig.~\ref{fig:mchi_ligo}.\label{fig:mixed_contour_overtone_t0.77}]{\includegraphics[width=\columnwidth,clip=true]{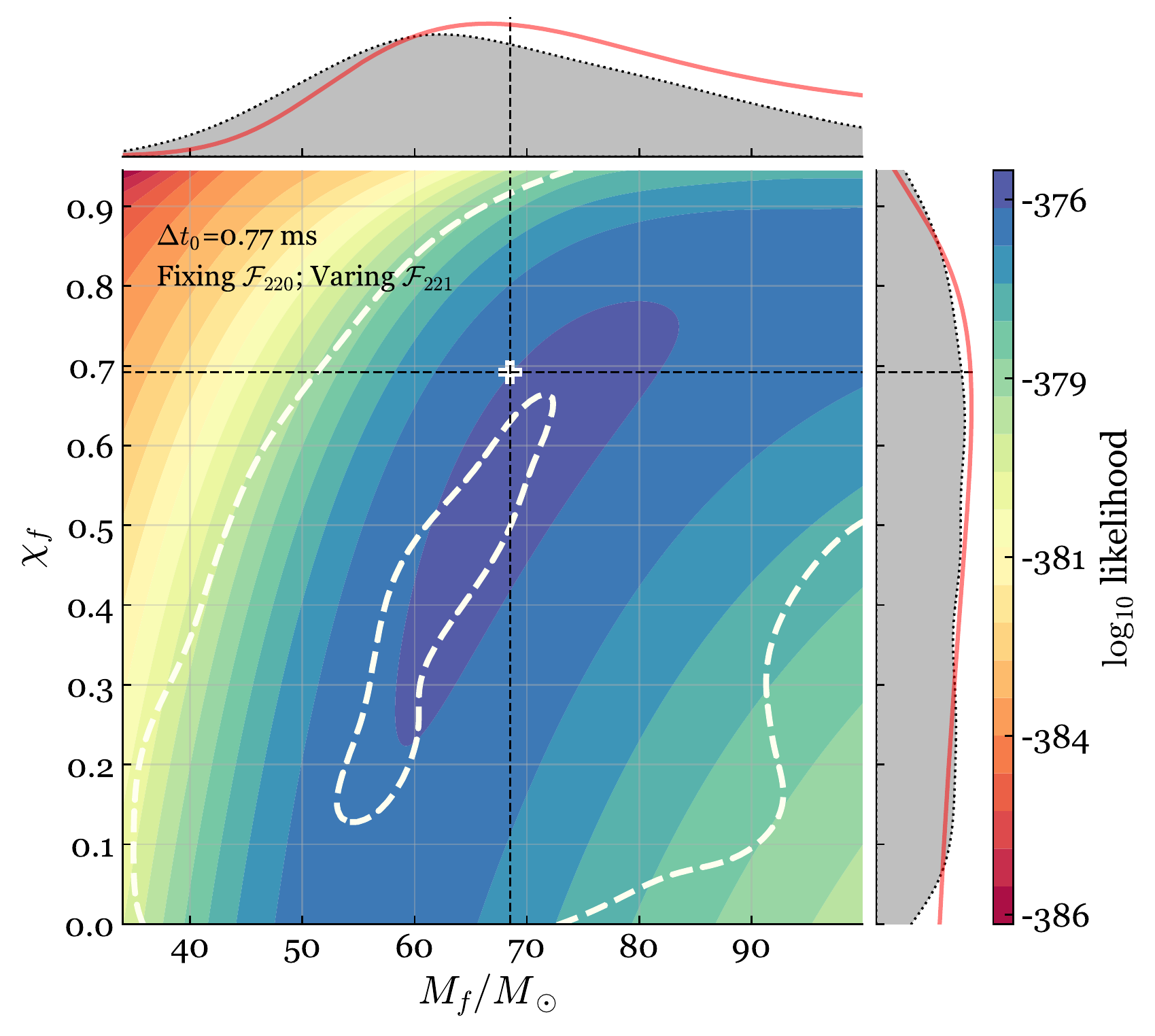}} \\
        \subfloat[Reproduce the red contours in Fig.~\ref{fig:mchi_ligo_1em1}.\label{fig:mixed_contour_t0.1}]{\includegraphics[width=\columnwidth,clip=true]{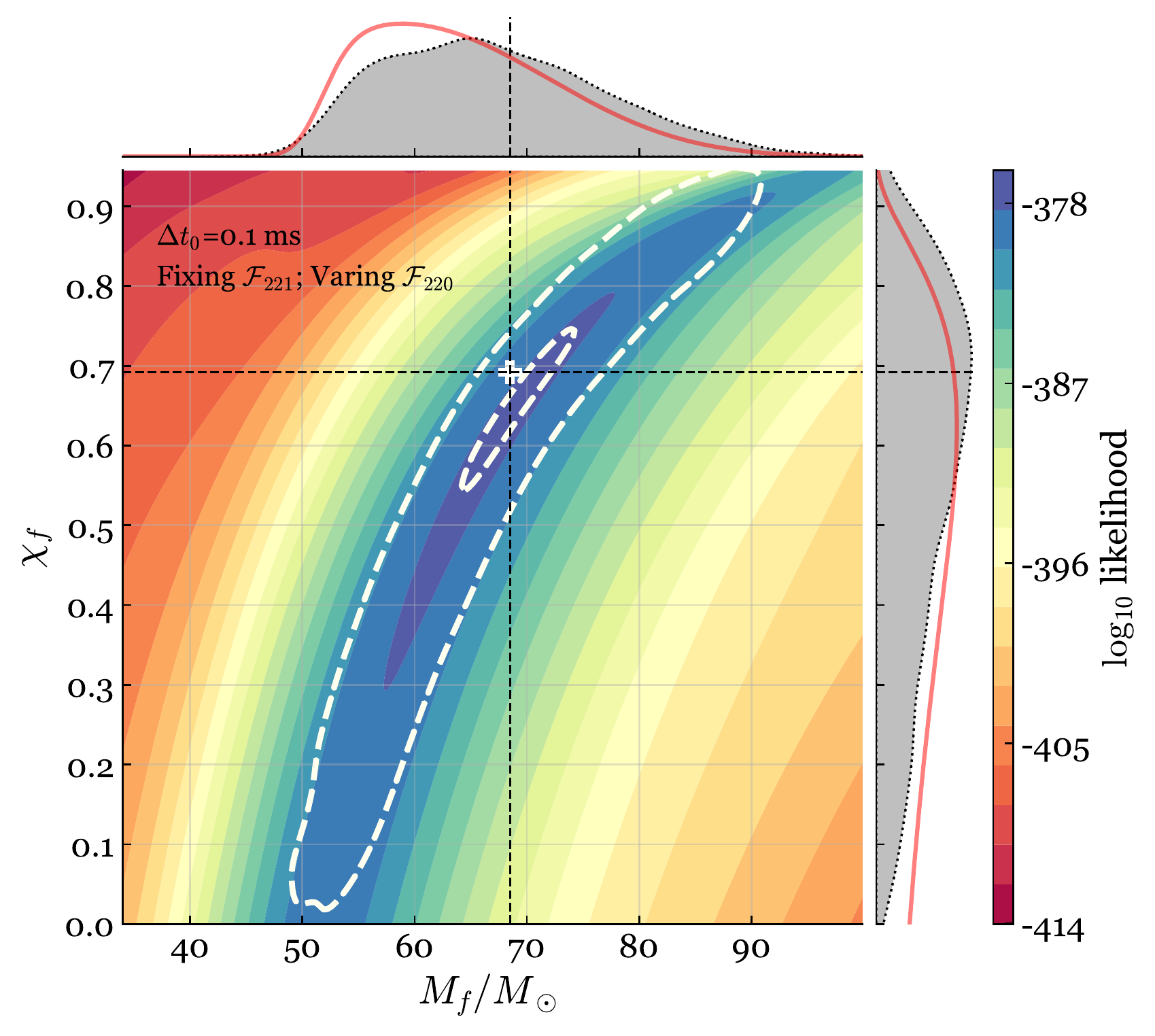}}
        \subfloat[Reproduce the yellow contours in Fig.~\ref{fig:mchi_ligo_1em1}.\label{fig:mixed_contour_overtone_t0.1}]{\includegraphics[width=\columnwidth,clip=true]{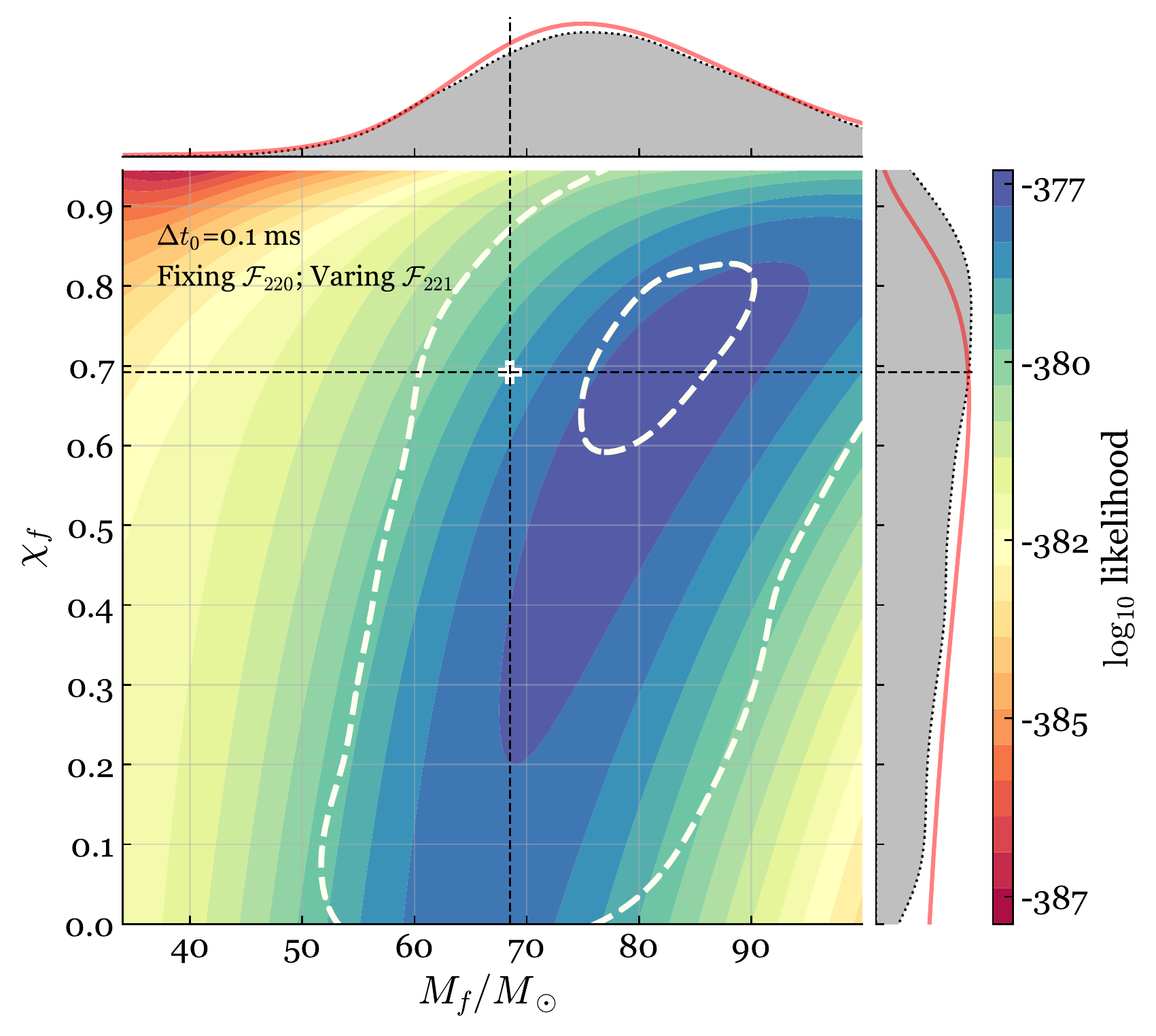}}
  \caption{(Similar to Fig.~\ref{fig:mixed_injection_contour}) Reproduce the estimates of $M_f$ and $\chi_f$ in Figs.~\ref{fig:mchi_ligo} and \ref{fig:mchi_ligo_1em1}, using the filters, for GW150914. See Fig.~\ref{fig:mixed_injection_contour} caption for detailed descriptions. The top and bottom panels are for $\Delta t_0 = 0.77$ ms and 0.1 ms, respectively. The filters used to remove either the first overtone in the left panels or the fundamental mode in the right panels are built with the estimated $M_f$ and $\chi_f$ from the IMR signal.}
 \label{fig:mixed_contour_150914}
\end{figure*}

\begin{figure*}
        \subfloat[$\Delta t_0=0.0 \,{\rm ms}$]{\includegraphics[width=\columnwidth,clip=true]{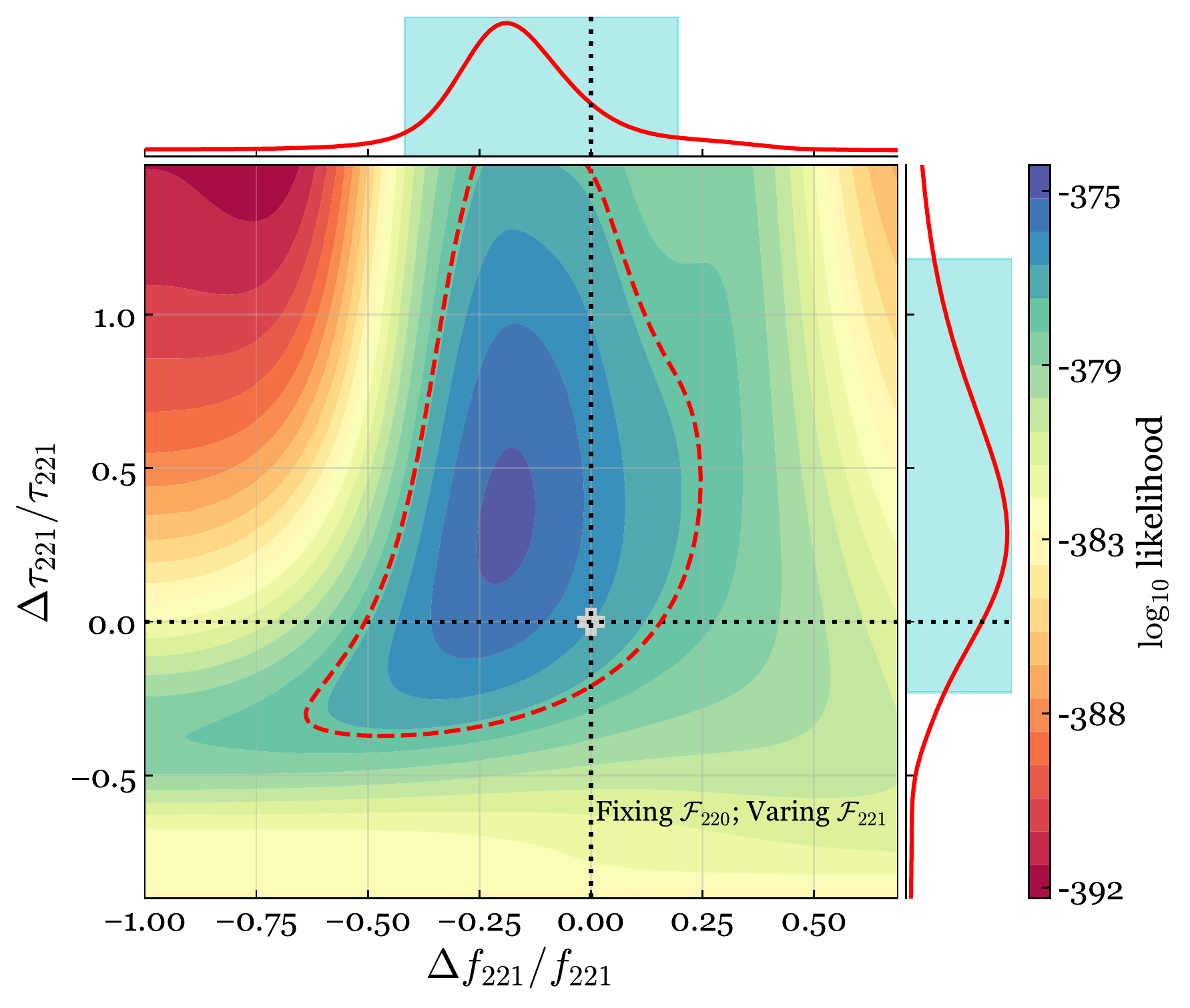}}
        \subfloat[$\Delta t_0=0.1 \,{\rm ms}$]{\includegraphics[width=\columnwidth,clip=true]{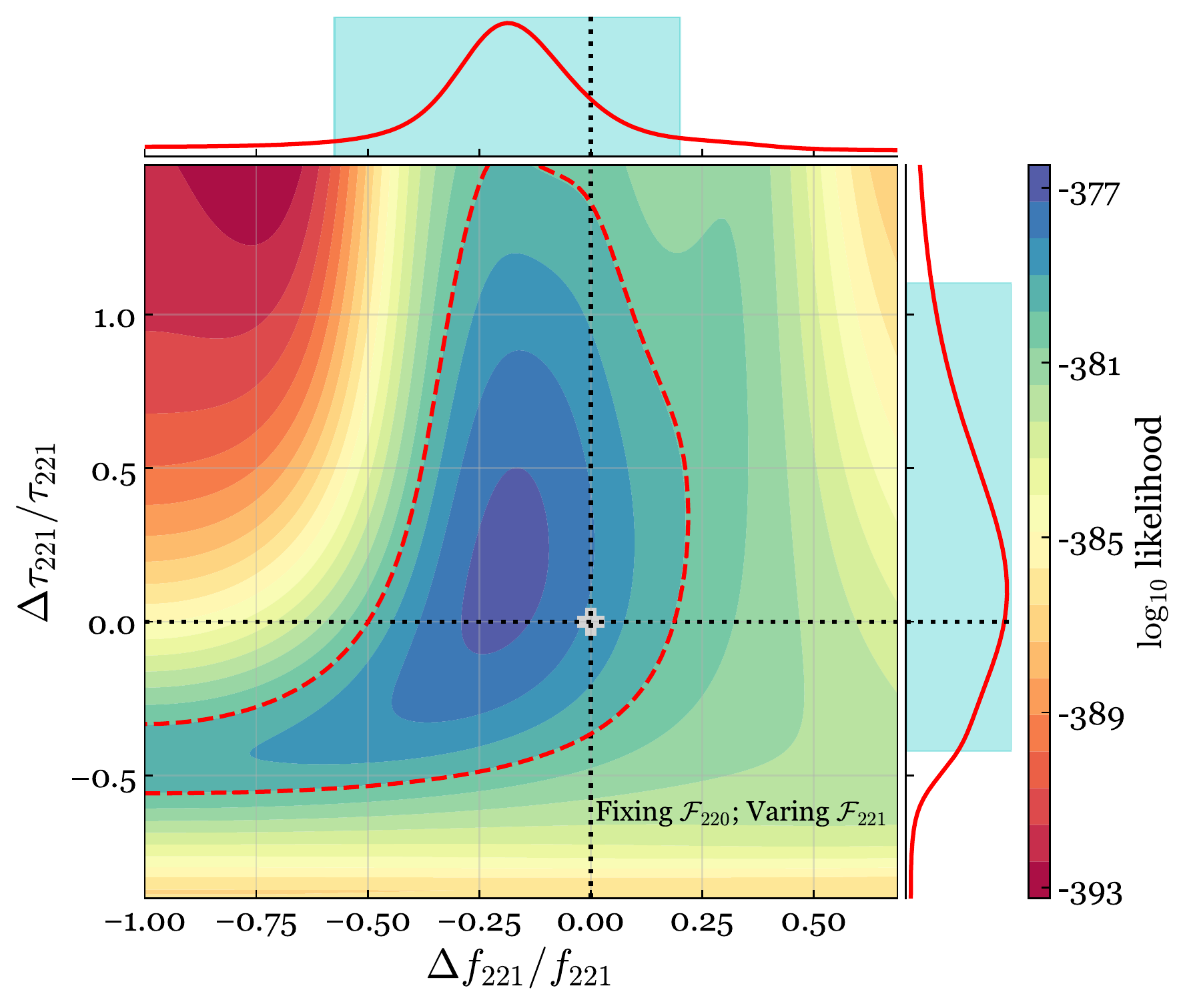}}\\
        \subfloat[$\Delta t_0=0.4 \,{\rm ms}$]{\includegraphics[width=\columnwidth,clip=true]{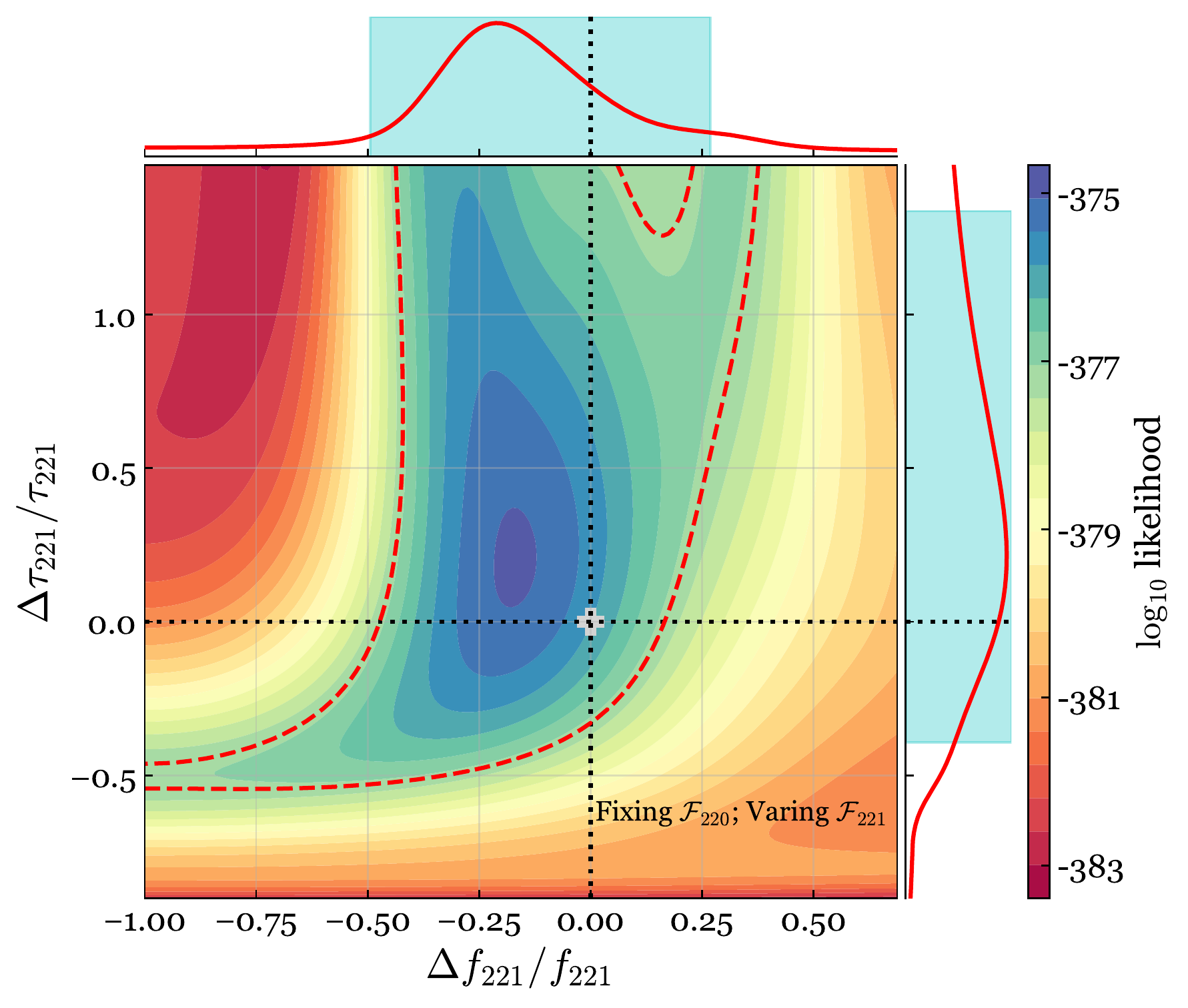}}
        \subfloat[$\Delta t_0=0.77 \,{\rm ms}$]{\includegraphics[width=\columnwidth,clip=true]{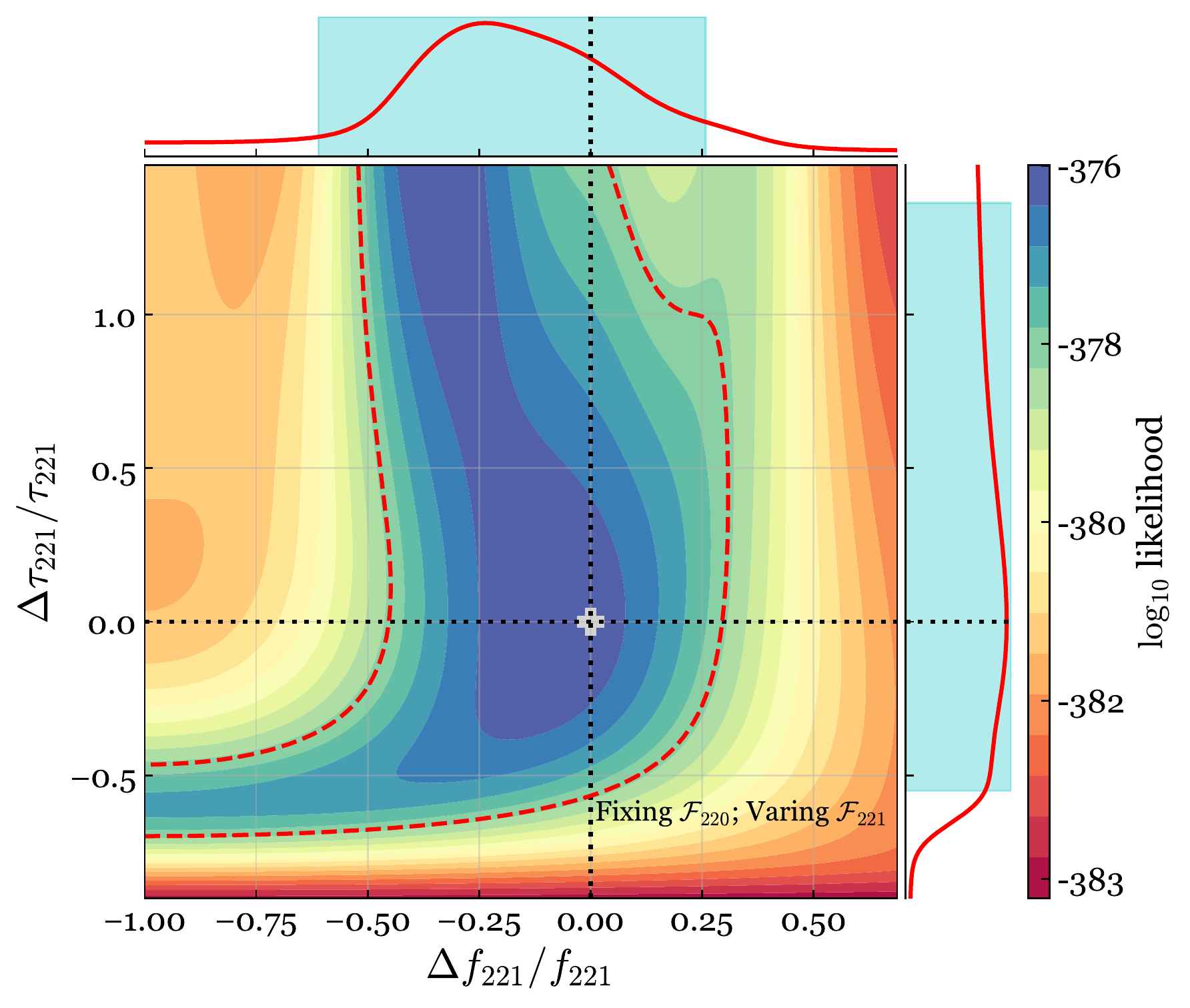}}
  \caption{Contours of likelihoods as a function of fractional deviations in the frequency ($\Delta f_{221}/f_{221}$) and decay rate ($\Delta \tau_{221}/\tau_{221}$) of the first overtone after the fundamental mode is removed from GW150914 [different $\Delta t_0$ times from (a) to (d)]. The $\mathcal{F}_{220}$ filter used to clean the fundamental mode is built using the IMR-estimated $M_f$ and $\chi_f$. The fiducial values to evaluate the fractional deviations are set to the IMR results. The red-dashed contours enclose the 2D 90\% credible region. The cyan-shaded regions on the side stand for the 1D 90\%-credible ranges of $\Delta f_{221}/f_{221}$ and $\Delta \tau_{221}/\tau_{221}$.}
 \label{fig:mix_contour_reim}
\end{figure*}

\section{Reproducing posteriors of $M_f$ and $\chi_f$ in Sec.~\ref{sec:mixed_appraoch} via a varying filter}
\label{app:sec:varying_filter}
In Sec.~\ref{sec:mixed_appraoch}, we discussed a mixed approach for BH spectroscopy, namely fitting the filtered data with MCMC. We have demonstrated that this hybrid method yields more information than either the full-RD MCMC or the pure filter method: Compared to the full-RD MCMC approach, the filter allows us to study subdominant QNMs by excluding the impact from dominant modes; compared to the pure filter method, we can still obtain the information about mode amplitudes.

With this mixed approach, we can infer the remnant properties ($M_f$ and $\chi_f$) exclusively from every single mode, e.g., Figs.~\ref{fig:low_res_1.5_mchi_ligo}, \ref{fig:low_res_1.0_mchi_ligo}, \ref{fig:mchi_ligo} and \ref{fig:mchi_ligo_1em1}. According to our discussions in Sec.~\ref{sec:filter_bayes} and the Supplemental Material of our companion paper \cite{Ma_prl}, the posteriors of $M_f$ and $\chi_f$ obtained via MCMC can be reproduced by purely using the rational filter. To demonstrate the equivalence between MCMC and pure filtering, here we use a fixed filter (built with the true values of $M_f$ and $\chi_f$) to remove one mode while varying $(M_f, \chi_f)$ in the other filter to find the best fit for the remaining mode --- this is different from the study in Sec.~\ref{sec:filter_bayes}, where all the filters are built from the same set of $(M_f, \chi_f)$. For example, in Figs.~\ref{fig:mixed_injection_contour_t1.5} and \ref{fig:mixed_injection_contour_t1.0} we reproduce the ``$\mathcal{F}_{221}+$one-QNM'' results in Figs.~\ref{fig:low_res_1.5_mchi_ligo} and \ref{fig:low_res_1.0_mchi_ligo} (red contours), where we vary $\mathcal{F}_{220}$ while fixing $\mathcal{F}_{221}$ to the injected true value. We can see the colored contours obtained by varying the filter are in agreement with the MCMC one (white dashed contours). Similarly, we vary $\mathcal{F}_{221}$ but fix $\mathcal{F}_{220}$ to reproduce the ``$\mathcal{F}_{220}+$one-QNM'' results (yellow contours in Figs.~\ref{fig:low_res_1.5_mchi_ligo} and \ref{fig:low_res_1.0_mchi_ligo}) in Figs.~\ref{fig:mixed_injection_contour_overtone_t1.5} and ~\ref{fig:mixed_injection_contour_overtone_t1.0}. The comparisons for GW150914 are similar and the results can be found in Fig.~\ref{fig:mixed_contour_150914}.

\begin{table}[tbh]
    \centering
    \setlength{\tabcolsep}{3pt}
	\renewcommand\arraystretch{1.2}
    \caption{The measurements of $\Delta f_{221}/f_{221}$ and $\Delta \tau_{221}/\tau_{221}$ (68\% credible intervals) with different choices of $\Delta t_0$, based on the ``$\mathcal{F}_{220}+$one-QNM'' scheme and the results in Fig.~\ref{fig:mix_contour_reim}. The fiducial values are set to the IMR results.}
    \begin{tabular}{c c c c c} \hline\hline
 $\Delta t_0$ (ms) &0.0 & 0.1 & 0.4 & 0.77 \\ \hline
$\Delta f_{221}/f_{221}$ & $-0.17^{+0.17}_{-0.13}$& $-0.18^{+0.18}_{-0.17}$ & $-0.17^{+0.22}_{-0.17}$ & $-0.18^{+0.25}_{-0.22}$   \\ \hline
$\Delta \tau_{221}/\tau_{221}$ & $0.38^{+0.48}_{-0.39}$ & $0.20^{+0.52}_{-0.42}$ &$0.40^{+0.64}_{-0.53}$ &$0.33^{+0.76}_{-0.64}$ \\ \hline\hline
     \end{tabular}
     \label{table:app:measure_kerr}
\end{table}

\section{Deviation from the Kerr assumption}
\label{app:sec:devitation_from_kerr}
The ``$\mathcal{F}_{220}+$one-QNM'' study in Appendix \ref{app:sec:varying_filter} (Figs.~\ref{fig:mixed_contour_overtone_t0.77} and \ref{fig:mixed_contour_overtone_t0.1}) is closely related to the beyond-Kerr fit discussed in Refs.~\cite{Isi:2019aib,Finch:2022ynt}, where the frequency and decay rate of the first overtone are allowed to differ from the no-hair values. Here we could do a similar thing by replacing the parameters, $M_f$ and $\chi_f$, with fractional deviations, $\Delta f_{221}/f_{221}$ and $\Delta \tau_{221}/\tau_{221}$, while varying $\mathcal{F}_{221}$.
We still build a fixed filter $\mathcal{F}_{220}$ using the IMR-estimated $M_f$ and $\chi_f$, and use these IMR results as the fiducial values. The resulting posterior distributions are shown in  Fig.~\ref{fig:mix_contour_reim}, evaluated at different $\Delta t_0$ times. The measurements of $\Delta f_{221}/f_{221}$ and $\Delta \tau_{221}/\tau_{221}$ are summarized in Table \ref{table:app:measure_kerr}, with 68\% credibility.


\def\bibsection{\section*{References}}
\bibliography{References}

\begin{thebibliography}{82}%
\makeatletter
\providecommand \@ifxundefined [1]{%
 \@ifx{#1\undefined}
}%
\providecommand \@ifnum [1]{%
 \ifnum #1\expandafter \@firstoftwo
 \else \expandafter \@secondoftwo
 \fi
}%
\providecommand \@ifx [1]{%
 \ifx #1\expandafter \@firstoftwo
 \else \expandafter \@secondoftwo
 \fi
}%
\providecommand \natexlab [1]{#1}%
\providecommand \enquote  [1]{``#1''}%
\providecommand \bibnamefont  [1]{#1}%
\providecommand \bibfnamefont [1]{#1}%
\providecommand \citenamefont [1]{#1}%
\providecommand \href@noop [0]{\@secondoftwo}%
\providecommand \href [0]{\begingroup \@sanitize@url \@href}%
\providecommand \@href[1]{\@@startlink{#1}\@@href}%
\providecommand \@@href[1]{\endgroup#1\@@endlink}%
\providecommand \@sanitize@url [0]{\catcode `\\12\catcode `\$12\catcode
  `\&12\catcode `\#12\catcode `\^12\catcode `\_12\catcode `\%12\relax}%
\providecommand \@@startlink[1]{}%
\providecommand \@@endlink[0]{}%
\providecommand \url  [0]{\begingroup\@sanitize@url \@url }%
\providecommand \@url [1]{\endgroup\@href {#1}{\urlprefix }}%
\providecommand \urlprefix  [0]{URL }%
\providecommand \Eprint [0]{\href }%
\providecommand \doibase [0]{http://dx.doi.org/}%
\providecommand \selectlanguage [0]{\@gobble}%
\providecommand \bibinfo  [0]{\@secondoftwo}%
\providecommand \bibfield  [0]{\@secondoftwo}%
\providecommand \translation [1]{[#1]}%
\providecommand \BibitemOpen [0]{}%
\providecommand \bibitemStop [0]{}%
\providecommand \bibitemNoStop [0]{.\EOS\space}%
\providecommand \EOS [0]{\spacefactor3000\relax}%
\providecommand \BibitemShut  [1]{\csname bibitem#1\endcsname}%
\let\auto@bib@innerbib\@empty
\bibitem [{\citenamefont {Kokkotas}\ and\ \citenamefont
  {Schmidt}(1999)}]{Kokkotas:1999bd}%
  \BibitemOpen
  \bibfield  {author} {\bibinfo {author} {\bibfnamefont {Kostas~D.}\
  \bibnamefont {Kokkotas}}\ and\ \bibinfo {author} {\bibfnamefont {Bernd~G.}\
  \bibnamefont {Schmidt}},\ }\bibfield  {title} {\enquote {\bibinfo {title}
  {{Quasinormal modes of stars and black holes}},}\ }\href {\doibase
  10.12942/lrr-1999-2} {\bibfield  {journal} {\bibinfo  {journal} {Living Rev.
  Rel.}\ }\textbf {\bibinfo {volume} {2}},\ \bibinfo {pages} {2} (\bibinfo
  {year} {1999})},\ \Eprint {http://arxiv.org/abs/gr-qc/9909058}
  {arXiv:gr-qc/9909058} \BibitemShut {NoStop}%
\bibitem [{\citenamefont {Nollert}(1999)}]{Nollert_1999}%
  \BibitemOpen
  \bibfield  {author} {\bibinfo {author} {\bibfnamefont {Hans-Peter}\
  \bibnamefont {Nollert}},\ }\bibfield  {title} {\enquote {\bibinfo {title}
  {Quasinormal modes: the characteristic
  {\textasciigrave}sound{\textquotesingle} of black holes and neutron stars},}\
  }\href {\doibase 10.1088/0264-9381/16/12/201} {\bibfield  {journal} {\bibinfo
   {journal} {Classical and Quantum Gravity}\ }\textbf {\bibinfo {volume}
  {16}},\ \bibinfo {pages} {R159--R216} (\bibinfo {year} {1999})}\BibitemShut
  {NoStop}%
\bibitem [{\citenamefont {Cardoso}\ and\ \citenamefont
  {Gualtieri}(2016)}]{Cardoso:2016ryw}%
  \BibitemOpen
  \bibfield  {author} {\bibinfo {author} {\bibfnamefont {Vitor}\ \bibnamefont
  {Cardoso}}\ and\ \bibinfo {author} {\bibfnamefont {Leonardo}\ \bibnamefont
  {Gualtieri}},\ }\bibfield  {title} {\enquote {\bibinfo {title} {{Testing the
  black hole \textquoteleft{}no-hair\textquoteright{} hypothesis}},}\ }\href
  {\doibase 10.1088/0264-9381/33/17/174001} {\bibfield  {journal} {\bibinfo
  {journal} {Class. Quant. Grav.}\ }\textbf {\bibinfo {volume} {33}},\ \bibinfo
  {pages} {174001} (\bibinfo {year} {2016})},\ \Eprint
  {http://arxiv.org/abs/1607.03133} {arXiv:1607.03133 [gr-qc]} \BibitemShut
  {NoStop}%
\bibitem [{\citenamefont {Berti}\ \emph {et~al.}(2009)\citenamefont {Berti},
  \citenamefont {Cardoso},\ and\ \citenamefont {Starinets}}]{Berti:2009kk}%
  \BibitemOpen
  \bibfield  {author} {\bibinfo {author} {\bibfnamefont {Emanuele}\
  \bibnamefont {Berti}}, \bibinfo {author} {\bibfnamefont {Vitor}\ \bibnamefont
  {Cardoso}}, \ and\ \bibinfo {author} {\bibfnamefont {Andrei~O.}\ \bibnamefont
  {Starinets}},\ }\bibfield  {title} {\enquote {\bibinfo {title} {{Quasinormal
  modes of black holes and black branes}},}\ }\href {\doibase
  10.1088/0264-9381/26/16/163001} {\bibfield  {journal} {\bibinfo  {journal}
  {Class. Quant. Grav.}\ }\textbf {\bibinfo {volume} {26}},\ \bibinfo {pages}
  {163001} (\bibinfo {year} {2009})},\ \Eprint {http://arxiv.org/abs/0905.2975}
  {arXiv:0905.2975 [gr-qc]} \BibitemShut {NoStop}%
\bibitem [{\citenamefont {Penrose}(2002)}]{penrose2002golden}%
  \BibitemOpen
  \bibfield  {author} {\bibinfo {author} {\bibfnamefont {Roger}\ \bibnamefont
  {Penrose}},\ }\bibfield  {title} {\enquote {\bibinfo {title} {“golden
  oldie”: Gravitational collapse: the role of general relativity},}\
  }\href@noop {} {\bibfield  {journal} {\bibinfo  {journal} {General Relativity
  and Gravitation}\ }\textbf {\bibinfo {volume} {34}},\ \bibinfo {pages}
  {1141--1165} (\bibinfo {year} {2002})}\BibitemShut {NoStop}%
\bibitem [{\citenamefont {Chrusciel}\ \emph {et~al.}(2012)\citenamefont
  {Chrusciel}, \citenamefont {Lopes~Costa},\ and\ \citenamefont
  {Heusler}}]{Chrusciel:2012jk}%
  \BibitemOpen
  \bibfield  {author} {\bibinfo {author} {\bibfnamefont {Piotr~T.}\
  \bibnamefont {Chrusciel}}, \bibinfo {author} {\bibfnamefont {Joao}\
  \bibnamefont {Lopes~Costa}}, \ and\ \bibinfo {author} {\bibfnamefont
  {Markus}\ \bibnamefont {Heusler}},\ }\bibfield  {title} {\enquote {\bibinfo
  {title} {{Stationary Black Holes: Uniqueness and Beyond}},}\ }\href {\doibase
  10.12942/lrr-2012-7} {\bibfield  {journal} {\bibinfo  {journal} {Living Rev.
  Rel.}\ }\textbf {\bibinfo {volume} {15}},\ \bibinfo {pages} {7} (\bibinfo
  {year} {2012})},\ \Eprint {http://arxiv.org/abs/1205.6112} {arXiv:1205.6112
  [gr-qc]} \BibitemShut {NoStop}%
\bibitem [{\citenamefont {Carter}(1971)}]{PhysRevLett.26.331}%
  \BibitemOpen
  \bibfield  {author} {\bibinfo {author} {\bibfnamefont {B.}~\bibnamefont
  {Carter}},\ }\bibfield  {title} {\enquote {\bibinfo {title} {Axisymmetric
  black hole has only two degrees of freedom},}\ }\href {\doibase
  10.1103/PhysRevLett.26.331} {\bibfield  {journal} {\bibinfo  {journal} {Phys.
  Rev. Lett.}\ }\textbf {\bibinfo {volume} {26}},\ \bibinfo {pages} {331--333}
  (\bibinfo {year} {1971})}\BibitemShut {NoStop}%
\bibitem [{\citenamefont {Israel}(1967)}]{PhysRev.164.1776}%
  \BibitemOpen
  \bibfield  {author} {\bibinfo {author} {\bibfnamefont {Werner}\ \bibnamefont
  {Israel}},\ }\bibfield  {title} {\enquote {\bibinfo {title} {Event horizons
  in static vacuum space-times},}\ }\href {\doibase 10.1103/PhysRev.164.1776}
  {\bibfield  {journal} {\bibinfo  {journal} {Phys. Rev.}\ }\textbf {\bibinfo
  {volume} {164}},\ \bibinfo {pages} {1776--1779} (\bibinfo {year}
  {1967})}\BibitemShut {NoStop}%
\bibitem [{\citenamefont {Echeverria}(1989)}]{PhysRevD.40.3194}%
  \BibitemOpen
  \bibfield  {author} {\bibinfo {author} {\bibfnamefont {Fernando}\
  \bibnamefont {Echeverria}},\ }\bibfield  {title} {\enquote {\bibinfo {title}
  {Gravitational-wave measurements of the mass and angular momentum of a black
  hole},}\ }\href {\doibase 10.1103/PhysRevD.40.3194} {\bibfield  {journal}
  {\bibinfo  {journal} {Phys. Rev. D}\ }\textbf {\bibinfo {volume} {40}},\
  \bibinfo {pages} {3194--3203} (\bibinfo {year} {1989})}\BibitemShut {NoStop}%
\bibitem [{\citenamefont {Dreyer}\ \emph {et~al.}(2004)\citenamefont {Dreyer},
  \citenamefont {Kelly}, \citenamefont {Krishnan}, \citenamefont {Finn},
  \citenamefont {Garrison},\ and\ \citenamefont
  {Lopez-Aleman}}]{Dreyer:2003bv}%
  \BibitemOpen
  \bibfield  {author} {\bibinfo {author} {\bibfnamefont {Olaf}\ \bibnamefont
  {Dreyer}}, \bibinfo {author} {\bibfnamefont {Bernard~J.}\ \bibnamefont
  {Kelly}}, \bibinfo {author} {\bibfnamefont {Badri}\ \bibnamefont {Krishnan}},
  \bibinfo {author} {\bibfnamefont {Lee~Samuel}\ \bibnamefont {Finn}}, \bibinfo
  {author} {\bibfnamefont {David}\ \bibnamefont {Garrison}}, \ and\ \bibinfo
  {author} {\bibfnamefont {Ramon}\ \bibnamefont {Lopez-Aleman}},\ }\bibfield
  {title} {\enquote {\bibinfo {title} {{Black hole spectroscopy: Testing
  general relativity through gravitational wave observations}},}\ }\href
  {\doibase 10.1088/0264-9381/21/4/003} {\bibfield  {journal} {\bibinfo
  {journal} {Class. Quant. Grav.}\ }\textbf {\bibinfo {volume} {21}},\ \bibinfo
  {pages} {787--804} (\bibinfo {year} {2004})},\ \Eprint
  {http://arxiv.org/abs/gr-qc/0309007} {arXiv:gr-qc/0309007 [gr-qc]}
  \BibitemShut {NoStop}%
\bibitem [{\citenamefont {Berti}\ \emph {et~al.}(2006)\citenamefont {Berti},
  \citenamefont {Cardoso},\ and\ \citenamefont {Will}}]{Berti:2005ys}%
  \BibitemOpen
  \bibfield  {author} {\bibinfo {author} {\bibfnamefont {Emanuele}\
  \bibnamefont {Berti}}, \bibinfo {author} {\bibfnamefont {Vitor}\ \bibnamefont
  {Cardoso}}, \ and\ \bibinfo {author} {\bibfnamefont {Clifford~M.}\
  \bibnamefont {Will}},\ }\bibfield  {title} {\enquote {\bibinfo {title} {{On
  gravitational-wave spectroscopy of massive black holes with the space
  interferometer LISA}},}\ }\href {\doibase 10.1103/PhysRevD.73.064030}
  {\bibfield  {journal} {\bibinfo  {journal} {Phys. Rev.}\ }\textbf {\bibinfo
  {volume} {D73}},\ \bibinfo {pages} {064030} (\bibinfo {year} {2006})},\
  \Eprint {http://arxiv.org/abs/gr-qc/0512160} {arXiv:gr-qc/0512160 [gr-qc]}
  \BibitemShut {NoStop}%
\bibitem [{\citenamefont {Berti}\ \emph
  {et~al.}(2007{\natexlab{a}})\citenamefont {Berti}, \citenamefont {Cardoso},
  \citenamefont {Cardoso},\ and\ \citenamefont {Cavaglia}}]{Berti:2007zu}%
  \BibitemOpen
  \bibfield  {author} {\bibinfo {author} {\bibfnamefont {Emanuele}\
  \bibnamefont {Berti}}, \bibinfo {author} {\bibfnamefont {Jaime}\ \bibnamefont
  {Cardoso}}, \bibinfo {author} {\bibfnamefont {Vitor}\ \bibnamefont
  {Cardoso}}, \ and\ \bibinfo {author} {\bibfnamefont {Marco}\ \bibnamefont
  {Cavaglia}},\ }\bibfield  {title} {\enquote {\bibinfo {title}
  {{Matched-filtering and parameter estimation of ringdown waveforms}},}\
  }\href {\doibase 10.1103/PhysRevD.76.104044} {\bibfield  {journal} {\bibinfo
  {journal} {Phys. Rev. D}\ }\textbf {\bibinfo {volume} {76}},\ \bibinfo
  {pages} {104044} (\bibinfo {year} {2007}{\natexlab{a}})},\ \Eprint
  {http://arxiv.org/abs/0707.1202} {arXiv:0707.1202 [gr-qc]} \BibitemShut
  {NoStop}%
\bibitem [{\citenamefont {Gossan}\ \emph {et~al.}(2012)\citenamefont {Gossan},
  \citenamefont {Veitch},\ and\ \citenamefont {Sathyaprakash}}]{Gossan:2011ha}%
  \BibitemOpen
  \bibfield  {author} {\bibinfo {author} {\bibfnamefont {S.}~\bibnamefont
  {Gossan}}, \bibinfo {author} {\bibfnamefont {J.}~\bibnamefont {Veitch}}, \
  and\ \bibinfo {author} {\bibfnamefont {B.~S.}\ \bibnamefont
  {Sathyaprakash}},\ }\bibfield  {title} {\enquote {\bibinfo {title} {{Bayesian
  model selection for testing the no-hair theorem with black hole
  ringdowns}},}\ }\href {\doibase 10.1103/PhysRevD.85.124056} {\bibfield
  {journal} {\bibinfo  {journal} {Phys. Rev.}\ }\textbf {\bibinfo {volume}
  {D85}},\ \bibinfo {pages} {124056} (\bibinfo {year} {2012})},\ \Eprint
  {http://arxiv.org/abs/1111.5819} {arXiv:1111.5819 [gr-qc]} \BibitemShut
  {NoStop}%
\bibitem [{\citenamefont {Caudill}\ \emph {et~al.}(2012)\citenamefont
  {Caudill}, \citenamefont {Field}, \citenamefont {Galley}, \citenamefont
  {Herrmann},\ and\ \citenamefont {Tiglio}}]{Caudill:2011kv}%
  \BibitemOpen
  \bibfield  {author} {\bibinfo {author} {\bibfnamefont {Sarah}\ \bibnamefont
  {Caudill}}, \bibinfo {author} {\bibfnamefont {Scott~E.}\ \bibnamefont
  {Field}}, \bibinfo {author} {\bibfnamefont {Chad~R.}\ \bibnamefont {Galley}},
  \bibinfo {author} {\bibfnamefont {Frank}\ \bibnamefont {Herrmann}}, \ and\
  \bibinfo {author} {\bibfnamefont {Manuel}\ \bibnamefont {Tiglio}},\
  }\bibfield  {title} {\enquote {\bibinfo {title} {{Reduced Basis
  representations of multi-mode black hole ringdown gravitational waves}},}\
  }\href {\doibase 10.1088/0264-9381/29/9/095016} {\bibfield  {journal}
  {\bibinfo  {journal} {Class. Quant. Grav.}\ }\textbf {\bibinfo {volume}
  {29}},\ \bibinfo {pages} {095016} (\bibinfo {year} {2012})},\ \Eprint
  {http://arxiv.org/abs/1109.5642} {arXiv:1109.5642 [gr-qc]} \BibitemShut
  {NoStop}%
\bibitem [{\citenamefont {Meidam}\ \emph {et~al.}(2014)\citenamefont {Meidam},
  \citenamefont {Agathos}, \citenamefont {Van Den~Broeck}, \citenamefont
  {Veitch},\ and\ \citenamefont {Sathyaprakash}}]{Meidam:2014jpa}%
  \BibitemOpen
  \bibfield  {author} {\bibinfo {author} {\bibfnamefont {J.}~\bibnamefont
  {Meidam}}, \bibinfo {author} {\bibfnamefont {M.}~\bibnamefont {Agathos}},
  \bibinfo {author} {\bibfnamefont {C.}~\bibnamefont {Van Den~Broeck}},
  \bibinfo {author} {\bibfnamefont {J.}~\bibnamefont {Veitch}}, \ and\ \bibinfo
  {author} {\bibfnamefont {B.~S.}\ \bibnamefont {Sathyaprakash}},\ }\bibfield
  {title} {\enquote {\bibinfo {title} {{Testing the no-hair theorem with black
  hole ringdowns using TIGER}},}\ }\href {\doibase 10.1103/PhysRevD.90.064009}
  {\bibfield  {journal} {\bibinfo  {journal} {Phys. Rev.}\ }\textbf {\bibinfo
  {volume} {D90}},\ \bibinfo {pages} {064009} (\bibinfo {year} {2014})},\
  \Eprint {http://arxiv.org/abs/1406.3201} {arXiv:1406.3201 [gr-qc]}
  \BibitemShut {NoStop}%
\bibitem [{\citenamefont {Bhagwat}\ \emph {et~al.}(2016)\citenamefont
  {Bhagwat}, \citenamefont {Brown},\ and\ \citenamefont
  {Ballmer}}]{Bhagwat:2016ntk}%
  \BibitemOpen
  \bibfield  {author} {\bibinfo {author} {\bibfnamefont {Swetha}\ \bibnamefont
  {Bhagwat}}, \bibinfo {author} {\bibfnamefont {Duncan~A.}\ \bibnamefont
  {Brown}}, \ and\ \bibinfo {author} {\bibfnamefont {Stefan~W.}\ \bibnamefont
  {Ballmer}},\ }\bibfield  {title} {\enquote {\bibinfo {title} {{Spectroscopic
  analysis of stellar mass black-hole mergers in our local universe with
  ground-based gravitational wave detectors}},}\ }\href {\doibase
  10.1103/PhysRevD.94.084024} {\bibfield  {journal} {\bibinfo  {journal} {Phys.
  Rev. D}\ }\textbf {\bibinfo {volume} {94}},\ \bibinfo {pages} {084024}
  (\bibinfo {year} {2016})},\ \bibinfo {note} {[Erratum: Phys.Rev.D 95, 069906
  (2017)]},\ \Eprint {http://arxiv.org/abs/1607.07845} {arXiv:1607.07845
  [gr-qc]} \BibitemShut {NoStop}%
\bibitem [{\citenamefont {Berti}\ \emph {et~al.}(2016)\citenamefont {Berti},
  \citenamefont {Sesana}, \citenamefont {Barausse}, \citenamefont {Cardoso},\
  and\ \citenamefont {Belczynski}}]{Berti:2016lat}%
  \BibitemOpen
  \bibfield  {author} {\bibinfo {author} {\bibfnamefont {Emanuele}\
  \bibnamefont {Berti}}, \bibinfo {author} {\bibfnamefont {Alberto}\
  \bibnamefont {Sesana}}, \bibinfo {author} {\bibfnamefont {Enrico}\
  \bibnamefont {Barausse}}, \bibinfo {author} {\bibfnamefont {Vitor}\
  \bibnamefont {Cardoso}}, \ and\ \bibinfo {author} {\bibfnamefont {Krzysztof}\
  \bibnamefont {Belczynski}},\ }\bibfield  {title} {\enquote {\bibinfo {title}
  {{Spectroscopy of Kerr black holes with Earth- and space-based
  interferometers}},}\ }\href {\doibase 10.1103/PhysRevLett.117.101102}
  {\bibfield  {journal} {\bibinfo  {journal} {Phys. Rev. Lett.}\ }\textbf
  {\bibinfo {volume} {117}},\ \bibinfo {pages} {101102} (\bibinfo {year}
  {2016})},\ \Eprint {http://arxiv.org/abs/1605.09286} {arXiv:1605.09286
  [gr-qc]} \BibitemShut {NoStop}%
\bibitem [{\citenamefont {Baibhav}\ \emph {et~al.}(2018)\citenamefont
  {Baibhav}, \citenamefont {Berti}, \citenamefont {Cardoso},\ and\
  \citenamefont {Khanna}}]{Baibhav:2017jhs}%
  \BibitemOpen
  \bibfield  {author} {\bibinfo {author} {\bibfnamefont {Vishal}\ \bibnamefont
  {Baibhav}}, \bibinfo {author} {\bibfnamefont {Emanuele}\ \bibnamefont
  {Berti}}, \bibinfo {author} {\bibfnamefont {Vitor}\ \bibnamefont {Cardoso}},
  \ and\ \bibinfo {author} {\bibfnamefont {Gaurav}\ \bibnamefont {Khanna}},\
  }\bibfield  {title} {\enquote {\bibinfo {title} {{Black Hole Spectroscopy:
  Systematic Errors and Ringdown Energy Estimates}},}\ }\href {\doibase
  10.1103/PhysRevD.97.044048} {\bibfield  {journal} {\bibinfo  {journal} {Phys.
  Rev. D}\ }\textbf {\bibinfo {volume} {97}},\ \bibinfo {pages} {044048}
  (\bibinfo {year} {2018})},\ \Eprint {http://arxiv.org/abs/1710.02156}
  {arXiv:1710.02156 [gr-qc]} \BibitemShut {NoStop}%
\bibitem [{\citenamefont {Maselli}\ \emph {et~al.}(2017)\citenamefont
  {Maselli}, \citenamefont {Kokkotas},\ and\ \citenamefont
  {Laguna}}]{Maselli:2017kvl}%
  \BibitemOpen
  \bibfield  {author} {\bibinfo {author} {\bibfnamefont {Andrea}\ \bibnamefont
  {Maselli}}, \bibinfo {author} {\bibfnamefont {Kostas}\ \bibnamefont
  {Kokkotas}}, \ and\ \bibinfo {author} {\bibfnamefont {Pablo}\ \bibnamefont
  {Laguna}},\ }\bibfield  {title} {\enquote {\bibinfo {title} {{Observing
  binary black hole ringdowns by advanced gravitational wave detectors}},}\
  }\href {\doibase 10.1103/PhysRevD.95.104026} {\bibfield  {journal} {\bibinfo
  {journal} {Phys. Rev. D}\ }\textbf {\bibinfo {volume} {95}},\ \bibinfo
  {pages} {104026} (\bibinfo {year} {2017})},\ \Eprint
  {http://arxiv.org/abs/1702.01110} {arXiv:1702.01110 [gr-qc]} \BibitemShut
  {NoStop}%
\bibitem [{\citenamefont {Yang}\ \emph {et~al.}(2017)\citenamefont {Yang},
  \citenamefont {Yagi}, \citenamefont {Blackman}, \citenamefont {Lehner},
  \citenamefont {Paschalidis}, \citenamefont {Pretorius},\ and\ \citenamefont
  {Yunes}}]{Yang:2017zxs}%
  \BibitemOpen
  \bibfield  {author} {\bibinfo {author} {\bibfnamefont {Huan}\ \bibnamefont
  {Yang}}, \bibinfo {author} {\bibfnamefont {Kent}\ \bibnamefont {Yagi}},
  \bibinfo {author} {\bibfnamefont {Jonathan}\ \bibnamefont {Blackman}},
  \bibinfo {author} {\bibfnamefont {Luis}\ \bibnamefont {Lehner}}, \bibinfo
  {author} {\bibfnamefont {Vasileios}\ \bibnamefont {Paschalidis}}, \bibinfo
  {author} {\bibfnamefont {Frans}\ \bibnamefont {Pretorius}}, \ and\ \bibinfo
  {author} {\bibfnamefont {Nicolás}\ \bibnamefont {Yunes}},\ }\bibfield
  {title} {\enquote {\bibinfo {title} {{Black hole spectroscopy with coherent
  mode stacking}},}\ }\href {\doibase 10.1103/PhysRevLett.118.161101}
  {\bibfield  {journal} {\bibinfo  {journal} {Phys. Rev. Lett.}\ }\textbf
  {\bibinfo {volume} {118}},\ \bibinfo {pages} {161101} (\bibinfo {year}
  {2017})},\ \Eprint {http://arxiv.org/abs/1701.05808} {arXiv:1701.05808
  [gr-qc]} \BibitemShut {NoStop}%
\bibitem [{\citenamefont {Da~Silva~Costa}\ \emph {et~al.}(2018)\citenamefont
  {Da~Silva~Costa}, \citenamefont {Tiwari}, \citenamefont {Klimenko},\ and\
  \citenamefont {Salemi}}]{DaSilvaCosta:2017njq}%
  \BibitemOpen
  \bibfield  {author} {\bibinfo {author} {\bibfnamefont {C.~F.}\ \bibnamefont
  {Da~Silva~Costa}}, \bibinfo {author} {\bibfnamefont {S.}~\bibnamefont
  {Tiwari}}, \bibinfo {author} {\bibfnamefont {S.}~\bibnamefont {Klimenko}}, \
  and\ \bibinfo {author} {\bibfnamefont {F.}~\bibnamefont {Salemi}},\
  }\bibfield  {title} {\enquote {\bibinfo {title} {{Detection of (2,2)
  quasinormal mode from a population of black holes with a constructive
  summation method}},}\ }\href {\doibase 10.1103/PhysRevD.98.024052} {\bibfield
   {journal} {\bibinfo  {journal} {Phys. Rev. D}\ }\textbf {\bibinfo {volume}
  {98}},\ \bibinfo {pages} {024052} (\bibinfo {year} {2018})},\ \Eprint
  {http://arxiv.org/abs/1711.00551} {arXiv:1711.00551 [gr-qc]} \BibitemShut
  {NoStop}%
\bibitem [{\citenamefont {Baibhav}\ and\ \citenamefont
  {Berti}(2019)}]{Baibhav:2018rfk}%
  \BibitemOpen
  \bibfield  {author} {\bibinfo {author} {\bibfnamefont {Vishal}\ \bibnamefont
  {Baibhav}}\ and\ \bibinfo {author} {\bibfnamefont {Emanuele}\ \bibnamefont
  {Berti}},\ }\bibfield  {title} {\enquote {\bibinfo {title} {{Multimode black
  hole spectroscopy}},}\ }\href {\doibase 10.1103/PhysRevD.99.024005}
  {\bibfield  {journal} {\bibinfo  {journal} {Phys. Rev. D}\ }\textbf {\bibinfo
  {volume} {99}},\ \bibinfo {pages} {024005} (\bibinfo {year} {2019})},\
  \Eprint {http://arxiv.org/abs/1809.03500} {arXiv:1809.03500 [gr-qc]}
  \BibitemShut {NoStop}%
\bibitem [{\citenamefont {Carullo}\ \emph {et~al.}(2018)\citenamefont {Carullo}
  \emph {et~al.}}]{Carullo:2018sfu}%
  \BibitemOpen
  \bibfield  {author} {\bibinfo {author} {\bibfnamefont {Gregorio}\
  \bibnamefont {Carullo}} \emph {et~al.},\ }\bibfield  {title} {\enquote
  {\bibinfo {title} {{Empirical tests of the black hole no-hair conjecture
  using gravitational-wave observations}},}\ }\href {\doibase
  10.1103/PhysRevD.98.104020} {\bibfield  {journal} {\bibinfo  {journal} {Phys.
  Rev.}\ }\textbf {\bibinfo {volume} {D98}},\ \bibinfo {pages} {104020}
  (\bibinfo {year} {2018})},\ \Eprint {http://arxiv.org/abs/1805.04760}
  {arXiv:1805.04760 [gr-qc]} \BibitemShut {NoStop}%
\bibitem [{\citenamefont {Brito}\ \emph {et~al.}(2018)\citenamefont {Brito},
  \citenamefont {Buonanno},\ and\ \citenamefont {Raymond}}]{Brito:2018rfr}%
  \BibitemOpen
  \bibfield  {author} {\bibinfo {author} {\bibfnamefont {Richard}\ \bibnamefont
  {Brito}}, \bibinfo {author} {\bibfnamefont {Alessandra}\ \bibnamefont
  {Buonanno}}, \ and\ \bibinfo {author} {\bibfnamefont {Vivien}\ \bibnamefont
  {Raymond}},\ }\bibfield  {title} {\enquote {\bibinfo {title} {{Black-hole
  Spectroscopy by Making Full Use of Gravitational-Wave Modeling}},}\ }\href
  {\doibase 10.1103/PhysRevD.98.084038} {\bibfield  {journal} {\bibinfo
  {journal} {Phys. Rev.}\ }\textbf {\bibinfo {volume} {D98}},\ \bibinfo {pages}
  {084038} (\bibinfo {year} {2018})},\ \Eprint
  {http://arxiv.org/abs/1805.00293} {arXiv:1805.00293 [gr-qc]} \BibitemShut
  {NoStop}%
\bibitem [{\citenamefont {Nakano}\ \emph {et~al.}(2019)\citenamefont {Nakano},
  \citenamefont {Narikawa}, \citenamefont {Oohara}, \citenamefont {Sakai},
  \citenamefont {Shinkai}, \citenamefont {Takahashi}, \citenamefont {Tanaka},
  \citenamefont {Uchikata}, \citenamefont {Yamamoto},\ and\ \citenamefont
  {Yamamoto}}]{Nakano:2018vay}%
  \BibitemOpen
  \bibfield  {author} {\bibinfo {author} {\bibfnamefont {Hiroyuki}\
  \bibnamefont {Nakano}}, \bibinfo {author} {\bibfnamefont {Tatsuya}\
  \bibnamefont {Narikawa}}, \bibinfo {author} {\bibfnamefont {Ken-ichi}\
  \bibnamefont {Oohara}}, \bibinfo {author} {\bibfnamefont {Kazuki}\
  \bibnamefont {Sakai}}, \bibinfo {author} {\bibfnamefont {Hisa-aki}\
  \bibnamefont {Shinkai}}, \bibinfo {author} {\bibfnamefont {Hirotaka}\
  \bibnamefont {Takahashi}}, \bibinfo {author} {\bibfnamefont {Takahiro}\
  \bibnamefont {Tanaka}}, \bibinfo {author} {\bibfnamefont {Nami}\ \bibnamefont
  {Uchikata}}, \bibinfo {author} {\bibfnamefont {Shun}\ \bibnamefont
  {Yamamoto}}, \ and\ \bibinfo {author} {\bibfnamefont {Takahiro~S.}\
  \bibnamefont {Yamamoto}},\ }\bibfield  {title} {\enquote {\bibinfo {title}
  {{Comparison of various methods to extract ringdown frequency from
  gravitational wave data}},}\ }\href {\doibase 10.1103/PhysRevD.99.124032}
  {\bibfield  {journal} {\bibinfo  {journal} {Phys. Rev. D}\ }\textbf {\bibinfo
  {volume} {99}},\ \bibinfo {pages} {124032} (\bibinfo {year} {2019})},\
  \Eprint {http://arxiv.org/abs/1811.06443} {arXiv:1811.06443 [gr-qc]}
  \BibitemShut {NoStop}%
\bibitem [{\citenamefont {Cabero}\ \emph {et~al.}(2020)\citenamefont {Cabero},
  \citenamefont {Westerweck}, \citenamefont {Capano}, \citenamefont {Kumar},
  \citenamefont {Nielsen},\ and\ \citenamefont {Krishnan}}]{Cabero:2019zyt}%
  \BibitemOpen
  \bibfield  {author} {\bibinfo {author} {\bibfnamefont {Miriam}\ \bibnamefont
  {Cabero}}, \bibinfo {author} {\bibfnamefont {Julian}\ \bibnamefont
  {Westerweck}}, \bibinfo {author} {\bibfnamefont {Collin~D.}\ \bibnamefont
  {Capano}}, \bibinfo {author} {\bibfnamefont {Sumit}\ \bibnamefont {Kumar}},
  \bibinfo {author} {\bibfnamefont {Alex~B.}\ \bibnamefont {Nielsen}}, \ and\
  \bibinfo {author} {\bibfnamefont {Badri}\ \bibnamefont {Krishnan}},\
  }\bibfield  {title} {\enquote {\bibinfo {title} {{Black hole spectroscopy in
  the next decade}},}\ }\href {\doibase 10.1103/PhysRevD.101.064044} {\bibfield
   {journal} {\bibinfo  {journal} {Phys. Rev. D}\ }\textbf {\bibinfo {volume}
  {101}},\ \bibinfo {pages} {064044} (\bibinfo {year} {2020})},\ \Eprint
  {http://arxiv.org/abs/1911.01361} {arXiv:1911.01361 [gr-qc]} \BibitemShut
  {NoStop}%
\bibitem [{\citenamefont {Bhagwat}\ \emph
  {et~al.}(2020{\natexlab{a}})\citenamefont {Bhagwat}, \citenamefont {Cabero},
  \citenamefont {Capano}, \citenamefont {Krishnan},\ and\ \citenamefont
  {Brown}}]{Bhagwat:2019bwv}%
  \BibitemOpen
  \bibfield  {author} {\bibinfo {author} {\bibfnamefont {Swetha}\ \bibnamefont
  {Bhagwat}}, \bibinfo {author} {\bibfnamefont {Miriam}\ \bibnamefont
  {Cabero}}, \bibinfo {author} {\bibfnamefont {Collin~D.}\ \bibnamefont
  {Capano}}, \bibinfo {author} {\bibfnamefont {Badri}\ \bibnamefont
  {Krishnan}}, \ and\ \bibinfo {author} {\bibfnamefont {Duncan~A.}\
  \bibnamefont {Brown}},\ }\bibfield  {title} {\enquote {\bibinfo {title}
  {{Detectability of the subdominant mode in a binary black hole ringdown}},}\
  }\href {\doibase 10.1103/PhysRevD.102.024023} {\bibfield  {journal} {\bibinfo
   {journal} {Phys. Rev. D}\ }\textbf {\bibinfo {volume} {102}},\ \bibinfo
  {pages} {024023} (\bibinfo {year} {2020}{\natexlab{a}})},\ \Eprint
  {http://arxiv.org/abs/1910.13203} {arXiv:1910.13203 [gr-qc]} \BibitemShut
  {NoStop}%
\bibitem [{\citenamefont {Ota}\ and\ \citenamefont
  {Chirenti}(2020)}]{Ota:2019bzl}%
  \BibitemOpen
  \bibfield  {author} {\bibinfo {author} {\bibfnamefont {Iara}\ \bibnamefont
  {Ota}}\ and\ \bibinfo {author} {\bibfnamefont {Cecilia}\ \bibnamefont
  {Chirenti}},\ }\bibfield  {title} {\enquote {\bibinfo {title} {{Overtones or
  higher harmonics? Prospects for testing the no-hair theorem with
  gravitational wave detections}},}\ }\href {\doibase
  10.1103/PhysRevD.101.104005} {\bibfield  {journal} {\bibinfo  {journal}
  {Phys. Rev. D}\ }\textbf {\bibinfo {volume} {101}},\ \bibinfo {pages}
  {104005} (\bibinfo {year} {2020})},\ \Eprint
  {http://arxiv.org/abs/1911.00440} {arXiv:1911.00440 [gr-qc]} \BibitemShut
  {NoStop}%
\bibitem [{\citenamefont {Bustillo}\ \emph {et~al.}(2021)\citenamefont
  {Bustillo}, \citenamefont {Lasky},\ and\ \citenamefont
  {Thrane}}]{Bustillo:2020buq}%
  \BibitemOpen
  \bibfield  {author} {\bibinfo {author} {\bibfnamefont {Juan~Calder\'on}\
  \bibnamefont {Bustillo}}, \bibinfo {author} {\bibfnamefont {Paul~D.}\
  \bibnamefont {Lasky}}, \ and\ \bibinfo {author} {\bibfnamefont {Eric}\
  \bibnamefont {Thrane}},\ }\bibfield  {title} {\enquote {\bibinfo {title}
  {{Black-hole spectroscopy, the no-hair theorem, and GW150914: Kerr versus
  Occam}},}\ }\href {\doibase 10.1103/PhysRevD.103.024041} {\bibfield
  {journal} {\bibinfo  {journal} {Phys. Rev. D}\ }\textbf {\bibinfo {volume}
  {103}},\ \bibinfo {pages} {024041} (\bibinfo {year} {2021})},\ \Eprint
  {http://arxiv.org/abs/2010.01857} {arXiv:2010.01857 [gr-qc]} \BibitemShut
  {NoStop}%
\bibitem [{\citenamefont {Jim\'enez~Forteza}\ \emph {et~al.}(2020)\citenamefont
  {Jim\'enez~Forteza}, \citenamefont {Bhagwat}, \citenamefont {Pani},\ and\
  \citenamefont {Ferrari}}]{JimenezForteza:2020cve}%
  \BibitemOpen
  \bibfield  {author} {\bibinfo {author} {\bibfnamefont {Xisco}\ \bibnamefont
  {Jim\'enez~Forteza}}, \bibinfo {author} {\bibfnamefont {Swetha}\ \bibnamefont
  {Bhagwat}}, \bibinfo {author} {\bibfnamefont {Paolo}\ \bibnamefont {Pani}}, \
  and\ \bibinfo {author} {\bibfnamefont {Valeria}\ \bibnamefont {Ferrari}},\
  }\bibfield  {title} {\enquote {\bibinfo {title} {{Spectroscopy of binary
  black hole ringdown using overtones and angular modes}},}\ }\href {\doibase
  10.1103/PhysRevD.102.044053} {\bibfield  {journal} {\bibinfo  {journal}
  {Phys. Rev. D}\ }\textbf {\bibinfo {volume} {102}},\ \bibinfo {pages}
  {044053} (\bibinfo {year} {2020})},\ \Eprint
  {http://arxiv.org/abs/2005.03260} {arXiv:2005.03260 [gr-qc]} \BibitemShut
  {NoStop}%
\bibitem [{\citenamefont {Isi}\ and\ \citenamefont
  {Farr}(2021{\natexlab{a}})}]{Isi:2021iql}%
  \BibitemOpen
  \bibfield  {author} {\bibinfo {author} {\bibfnamefont {Maximiliano}\
  \bibnamefont {Isi}}\ and\ \bibinfo {author} {\bibfnamefont {Will~M.}\
  \bibnamefont {Farr}},\ }\bibfield  {title} {\enquote {\bibinfo {title}
  {{Analyzing black-hole ringdowns}},}\ }\href@noop {} {\  (\bibinfo {year}
  {2021}{\natexlab{a}})},\ \Eprint {http://arxiv.org/abs/2107.05609}
  {arXiv:2107.05609 [gr-qc]} \BibitemShut {NoStop}%
\bibitem [{\citenamefont {Finch}\ and\ \citenamefont
  {Moore}(2021{\natexlab{a}})}]{Finch:2021qph}%
  \BibitemOpen
  \bibfield  {author} {\bibinfo {author} {\bibfnamefont {Eliot}\ \bibnamefont
  {Finch}}\ and\ \bibinfo {author} {\bibfnamefont {Christopher~J.}\
  \bibnamefont {Moore}},\ }\bibfield  {title} {\enquote {\bibinfo {title}
  {{Frequency-domain analysis of black-hole ringdowns}},}\ }\href {\doibase
  10.1103/PhysRevD.104.123034} {\bibfield  {journal} {\bibinfo  {journal}
  {Phys. Rev. D}\ }\textbf {\bibinfo {volume} {104}},\ \bibinfo {pages}
  {123034} (\bibinfo {year} {2021}{\natexlab{a}})},\ \Eprint
  {http://arxiv.org/abs/2108.09344} {arXiv:2108.09344 [gr-qc]} \BibitemShut
  {NoStop}%
\bibitem [{\citenamefont {Isi}\ and\ \citenamefont {Farr}(2022)}]{Isi:2022mhy}%
  \BibitemOpen
  \bibfield  {author} {\bibinfo {author} {\bibfnamefont {Maximiliano}\
  \bibnamefont {Isi}}\ and\ \bibinfo {author} {\bibfnamefont {Will~M.}\
  \bibnamefont {Farr}},\ }\bibfield  {title} {\enquote {\bibinfo {title}
  {{Revisiting the ringdown of GW150914}},}\ }\href@noop {} {\  (\bibinfo
  {year} {2022})},\ \Eprint {http://arxiv.org/abs/2202.02941} {arXiv:2202.02941
  [gr-qc]} \BibitemShut {NoStop}%
\bibitem [{\citenamefont {Finch}\ and\ \citenamefont
  {Moore}(2022)}]{Finch:2022ynt}%
  \BibitemOpen
  \bibfield  {author} {\bibinfo {author} {\bibfnamefont {Eliot}\ \bibnamefont
  {Finch}}\ and\ \bibinfo {author} {\bibfnamefont {Christopher~J.}\
  \bibnamefont {Moore}},\ }\bibfield  {title} {\enquote {\bibinfo {title}
  {{Searching for a ringdown overtone in GW150914}},}\ }\href {\doibase
  10.1103/PhysRevD.106.043005} {\bibfield  {journal} {\bibinfo  {journal}
  {Phys. Rev. D}\ }\textbf {\bibinfo {volume} {106}},\ \bibinfo {pages}
  {043005} (\bibinfo {year} {2022})},\ \Eprint
  {http://arxiv.org/abs/2205.07809} {arXiv:2205.07809 [gr-qc]} \BibitemShut
  {NoStop}%
\bibitem [{\citenamefont {Cotesta}\ \emph {et~al.}(2022)\citenamefont
  {Cotesta}, \citenamefont {Carullo}, \citenamefont {Berti},\ and\
  \citenamefont {Cardoso}}]{Cotesta:2022pci}%
  \BibitemOpen
  \bibfield  {author} {\bibinfo {author} {\bibfnamefont {Roberto}\ \bibnamefont
  {Cotesta}}, \bibinfo {author} {\bibfnamefont {Gregorio}\ \bibnamefont
  {Carullo}}, \bibinfo {author} {\bibfnamefont {Emanuele}\ \bibnamefont
  {Berti}}, \ and\ \bibinfo {author} {\bibfnamefont {Vitor}\ \bibnamefont
  {Cardoso}},\ }\bibfield  {title} {\enquote {\bibinfo {title} {{On the
  detection of ringdown overtones in GW150914}},}\ }\href@noop {} {\  (\bibinfo
  {year} {2022})},\ \Eprint {http://arxiv.org/abs/2201.00822} {arXiv:2201.00822
  [gr-qc]} \BibitemShut {NoStop}%
\bibitem [{\citenamefont {Abbott}\ \emph
  {et~al.}(2016{\natexlab{a}})\citenamefont {Abbott} \emph
  {et~al.}}]{TheLIGOScientific:2016src}%
  \BibitemOpen
  \bibfield  {author} {\bibinfo {author} {\bibfnamefont {B.~P.}\ \bibnamefont
  {Abbott}} \emph {et~al.} (\bibinfo {collaboration} {LIGO Scientific,
  Virgo}),\ }\bibfield  {title} {\enquote {\bibinfo {title} {{Tests of general
  relativity with GW150914}},}\ }\href {\doibase
  10.1103/PhysRevLett.116.221101} {\bibfield  {journal} {\bibinfo  {journal}
  {Phys. Rev. Lett.}\ }\textbf {\bibinfo {volume} {116}},\ \bibinfo {pages}
  {221101} (\bibinfo {year} {2016}{\natexlab{a}})},\ \bibinfo {note} {[Erratum:
  Phys. Rev. Lett.121,no.12,129902(2018)]},\ \Eprint
  {http://arxiv.org/abs/1602.03841} {arXiv:1602.03841 [gr-qc]} \BibitemShut
  {NoStop}%
\bibitem [{\citenamefont {Isi}\ \emph {et~al.}(2019)\citenamefont {Isi},
  \citenamefont {Giesler}, \citenamefont {Farr}, \citenamefont {Scheel},\ and\
  \citenamefont {Teukolsky}}]{Isi:2019aib}%
  \BibitemOpen
  \bibfield  {author} {\bibinfo {author} {\bibfnamefont {Maximiliano}\
  \bibnamefont {Isi}}, \bibinfo {author} {\bibfnamefont {Matthew}\ \bibnamefont
  {Giesler}}, \bibinfo {author} {\bibfnamefont {Will~M.}\ \bibnamefont {Farr}},
  \bibinfo {author} {\bibfnamefont {Mark~A.}\ \bibnamefont {Scheel}}, \ and\
  \bibinfo {author} {\bibfnamefont {Saul~A.}\ \bibnamefont {Teukolsky}},\
  }\bibfield  {title} {\enquote {\bibinfo {title} {{Testing the no-hair theorem
  with GW150914}},}\ }\href {\doibase 10.1103/PhysRevLett.123.111102}
  {\bibfield  {journal} {\bibinfo  {journal} {Phys. Rev. Lett.}\ }\textbf
  {\bibinfo {volume} {123}},\ \bibinfo {pages} {111102} (\bibinfo {year}
  {2019})},\ \Eprint {http://arxiv.org/abs/1905.00869} {arXiv:1905.00869
  [gr-qc]} \BibitemShut {NoStop}%
\bibitem [{\citenamefont {Capano}\ \emph {et~al.}(2021)\citenamefont {Capano},
  \citenamefont {Cabero}, \citenamefont {Westerweck}, \citenamefont {Abedi},
  \citenamefont {Kastha}, \citenamefont {Nitz}, \citenamefont {Nielsen},\ and\
  \citenamefont {Krishnan}}]{Capano:2021etf}%
  \BibitemOpen
  \bibfield  {author} {\bibinfo {author} {\bibfnamefont {Collin~D.}\
  \bibnamefont {Capano}}, \bibinfo {author} {\bibfnamefont {Miriam}\
  \bibnamefont {Cabero}}, \bibinfo {author} {\bibfnamefont {Julian}\
  \bibnamefont {Westerweck}}, \bibinfo {author} {\bibfnamefont {Jahed}\
  \bibnamefont {Abedi}}, \bibinfo {author} {\bibfnamefont {Shilpa}\
  \bibnamefont {Kastha}}, \bibinfo {author} {\bibfnamefont {Alexander~H.}\
  \bibnamefont {Nitz}}, \bibinfo {author} {\bibfnamefont {Alex~B.}\
  \bibnamefont {Nielsen}}, \ and\ \bibinfo {author} {\bibfnamefont {Badri}\
  \bibnamefont {Krishnan}},\ }\bibfield  {title} {\enquote {\bibinfo {title}
  {{Observation of a multimode quasi-normal spectrum from a perturbed black
  hole}},}\ }\href@noop {} {\  (\bibinfo {year} {2021})},\ \Eprint
  {http://arxiv.org/abs/2105.05238} {arXiv:2105.05238 [gr-qc]} \BibitemShut
  {NoStop}%
\bibitem [{\citenamefont {Capano}\ \emph {et~al.}(2022)\citenamefont {Capano},
  \citenamefont {Abedi}, \citenamefont {Kastha}, \citenamefont {Nitz},
  \citenamefont {Westerweck}, \citenamefont {Cabero}, \citenamefont {Nielsen},\
  and\ \citenamefont {Krishnan}}]{Capano:2022zqm}%
  \BibitemOpen
  \bibfield  {author} {\bibinfo {author} {\bibfnamefont {Collin~D.}\
  \bibnamefont {Capano}}, \bibinfo {author} {\bibfnamefont {Jahed}\
  \bibnamefont {Abedi}}, \bibinfo {author} {\bibfnamefont {Shilpa}\
  \bibnamefont {Kastha}}, \bibinfo {author} {\bibfnamefont {Alexander~H.}\
  \bibnamefont {Nitz}}, \bibinfo {author} {\bibfnamefont {Julian}\ \bibnamefont
  {Westerweck}}, \bibinfo {author} {\bibfnamefont {Miriam}\ \bibnamefont
  {Cabero}}, \bibinfo {author} {\bibfnamefont {Alex~B.}\ \bibnamefont
  {Nielsen}}, \ and\ \bibinfo {author} {\bibfnamefont {Badri}\ \bibnamefont
  {Krishnan}},\ }\bibfield  {title} {\enquote {\bibinfo {title} {{Statistical
  validation of the detection of a sub-dominant quasi-normal mode in
  GW190521}},}\ }\href@noop {} {\  (\bibinfo {year} {2022})},\ \Eprint
  {http://arxiv.org/abs/2209.00640} {arXiv:2209.00640 [gr-qc]} \BibitemShut
  {NoStop}%
\bibitem [{\citenamefont {Buonanno}\ \emph {et~al.}(2007)\citenamefont
  {Buonanno}, \citenamefont {Cook},\ and\ \citenamefont
  {Pretorius}}]{Buonanno:2006ui}%
  \BibitemOpen
  \bibfield  {author} {\bibinfo {author} {\bibfnamefont {Alessandra}\
  \bibnamefont {Buonanno}}, \bibinfo {author} {\bibfnamefont {Gregory~B.}\
  \bibnamefont {Cook}}, \ and\ \bibinfo {author} {\bibfnamefont {Frans}\
  \bibnamefont {Pretorius}},\ }\bibfield  {title} {\enquote {\bibinfo {title}
  {{Inspiral, merger and ring-down of equal-mass black-hole binaries}},}\
  }\href {\doibase 10.1103/PhysRevD.75.124018} {\bibfield  {journal} {\bibinfo
  {journal} {Phys. Rev.}\ }\textbf {\bibinfo {volume} {D75}},\ \bibinfo {pages}
  {124018} (\bibinfo {year} {2007})},\ \Eprint
  {http://arxiv.org/abs/gr-qc/0610122} {arXiv:gr-qc/0610122 [gr-qc]}
  \BibitemShut {NoStop}%
\bibitem [{\citenamefont {Berti}\ \emph
  {et~al.}(2007{\natexlab{b}})\citenamefont {Berti}, \citenamefont {Cardoso},
  \citenamefont {Gonzalez}, \citenamefont {Sperhake}, \citenamefont {Hannam},
  \citenamefont {Husa},\ and\ \citenamefont {Bruegmann}}]{Berti:2007fi}%
  \BibitemOpen
  \bibfield  {author} {\bibinfo {author} {\bibfnamefont {Emanuele}\
  \bibnamefont {Berti}}, \bibinfo {author} {\bibfnamefont {Vitor}\ \bibnamefont
  {Cardoso}}, \bibinfo {author} {\bibfnamefont {Jose~A.}\ \bibnamefont
  {Gonzalez}}, \bibinfo {author} {\bibfnamefont {Ulrich}\ \bibnamefont
  {Sperhake}}, \bibinfo {author} {\bibfnamefont {Mark}\ \bibnamefont {Hannam}},
  \bibinfo {author} {\bibfnamefont {Sascha}\ \bibnamefont {Husa}}, \ and\
  \bibinfo {author} {\bibfnamefont {Bernd}\ \bibnamefont {Bruegmann}},\
  }\bibfield  {title} {\enquote {\bibinfo {title} {{Inspiral, merger and
  ringdown of unequal mass black hole binaries: A Multipolar analysis}},}\
  }\href {\doibase 10.1103/PhysRevD.76.064034} {\bibfield  {journal} {\bibinfo
  {journal} {Phys. Rev. D}\ }\textbf {\bibinfo {volume} {76}},\ \bibinfo
  {pages} {064034} (\bibinfo {year} {2007}{\natexlab{b}})},\ \Eprint
  {http://arxiv.org/abs/gr-qc/0703053} {arXiv:gr-qc/0703053} \BibitemShut
  {NoStop}%
\bibitem [{\citenamefont {Berti}\ \emph
  {et~al.}(2007{\natexlab{c}})\citenamefont {Berti}, \citenamefont {Cardoso},
  \citenamefont {Gonzalez},\ and\ \citenamefont {Sperhake}}]{Berti:2007dg}%
  \BibitemOpen
  \bibfield  {author} {\bibinfo {author} {\bibfnamefont {Emanuele}\
  \bibnamefont {Berti}}, \bibinfo {author} {\bibfnamefont {Vitor}\ \bibnamefont
  {Cardoso}}, \bibinfo {author} {\bibfnamefont {Jose~A.}\ \bibnamefont
  {Gonzalez}}, \ and\ \bibinfo {author} {\bibfnamefont {Ulrich}\ \bibnamefont
  {Sperhake}},\ }\bibfield  {title} {\enquote {\bibinfo {title} {{Mining
  information from binary black hole mergers: A Comparison of estimation
  methods for complex exponentials in noise}},}\ }\href {\doibase
  10.1103/PhysRevD.75.124017} {\bibfield  {journal} {\bibinfo  {journal} {Phys.
  Rev. D}\ }\textbf {\bibinfo {volume} {75}},\ \bibinfo {pages} {124017}
  (\bibinfo {year} {2007}{\natexlab{c}})},\ \Eprint
  {http://arxiv.org/abs/gr-qc/0701086} {arXiv:gr-qc/0701086} \BibitemShut
  {NoStop}%
\bibitem [{\citenamefont {Kamaretsos}\ \emph {et~al.}(2012)\citenamefont
  {Kamaretsos}, \citenamefont {Hannam}, \citenamefont {Husa},\ and\
  \citenamefont {Sathyaprakash}}]{Kamaretsos:2011um}%
  \BibitemOpen
  \bibfield  {author} {\bibinfo {author} {\bibfnamefont {Ioannis}\ \bibnamefont
  {Kamaretsos}}, \bibinfo {author} {\bibfnamefont {Mark}\ \bibnamefont
  {Hannam}}, \bibinfo {author} {\bibfnamefont {Sascha}\ \bibnamefont {Husa}}, \
  and\ \bibinfo {author} {\bibfnamefont {B.~S.}\ \bibnamefont
  {Sathyaprakash}},\ }\bibfield  {title} {\enquote {\bibinfo {title}
  {{Black-hole hair loss: learning about binary progenitors from ringdown
  signals}},}\ }\href {\doibase 10.1103/PhysRevD.85.024018} {\bibfield
  {journal} {\bibinfo  {journal} {Phys. Rev. D}\ }\textbf {\bibinfo {volume}
  {85}},\ \bibinfo {pages} {024018} (\bibinfo {year} {2012})},\ \Eprint
  {http://arxiv.org/abs/1107.0854} {arXiv:1107.0854 [gr-qc]} \BibitemShut
  {NoStop}%
\bibitem [{\citenamefont {London}\ \emph {et~al.}(2014)\citenamefont {London},
  \citenamefont {Shoemaker},\ and\ \citenamefont {Healy}}]{London:2014cma}%
  \BibitemOpen
  \bibfield  {author} {\bibinfo {author} {\bibfnamefont {Lionel}\ \bibnamefont
  {London}}, \bibinfo {author} {\bibfnamefont {Deirdre}\ \bibnamefont
  {Shoemaker}}, \ and\ \bibinfo {author} {\bibfnamefont {James}\ \bibnamefont
  {Healy}},\ }\bibfield  {title} {\enquote {\bibinfo {title} {{Modeling
  ringdown: Beyond the fundamental quasinormal modes}},}\ }\href {\doibase
  10.1103/PhysRevD.90.124032} {\bibfield  {journal} {\bibinfo  {journal} {Phys.
  Rev. D}\ }\textbf {\bibinfo {volume} {90}},\ \bibinfo {pages} {124032}
  (\bibinfo {year} {2014})},\ \bibinfo {note} {[Erratum: Phys.Rev.D 94, 069902
  (2016)]},\ \Eprint {http://arxiv.org/abs/1404.3197} {arXiv:1404.3197 [gr-qc]}
  \BibitemShut {NoStop}%
\bibitem [{\citenamefont {Thrane}\ \emph {et~al.}(2017)\citenamefont {Thrane},
  \citenamefont {Lasky},\ and\ \citenamefont {Levin}}]{Thrane:2017lqn}%
  \BibitemOpen
  \bibfield  {author} {\bibinfo {author} {\bibfnamefont {Eric}\ \bibnamefont
  {Thrane}}, \bibinfo {author} {\bibfnamefont {Paul~D.}\ \bibnamefont {Lasky}},
  \ and\ \bibinfo {author} {\bibfnamefont {Yuri}\ \bibnamefont {Levin}},\
  }\bibfield  {title} {\enquote {\bibinfo {title} {{Challenges testing the
  no-hair theorem with gravitational waves}},}\ }\href {\doibase
  10.1103/PhysRevD.96.102004} {\bibfield  {journal} {\bibinfo  {journal} {Phys.
  Rev. D}\ }\textbf {\bibinfo {volume} {96}},\ \bibinfo {pages} {102004}
  (\bibinfo {year} {2017})},\ \Eprint {http://arxiv.org/abs/1706.05152}
  {arXiv:1706.05152 [gr-qc]} \BibitemShut {NoStop}%
\bibitem [{\citenamefont {Giesler}\ \emph {et~al.}(2019)\citenamefont
  {Giesler}, \citenamefont {Isi}, \citenamefont {Scheel},\ and\ \citenamefont
  {Teukolsky}}]{Giesler:2019uxc}%
  \BibitemOpen
  \bibfield  {author} {\bibinfo {author} {\bibfnamefont {Matthew}\ \bibnamefont
  {Giesler}}, \bibinfo {author} {\bibfnamefont {Maximiliano}\ \bibnamefont
  {Isi}}, \bibinfo {author} {\bibfnamefont {Mark}\ \bibnamefont {Scheel}}, \
  and\ \bibinfo {author} {\bibfnamefont {Saul}\ \bibnamefont {Teukolsky}},\
  }\bibfield  {title} {\enquote {\bibinfo {title} {{Black hole ringdown: the
  importance of overtones}},}\ }\href {\doibase 10.1103/PhysRevX.9.041060}
  {\bibfield  {journal} {\bibinfo  {journal} {Phys. Rev.}\ }\textbf {\bibinfo
  {volume} {X9}},\ \bibinfo {pages} {041060} (\bibinfo {year} {2019})},\
  \Eprint {http://arxiv.org/abs/1903.08284} {arXiv:1903.08284 [gr-qc]}
  \BibitemShut {NoStop}%
\bibitem [{\citenamefont {Cook}(2020)}]{Cook:2020otn}%
  \BibitemOpen
  \bibfield  {author} {\bibinfo {author} {\bibfnamefont {Gregory~B.}\
  \bibnamefont {Cook}},\ }\bibfield  {title} {\enquote {\bibinfo {title}
  {{Aspects of multimode Kerr ringdown fitting}},}\ }\href {\doibase
  10.1103/PhysRevD.102.024027} {\bibfield  {journal} {\bibinfo  {journal}
  {Phys. Rev. D}\ }\textbf {\bibinfo {volume} {102}},\ \bibinfo {pages}
  {024027} (\bibinfo {year} {2020})},\ \Eprint
  {http://arxiv.org/abs/2004.08347} {arXiv:2004.08347 [gr-qc]} \BibitemShut
  {NoStop}%
\bibitem [{\citenamefont {Dhani}(2021)}]{Dhani:2020nik}%
  \BibitemOpen
  \bibfield  {author} {\bibinfo {author} {\bibfnamefont {Arnab}\ \bibnamefont
  {Dhani}},\ }\bibfield  {title} {\enquote {\bibinfo {title} {{Importance of
  mirror modes in binary black hole ringdown waveform}},}\ }\href {\doibase
  10.1103/PhysRevD.103.104048} {\bibfield  {journal} {\bibinfo  {journal}
  {Phys. Rev. D}\ }\textbf {\bibinfo {volume} {103}},\ \bibinfo {pages}
  {104048} (\bibinfo {year} {2021})},\ \Eprint
  {http://arxiv.org/abs/2010.08602} {arXiv:2010.08602 [gr-qc]} \BibitemShut
  {NoStop}%
\bibitem [{\citenamefont {Dhani}\ and\ \citenamefont
  {Sathyaprakash}(2021)}]{Dhani:2021vac}%
  \BibitemOpen
  \bibfield  {author} {\bibinfo {author} {\bibfnamefont {Arnab}\ \bibnamefont
  {Dhani}}\ and\ \bibinfo {author} {\bibfnamefont {B.~S.}\ \bibnamefont
  {Sathyaprakash}},\ }\bibfield  {title} {\enquote {\bibinfo {title}
  {{Overtones, mirror modes, and mode-mixing in binary black hole mergers}},}\
  }\href@noop {} {\  (\bibinfo {year} {2021})},\ \Eprint
  {http://arxiv.org/abs/2107.14195} {arXiv:2107.14195 [gr-qc]} \BibitemShut
  {NoStop}%
\bibitem [{\citenamefont {Forteza}\ and\ \citenamefont
  {Mourier}(2021)}]{Forteza:2021wfq}%
  \BibitemOpen
  \bibfield  {author} {\bibinfo {author} {\bibfnamefont {Xisco~Jim\'enez}\
  \bibnamefont {Forteza}}\ and\ \bibinfo {author} {\bibfnamefont {Pierre}\
  \bibnamefont {Mourier}},\ }\bibfield  {title} {\enquote {\bibinfo {title}
  {{High-overtone fits to numerical relativity ringdowns: Beyond the dismissed
  n=8 special tone}},}\ }\href {\doibase 10.1103/PhysRevD.104.124072}
  {\bibfield  {journal} {\bibinfo  {journal} {Phys. Rev. D}\ }\textbf {\bibinfo
  {volume} {104}},\ \bibinfo {pages} {124072} (\bibinfo {year} {2021})},\
  \Eprint {http://arxiv.org/abs/2107.11829} {arXiv:2107.11829 [gr-qc]}
  \BibitemShut {NoStop}%
\bibitem [{\citenamefont {Finch}\ and\ \citenamefont
  {Moore}(2021{\natexlab{b}})}]{Finch:2021iip}%
  \BibitemOpen
  \bibfield  {author} {\bibinfo {author} {\bibfnamefont {Eliot}\ \bibnamefont
  {Finch}}\ and\ \bibinfo {author} {\bibfnamefont {Christopher~J.}\
  \bibnamefont {Moore}},\ }\bibfield  {title} {\enquote {\bibinfo {title}
  {{Modeling the ringdown from precessing black hole binaries}},}\ }\href
  {\doibase 10.1103/PhysRevD.103.084048} {\bibfield  {journal} {\bibinfo
  {journal} {Phys. Rev. D}\ }\textbf {\bibinfo {volume} {103}},\ \bibinfo
  {pages} {084048} (\bibinfo {year} {2021}{\natexlab{b}})},\ \Eprint
  {http://arxiv.org/abs/2102.07794} {arXiv:2102.07794 [gr-qc]} \BibitemShut
  {NoStop}%
\bibitem [{\citenamefont {Li}\ \emph {et~al.}(2022)\citenamefont {Li},
  \citenamefont {Sun}, \citenamefont {Lo}, \citenamefont {Payne},\ and\
  \citenamefont {Chen}}]{Li:2021wgz}%
  \BibitemOpen
  \bibfield  {author} {\bibinfo {author} {\bibfnamefont {Xiang}\ \bibnamefont
  {Li}}, \bibinfo {author} {\bibfnamefont {Ling}\ \bibnamefont {Sun}}, \bibinfo
  {author} {\bibfnamefont {Rico Ka~Lok}\ \bibnamefont {Lo}}, \bibinfo {author}
  {\bibfnamefont {Ethan}\ \bibnamefont {Payne}}, \ and\ \bibinfo {author}
  {\bibfnamefont {Yanbei}\ \bibnamefont {Chen}},\ }\bibfield  {title} {\enquote
  {\bibinfo {title} {{Angular emission patterns of remnant black holes}},}\
  }\href {\doibase 10.1103/PhysRevD.105.024016} {\bibfield  {journal} {\bibinfo
   {journal} {Phys. Rev. D}\ }\textbf {\bibinfo {volume} {105}},\ \bibinfo
  {pages} {024016} (\bibinfo {year} {2022})},\ \Eprint
  {http://arxiv.org/abs/2110.03116} {arXiv:2110.03116 [gr-qc]} \BibitemShut
  {NoStop}%
\bibitem [{\citenamefont {Maga\~na Zertuche}\ \emph {et~al.}(2022)\citenamefont
  {Maga\~na Zertuche} \emph {et~al.}}]{MaganaZertuche:2021syq}%
  \BibitemOpen
  \bibfield  {author} {\bibinfo {author} {\bibfnamefont {Lorena}\ \bibnamefont
  {Maga\~na Zertuche}} \emph {et~al.},\ }\bibfield  {title} {\enquote {\bibinfo
  {title} {{High Precision Ringdown Modeling: Multimode Fits and BMS
  Frames}},}\ }\href {\doibase 10.1103/PhysRevD.105.104015} {\bibfield
  {journal} {\bibinfo  {journal} {Phys. Rev. D}\ }\textbf {\bibinfo {volume}
  {105}},\ \bibinfo {pages} {104015} (\bibinfo {year} {2022})},\ \Eprint
  {http://arxiv.org/abs/2110.15922} {arXiv:2110.15922 [gr-qc]} \BibitemShut
  {NoStop}%
\bibitem [{\citenamefont {Ma}\ \emph {et~al.}(2021)\citenamefont {Ma},
  \citenamefont {Giesler}, \citenamefont {Varma}, \citenamefont {Scheel},\ and\
  \citenamefont {Chen}}]{Ma:2021znq}%
  \BibitemOpen
  \bibfield  {author} {\bibinfo {author} {\bibfnamefont {Sizheng}\ \bibnamefont
  {Ma}}, \bibinfo {author} {\bibfnamefont {Matthew}\ \bibnamefont {Giesler}},
  \bibinfo {author} {\bibfnamefont {Vijay}\ \bibnamefont {Varma}}, \bibinfo
  {author} {\bibfnamefont {Mark~A.}\ \bibnamefont {Scheel}}, \ and\ \bibinfo
  {author} {\bibfnamefont {Yanbei}\ \bibnamefont {Chen}},\ }\bibfield  {title}
  {\enquote {\bibinfo {title} {{Universal features of gravitational waves
  emitted by superkick binary black hole systems}},}\ }\href {\doibase
  10.1103/PhysRevD.104.084003} {\bibfield  {journal} {\bibinfo  {journal}
  {Phys. Rev. D}\ }\textbf {\bibinfo {volume} {104}},\ \bibinfo {pages}
  {084003} (\bibinfo {year} {2021})},\ \Eprint
  {http://arxiv.org/abs/2107.04890} {arXiv:2107.04890 [gr-qc]} \BibitemShut
  {NoStop}%
\bibitem [{\citenamefont {Mitman}\ \emph {et~al.}(2022)\citenamefont {Mitman}
  \emph {et~al.}}]{Mitman:2022qdl}%
  \BibitemOpen
  \bibfield  {author} {\bibinfo {author} {\bibfnamefont {Keefe}\ \bibnamefont
  {Mitman}} \emph {et~al.},\ }\bibfield  {title} {\enquote {\bibinfo {title}
  {{Nonlinearities in black hole ringdowns}},}\ }\href@noop {} {\  (\bibinfo
  {year} {2022})},\ \Eprint {http://arxiv.org/abs/2208.07380} {arXiv:2208.07380
  [gr-qc]} \BibitemShut {NoStop}%
\bibitem [{\citenamefont {Cheung}\ \emph {et~al.}(2022)\citenamefont {Cheung}
  \emph {et~al.}}]{Cheung:2022rbm}%
  \BibitemOpen
  \bibfield  {author} {\bibinfo {author} {\bibfnamefont {Mark Ho-Yeuk}\
  \bibnamefont {Cheung}} \emph {et~al.},\ }\bibfield  {title} {\enquote
  {\bibinfo {title} {{Nonlinear effects in black hole ringdown}},}\ }\href@noop
  {} {\  (\bibinfo {year} {2022})},\ \Eprint {http://arxiv.org/abs/2208.07374}
  {arXiv:2208.07374 [gr-qc]} \BibitemShut {NoStop}%
\bibitem [{\citenamefont {Lagos}\ and\ \citenamefont
  {Hui}(2022)}]{Lagos:2022otp}%
  \BibitemOpen
  \bibfield  {author} {\bibinfo {author} {\bibfnamefont {Macarena}\
  \bibnamefont {Lagos}}\ and\ \bibinfo {author} {\bibfnamefont {Lam}\
  \bibnamefont {Hui}},\ }\bibfield  {title} {\enquote {\bibinfo {title}
  {{Generation and propagation of nonlinear quasi-normal modes of a
  Schwarzschild black hole}},}\ }\href@noop {} {\  (\bibinfo {year} {2022})},\
  \Eprint {http://arxiv.org/abs/2208.07379} {arXiv:2208.07379 [gr-qc]}
  \BibitemShut {NoStop}%
\bibitem [{\citenamefont {Bhagwat}\ \emph {et~al.}(2018)\citenamefont
  {Bhagwat}, \citenamefont {Okounkova}, \citenamefont {Ballmer}, \citenamefont
  {Brown}, \citenamefont {Giesler}, \citenamefont {Scheel},\ and\ \citenamefont
  {Teukolsky}}]{Bhagwat:2017tkm}%
  \BibitemOpen
  \bibfield  {author} {\bibinfo {author} {\bibfnamefont {Swetha}\ \bibnamefont
  {Bhagwat}}, \bibinfo {author} {\bibfnamefont {Maria}\ \bibnamefont
  {Okounkova}}, \bibinfo {author} {\bibfnamefont {Stefan~W.}\ \bibnamefont
  {Ballmer}}, \bibinfo {author} {\bibfnamefont {Duncan~A.}\ \bibnamefont
  {Brown}}, \bibinfo {author} {\bibfnamefont {Matthew}\ \bibnamefont
  {Giesler}}, \bibinfo {author} {\bibfnamefont {Mark~A.}\ \bibnamefont
  {Scheel}}, \ and\ \bibinfo {author} {\bibfnamefont {Saul~A.}\ \bibnamefont
  {Teukolsky}},\ }\bibfield  {title} {\enquote {\bibinfo {title} {{On choosing
  the start time of binary black hole ringdowns}},}\ }\href {\doibase
  10.1103/PhysRevD.97.104065} {\bibfield  {journal} {\bibinfo  {journal} {Phys.
  Rev.}\ }\textbf {\bibinfo {volume} {D97}},\ \bibinfo {pages} {104065}
  (\bibinfo {year} {2018})},\ \Eprint {http://arxiv.org/abs/1711.00926}
  {arXiv:1711.00926 [gr-qc]} \BibitemShut {NoStop}%
\bibitem [{\citenamefont {Bhagwat}\ \emph
  {et~al.}(2020{\natexlab{b}})\citenamefont {Bhagwat}, \citenamefont {Forteza},
  \citenamefont {Pani},\ and\ \citenamefont {Ferrari}}]{Bhagwat:2019dtm}%
  \BibitemOpen
  \bibfield  {author} {\bibinfo {author} {\bibfnamefont {Swetha}\ \bibnamefont
  {Bhagwat}}, \bibinfo {author} {\bibfnamefont {Xisco~Jimenez}\ \bibnamefont
  {Forteza}}, \bibinfo {author} {\bibfnamefont {Paolo}\ \bibnamefont {Pani}}, \
  and\ \bibinfo {author} {\bibfnamefont {Valeria}\ \bibnamefont {Ferrari}},\
  }\bibfield  {title} {\enquote {\bibinfo {title} {{Ringdown overtones, black
  hole spectroscopy, and no-hair theorem tests}},}\ }\href {\doibase
  10.1103/PhysRevD.101.044033} {\bibfield  {journal} {\bibinfo  {journal}
  {Phys. Rev. D}\ }\textbf {\bibinfo {volume} {101}},\ \bibinfo {pages}
  {044033} (\bibinfo {year} {2020}{\natexlab{b}})},\ \Eprint
  {http://arxiv.org/abs/1910.08708} {arXiv:1910.08708 [gr-qc]} \BibitemShut
  {NoStop}%
\bibitem [{\citenamefont {Okounkova}(2020)}]{Okounkova:2020vwu}%
  \BibitemOpen
  \bibfield  {author} {\bibinfo {author} {\bibfnamefont {Maria}\ \bibnamefont
  {Okounkova}},\ }\bibfield  {title} {\enquote {\bibinfo {title} {{Revisiting
  non-linearity in binary black hole mergers}},}\ }\href@noop {} {\  (\bibinfo
  {year} {2020})},\ \Eprint {http://arxiv.org/abs/2004.00671} {arXiv:2004.00671
  [gr-qc]} \BibitemShut {NoStop}%
\bibitem [{\citenamefont {Lovelace}\ \emph {et~al.}(2016)\citenamefont
  {Lovelace}, \citenamefont {Lousto}, \citenamefont {Healy}, \citenamefont
  {Scheel}, \citenamefont {Garcia}, \citenamefont {O'Shaughnessy},
  \citenamefont {Boyle}, \citenamefont {Campanelli}, \citenamefont {Hemberger},
  \citenamefont {Kidder}, \citenamefont {Pfeiffer}, \citenamefont {Szilagyi},
  \citenamefont {Teukolsky},\ and\ \citenamefont
  {Zlochower}}]{Lovelace:2016uwp}%
  \BibitemOpen
  \bibfield  {author} {\bibinfo {author} {\bibfnamefont {Geoffrey}\
  \bibnamefont {Lovelace}}, \bibinfo {author} {\bibfnamefont {Carlos~O.}\
  \bibnamefont {Lousto}}, \bibinfo {author} {\bibfnamefont {James}\
  \bibnamefont {Healy}}, \bibinfo {author} {\bibfnamefont {Mark~A.}\
  \bibnamefont {Scheel}}, \bibinfo {author} {\bibfnamefont {Alyssa}\
  \bibnamefont {Garcia}}, \bibinfo {author} {\bibfnamefont {Richard}\
  \bibnamefont {O'Shaughnessy}}, \bibinfo {author} {\bibfnamefont {Michael}\
  \bibnamefont {Boyle}}, \bibinfo {author} {\bibfnamefont {Manuela}\
  \bibnamefont {Campanelli}}, \bibinfo {author} {\bibfnamefont {Daniel~A.}\
  \bibnamefont {Hemberger}}, \bibinfo {author} {\bibfnamefont {Lawrence~E.}\
  \bibnamefont {Kidder}}, \bibinfo {author} {\bibfnamefont {Harald~P.}\
  \bibnamefont {Pfeiffer}}, \bibinfo {author} {\bibfnamefont {Bela}\
  \bibnamefont {Szilagyi}}, \bibinfo {author} {\bibfnamefont {Saul~A.}\
  \bibnamefont {Teukolsky}}, \ and\ \bibinfo {author} {\bibfnamefont {Yosef}\
  \bibnamefont {Zlochower}},\ }\bibfield  {title} {\enquote {\bibinfo {title}
  {{Modeling the source of GW150914 with targeted numerical-relativity
  simulations}},}\ }\href {\doibase 10.1088/0264-9381/33/24/244002} {\bibfield
  {journal} {\bibinfo  {journal} {Class. Quant. Grav.}\ }\textbf {\bibinfo
  {volume} {33}},\ \bibinfo {pages} {244002} (\bibinfo {year} {2016})},\
  \Eprint {http://arxiv.org/abs/1607.05377} {arXiv:1607.05377 [gr-qc]}
  \BibitemShut {NoStop}%
\bibitem [{\citenamefont {Carullo}\ \emph {et~al.}(2019)\citenamefont
  {Carullo}, \citenamefont {Del~Pozzo},\ and\ \citenamefont
  {Veitch}}]{Carullo:2019flw}%
  \BibitemOpen
  \bibfield  {author} {\bibinfo {author} {\bibfnamefont {Gregorio}\
  \bibnamefont {Carullo}}, \bibinfo {author} {\bibfnamefont {Walter}\
  \bibnamefont {Del~Pozzo}}, \ and\ \bibinfo {author} {\bibfnamefont {John}\
  \bibnamefont {Veitch}},\ }\bibfield  {title} {\enquote {\bibinfo {title}
  {{Observational Black Hole Spectroscopy: A time-domain multimode analysis of
  GW150914}},}\ }\href {\doibase 10.1103/PhysRevD.99.123029,
  10.1103/PhysRevD.100.089903} {\bibfield  {journal} {\bibinfo  {journal}
  {Phys. Rev.}\ }\textbf {\bibinfo {volume} {D99}},\ \bibinfo {pages} {123029}
  (\bibinfo {year} {2019})},\ \bibinfo {note} {[Erratum: Phys.
  Rev.D100,no.8,089903(2019)]},\ \Eprint {http://arxiv.org/abs/1902.07527}
  {arXiv:1902.07527 [gr-qc]} \BibitemShut {NoStop}%
\bibitem [{\citenamefont {Abbott}\ \emph
  {et~al.}(2016{\natexlab{b}})\citenamefont {Abbott} \emph
  {et~al.}}]{LIGOScientific:2016aoc}%
  \BibitemOpen
  \bibfield  {author} {\bibinfo {author} {\bibfnamefont {B.~P.}\ \bibnamefont
  {Abbott}} \emph {et~al.} (\bibinfo {collaboration} {LIGO Scientific,
  Virgo}),\ }\bibfield  {title} {\enquote {\bibinfo {title} {{Observation of
  Gravitational Waves from a Binary Black Hole Merger}},}\ }\href {\doibase
  10.1103/PhysRevLett.116.061102} {\bibfield  {journal} {\bibinfo  {journal}
  {Phys. Rev. Lett.}\ }\textbf {\bibinfo {volume} {116}},\ \bibinfo {pages}
  {061102} (\bibinfo {year} {2016}{\natexlab{b}})},\ \Eprint
  {http://arxiv.org/abs/1602.03837} {arXiv:1602.03837 [gr-qc]} \BibitemShut
  {NoStop}%
\bibitem [{\citenamefont {Ma}\ \emph {et~al.}(2022{\natexlab{a}})\citenamefont
  {Ma}, \citenamefont {Mitman}, \citenamefont {Sun}, \citenamefont {Deppe},
  \citenamefont {H\'ebert}, \citenamefont {Kidder}, \citenamefont {Moxon},
  \citenamefont {Throwe}, \citenamefont {Vu},\ and\ \citenamefont
  {Chen}}]{Ma:2022wpv}%
  \BibitemOpen
  \bibfield  {author} {\bibinfo {author} {\bibfnamefont {Sizheng}\ \bibnamefont
  {Ma}}, \bibinfo {author} {\bibfnamefont {Keefe}\ \bibnamefont {Mitman}},
  \bibinfo {author} {\bibfnamefont {Ling}\ \bibnamefont {Sun}}, \bibinfo
  {author} {\bibfnamefont {Nils}\ \bibnamefont {Deppe}}, \bibinfo {author}
  {\bibfnamefont {Fran\c{c}ois}\ \bibnamefont {H\'ebert}}, \bibinfo {author}
  {\bibfnamefont {Lawrence~E.}\ \bibnamefont {Kidder}}, \bibinfo {author}
  {\bibfnamefont {Jordan}\ \bibnamefont {Moxon}}, \bibinfo {author}
  {\bibfnamefont {William}\ \bibnamefont {Throwe}}, \bibinfo {author}
  {\bibfnamefont {Nils~L.}\ \bibnamefont {Vu}}, \ and\ \bibinfo {author}
  {\bibfnamefont {Yanbei}\ \bibnamefont {Chen}},\ }\bibfield  {title} {\enquote
  {\bibinfo {title} {{Quasinormal-mode filters: A new approach to analyze the
  gravitational-wave ringdown of binary black-hole mergers}},}\ }\href
  {\doibase 10.1103/PhysRevD.106.084036} {\bibfield  {journal} {\bibinfo
  {journal} {Phys. Rev. D}\ }\textbf {\bibinfo {volume} {106}},\ \bibinfo
  {pages} {084036} (\bibinfo {year} {2022}{\natexlab{a}})},\ \Eprint
  {http://arxiv.org/abs/2207.10870} {arXiv:2207.10870 [gr-qc]} \BibitemShut
  {NoStop}%
\bibitem [{\citenamefont {Nichols}\ and\ \citenamefont
  {Chen}(2010)}]{Nichols:2010qi}%
  \BibitemOpen
  \bibfield  {author} {\bibinfo {author} {\bibfnamefont {David~A.}\
  \bibnamefont {Nichols}}\ and\ \bibinfo {author} {\bibfnamefont {Yanbei}\
  \bibnamefont {Chen}},\ }\bibfield  {title} {\enquote {\bibinfo {title} {{A
  hybrid method for understanding black-hole mergers: head-on case}},}\ }\href
  {\doibase 10.1103/PhysRevD.82.104020} {\bibfield  {journal} {\bibinfo
  {journal} {Phys. Rev. D}\ }\textbf {\bibinfo {volume} {82}},\ \bibinfo
  {pages} {104020} (\bibinfo {year} {2010})},\ \Eprint
  {http://arxiv.org/abs/1007.2024} {arXiv:1007.2024 [gr-qc]} \BibitemShut
  {NoStop}%
\bibitem [{\citenamefont {Nichols}\ and\ \citenamefont
  {Chen}(2012)}]{Nichols:2011ih}%
  \BibitemOpen
  \bibfield  {author} {\bibinfo {author} {\bibfnamefont {David~A.}\
  \bibnamefont {Nichols}}\ and\ \bibinfo {author} {\bibfnamefont {Yanbei}\
  \bibnamefont {Chen}},\ }\bibfield  {title} {\enquote {\bibinfo {title}
  {{Hybrid method for understanding black-hole mergers: Inspiralling case}},}\
  }\href {\doibase 10.1103/PhysRevD.85.044035} {\bibfield  {journal} {\bibinfo
  {journal} {Phys. Rev. D}\ }\textbf {\bibinfo {volume} {85}},\ \bibinfo
  {pages} {044035} (\bibinfo {year} {2012})},\ \Eprint
  {http://arxiv.org/abs/1109.0081} {arXiv:1109.0081 [gr-qc]} \BibitemShut
  {NoStop}%
\bibitem [{\citenamefont {Ma}\ \emph {et~al.}(2022{\natexlab{b}})\citenamefont
  {Ma}, \citenamefont {Wang}, \citenamefont {Deppe}, \citenamefont {H\'ebert},
  \citenamefont {Kidder}, \citenamefont {Moxon}, \citenamefont {Throwe},
  \citenamefont {Vu}, \citenamefont {Scheel},\ and\ \citenamefont
  {Chen}}]{Ma:2022xmp}%
  \BibitemOpen
  \bibfield  {author} {\bibinfo {author} {\bibfnamefont {Sizheng}\ \bibnamefont
  {Ma}}, \bibinfo {author} {\bibfnamefont {Qingwen}\ \bibnamefont {Wang}},
  \bibinfo {author} {\bibfnamefont {Nils}\ \bibnamefont {Deppe}}, \bibinfo
  {author} {\bibfnamefont {Fran\c{c}ois}\ \bibnamefont {H\'ebert}}, \bibinfo
  {author} {\bibfnamefont {Lawrence~E.}\ \bibnamefont {Kidder}}, \bibinfo
  {author} {\bibfnamefont {Jordan}\ \bibnamefont {Moxon}}, \bibinfo {author}
  {\bibfnamefont {William}\ \bibnamefont {Throwe}}, \bibinfo {author}
  {\bibfnamefont {Nils~L.}\ \bibnamefont {Vu}}, \bibinfo {author}
  {\bibfnamefont {Mark~A.}\ \bibnamefont {Scheel}}, \ and\ \bibinfo {author}
  {\bibfnamefont {Yanbei}\ \bibnamefont {Chen}},\ }\bibfield  {title} {\enquote
  {\bibinfo {title} {{Gravitational-wave echoes from numerical-relativity
  waveforms via spacetime construction near merging compact objects}},}\ }\href
  {\doibase 10.1103/PhysRevD.105.104007} {\bibfield  {journal} {\bibinfo
  {journal} {Phys. Rev. D}\ }\textbf {\bibinfo {volume} {105}},\ \bibinfo
  {pages} {104007} (\bibinfo {year} {2022}{\natexlab{b}})},\ \Eprint
  {http://arxiv.org/abs/2203.03174} {arXiv:2203.03174 [gr-qc]} \BibitemShut
  {NoStop}%
\bibitem [{\citenamefont {Ma}\ \emph {et~al.}(2023)\citenamefont {Ma},
  \citenamefont {Sun},\ and\ \citenamefont {Chen}}]{Ma_prl}%
  \BibitemOpen
  \bibfield  {author} {\bibinfo {author} {\bibfnamefont {Sizheng}\ \bibnamefont
  {Ma}}, \bibinfo {author} {\bibfnamefont {Ling}\ \bibnamefont {Sun}}, \ and\
  \bibinfo {author} {\bibfnamefont {Yanbei}\ \bibnamefont {Chen}},\ }\bibfield
  {title} {\enquote {\bibinfo {title} {{Black hole spectroscopy by mode
  cleaning}},}\ }\href@noop {} {\  (\bibinfo {year} {2023})},\ \Eprint
  {http://arxiv.org/abs/2301.06705} {arXiv:2301.06705 [gr-qc]} \BibitemShut
  {NoStop}%
\bibitem [{\citenamefont {Stein}(2019)}]{Stein:2019mop}%
  \BibitemOpen
  \bibfield  {author} {\bibinfo {author} {\bibfnamefont {Leo~C.}\ \bibnamefont
  {Stein}},\ }\bibfield  {title} {\enquote {\bibinfo {title} {{qnm: A Python
  package for calculating Kerr quasinormal modes, separation constants, and
  spherical-spheroidal mixing coefficients}},}\ }\href {\doibase
  10.21105/joss.01683} {\bibfield  {journal} {\bibinfo  {journal} {J. Open
  Source Softw.}\ }\textbf {\bibinfo {volume} {4}},\ \bibinfo {pages} {1683}
  (\bibinfo {year} {2019})},\ \Eprint {http://arxiv.org/abs/1908.10377}
  {arXiv:1908.10377 [gr-qc]} \BibitemShut {NoStop}%
\bibitem [{\citenamefont {Ashton}\ \emph {et~al.}(2019)\citenamefont {Ashton}
  \emph {et~al.}}]{Ashton:2018jfp}%
  \BibitemOpen
  \bibfield  {author} {\bibinfo {author} {\bibfnamefont {Gregory}\ \bibnamefont
  {Ashton}} \emph {et~al.},\ }\bibfield  {title} {\enquote {\bibinfo {title}
  {{BILBY: A user-friendly Bayesian inference library for gravitational-wave
  astronomy}},}\ }\href {\doibase 10.3847/1538-4365/ab06fc} {\bibfield
  {journal} {\bibinfo  {journal} {Astrophys. J. Suppl.}\ }\textbf {\bibinfo
  {volume} {241}},\ \bibinfo {pages} {27} (\bibinfo {year} {2019})},\ \Eprint
  {http://arxiv.org/abs/1811.02042} {arXiv:1811.02042 [astro-ph.IM]}
  \BibitemShut {NoStop}%
\bibitem [{\citenamefont {Romero-Shaw}\ \emph {et~al.}(2020)\citenamefont
  {Romero-Shaw} \emph {et~al.}}]{Romero-Shaw:2020owr}%
  \BibitemOpen
  \bibfield  {author} {\bibinfo {author} {\bibfnamefont {I.~M.}\ \bibnamefont
  {Romero-Shaw}} \emph {et~al.},\ }\bibfield  {title} {\enquote {\bibinfo
  {title} {{Bayesian inference for compact binary coalescences with bilby:
  validation and application to the first LIGO\textendash{}Virgo
  gravitational-wave transient catalogue}},}\ }\href {\doibase
  10.1093/mnras/staa2850} {\bibfield  {journal} {\bibinfo  {journal} {Mon. Not.
  Roy. Astron. Soc.}\ }\textbf {\bibinfo {volume} {499}},\ \bibinfo {pages}
  {3295--3319} (\bibinfo {year} {2020})},\ \Eprint
  {http://arxiv.org/abs/2006.00714} {arXiv:2006.00714 [astro-ph.IM]}
  \BibitemShut {NoStop}%
\bibitem [{\citenamefont {Isi}\ and\ \citenamefont
  {Farr}(2021{\natexlab{b}})}]{ringdown_isi}%
  \BibitemOpen
  \bibfield  {author} {\bibinfo {author} {\bibfnamefont {Maximiliano}\
  \bibnamefont {Isi}}\ and\ \bibinfo {author} {\bibfnamefont
  {Will~Meierjurgen}\ \bibnamefont {Farr}},\ }\href {\doibase
  10.5281/zenodo.5094068} {\enquote {\bibinfo {title} {maxisi/ringdown: Initial
  ringdown release},}\ } (\bibinfo {year} {2021}{\natexlab{b}})\BibitemShut
  {NoStop}%
\bibitem [{\citenamefont {Welch}(1967)}]{1161901}%
  \BibitemOpen
  \bibfield  {author} {\bibinfo {author} {\bibfnamefont {P.}~\bibnamefont
  {Welch}},\ }\bibfield  {title} {\enquote {\bibinfo {title} {The use of fast
  fourier transform for the estimation of power spectra: A method based on time
  averaging over short, modified periodograms},}\ }\href {\doibase
  10.1109/TAU.1967.1161901} {\bibfield  {journal} {\bibinfo  {journal} {IEEE
  Transactions on Audio and Electroacoustics}\ }\textbf {\bibinfo {volume}
  {15}},\ \bibinfo {pages} {70--73} (\bibinfo {year} {1967})}\BibitemShut
  {NoStop}%
\bibitem [{\citenamefont {Aasi}\ \emph {et~al.}(2015)\citenamefont {Aasi} \emph
  {et~al.}}]{TheLIGOScientific:2014jea}%
  \BibitemOpen
  \bibfield  {author} {\bibinfo {author} {\bibfnamefont {J.}~\bibnamefont
  {Aasi}} \emph {et~al.} (\bibinfo {collaboration} {LIGO Scientific}),\
  }\bibfield  {title} {\enquote {\bibinfo {title} {{Advanced LIGO}},}\ }\href
  {\doibase 10.1088/0264-9381/32/7/074001} {\bibfield  {journal} {\bibinfo
  {journal} {Class. Quant. Grav.}\ }\textbf {\bibinfo {volume} {32}},\ \bibinfo
  {pages} {074001} (\bibinfo {year} {2015})},\ \Eprint
  {http://arxiv.org/abs/1411.4547} {arXiv:1411.4547 [gr-qc]} \BibitemShut
  {NoStop}%
\bibitem [{\citenamefont {Abbott}\ \emph
  {et~al.}(2016{\natexlab{c}})\citenamefont {Abbott} \emph
  {et~al.}}]{Abbott:2016xvh}%
  \BibitemOpen
  \bibfield  {author} {\bibinfo {author} {\bibfnamefont {Benjamin~P.}\
  \bibnamefont {Abbott}} \emph {et~al.},\ }\bibfield  {title} {\enquote
  {\bibinfo {title} {{Sensitivity of the Advanced LIGO detectors at the
  beginning of gravitational wave astronomy}},}\ }\href {\doibase
  10.1103/PhysRevD.93.112004} {\bibfield  {journal} {\bibinfo  {journal} {Phys.
  Rev. D}\ }\textbf {\bibinfo {volume} {93}},\ \bibinfo {pages} {112004}
  (\bibinfo {year} {2016}{\natexlab{c}})},\ \bibinfo {note} {[Addendum:
  Phys.Rev.D 97, 059901 (2018)]},\ \Eprint {http://arxiv.org/abs/1604.00439}
  {arXiv:1604.00439 [astro-ph.IM]} \BibitemShut {NoStop}%
\bibitem [{\citenamefont {Abbott}\ \emph
  {et~al.}(2016{\natexlab{d}})\citenamefont {Abbott} \emph
  {et~al.}}]{TheLIGOScientific:2016agk}%
  \BibitemOpen
  \bibfield  {author} {\bibinfo {author} {\bibfnamefont {B.~P.}\ \bibnamefont
  {Abbott}} \emph {et~al.} (\bibinfo {collaboration} {LIGO Scientific,
  Virgo}),\ }\bibfield  {title} {\enquote {\bibinfo {title} {{GW150914: The
  Advanced LIGO Detectors in the Era of First Discoveries}},}\ }\href {\doibase
  10.1103/PhysRevLett.116.131103} {\bibfield  {journal} {\bibinfo  {journal}
  {Phys. Rev. Lett.}\ }\textbf {\bibinfo {volume} {116}},\ \bibinfo {pages}
  {131103} (\bibinfo {year} {2016}{\natexlab{d}})},\ \Eprint
  {http://arxiv.org/abs/1602.03838} {arXiv:1602.03838 [gr-qc]} \BibitemShut
  {NoStop}%
\bibitem [{\citenamefont {Harry}(2010)}]{Harry:2010zz}%
  \BibitemOpen
  \bibfield  {author} {\bibinfo {author} {\bibfnamefont {Gregory~M.}\
  \bibnamefont {Harry}} (\bibinfo {collaboration} {LIGO Scientific}),\
  }\bibfield  {title} {\enquote {\bibinfo {title} {{Advanced LIGO: The next
  generation of gravitational wave detectors}},}\ }\href {\doibase
  10.1088/0264-9381/27/8/084006} {\bibfield  {journal} {\bibinfo  {journal}
  {Class. Quant. Grav.}\ }\textbf {\bibinfo {volume} {27}},\ \bibinfo {pages}
  {084006} (\bibinfo {year} {2010})}\BibitemShut {NoStop}%
\bibitem [{\citenamefont {Abbott}\ \emph
  {et~al.}(2016{\natexlab{e}})\citenamefont {Abbott} \emph
  {et~al.}}]{Abbott:2016blz}%
  \BibitemOpen
  \bibfield  {author} {\bibinfo {author} {\bibfnamefont {B.~P.}\ \bibnamefont
  {Abbott}} \emph {et~al.} (\bibinfo {collaboration} {LIGO Scientific,
  Virgo}),\ }\bibfield  {title} {\enquote {\bibinfo {title} {{Observation of
  Gravitational Waves from a Binary Black Hole Merger}},}\ }\href {\doibase
  10.1103/PhysRevLett.116.061102} {\bibfield  {journal} {\bibinfo  {journal}
  {Phys. Rev. Lett.}\ }\textbf {\bibinfo {volume} {116}},\ \bibinfo {pages}
  {061102} (\bibinfo {year} {2016}{\natexlab{e}})},\ \Eprint
  {http://arxiv.org/abs/1602.03837} {arXiv:1602.03837 [gr-qc]} \BibitemShut
  {NoStop}%
\bibitem [{\citenamefont {Abbott}\ \emph {et~al.}(2019)\citenamefont {Abbott}
  \emph {et~al.}}]{LIGOScientific:2018mvr}%
  \BibitemOpen
  \bibfield  {author} {\bibinfo {author} {\bibfnamefont {B.~P.}\ \bibnamefont
  {Abbott}} \emph {et~al.} (\bibinfo {collaboration} {LIGO Scientific,
  Virgo}),\ }\bibfield  {title} {\enquote {\bibinfo {title} {{GWTC-1: A
  Gravitational-Wave Transient Catalog of Compact Binary Mergers Observed by
  LIGO and Virgo during the First and Second Observing Runs}},}\ }\href
  {\doibase 10.1103/PhysRevX.9.031040} {\bibfield  {journal} {\bibinfo
  {journal} {Phys. Rev.}\ }\textbf {\bibinfo {volume} {X9}},\ \bibinfo {pages}
  {031040} (\bibinfo {year} {2019})},\ \Eprint
  {http://arxiv.org/abs/1811.12907} {arXiv:1811.12907 [astro-ph.HE]}
  \BibitemShut {NoStop}%
\bibitem [{\citenamefont {Collaboration}\ and\ \citenamefont
  {Collaboration}()}]{GW_open_science_center}%
  \BibitemOpen
  \bibfield  {author} {\bibinfo {author} {\bibfnamefont {LIGO~Scientific}\
  \bibnamefont {Collaboration}}\ and\ \bibinfo {author} {\bibfnamefont {Virgo}\
  \bibnamefont {Collaboration}},\ }\bibfield  {title} {\enquote {\bibinfo
  {title} {{Gravitational Wave Open Science Center}},}\ }\href@noop {} {\
  }\bibinfo {note} {\url{https://www.gw-openscience.org}}\BibitemShut {NoStop}%
\bibitem [{\citenamefont {Boyle}\ \emph {et~al.}(2019)\citenamefont {Boyle}
  \emph {et~al.}}]{Boyle:2019kee}%
  \BibitemOpen
  \bibfield  {author} {\bibinfo {author} {\bibfnamefont {Michael}\ \bibnamefont
  {Boyle}} \emph {et~al.},\ }\bibfield  {title} {\enquote {\bibinfo {title}
  {{The SXS Collaboration catalog of binary black hole simulations}},}\ }\href
  {\doibase 10.1088/1361-6382/ab34e2} {\bibfield  {journal} {\bibinfo
  {journal} {Class. Quant. Grav.}\ }\textbf {\bibinfo {volume} {36}},\ \bibinfo
  {pages} {195006} (\bibinfo {year} {2019})},\ \Eprint
  {http://arxiv.org/abs/1904.04831} {arXiv:1904.04831 [gr-qc]} \BibitemShut
  {NoStop}%
\bibitem [{\citenamefont {{SXS Collaboration}}()}]{SXSCatalog}%
  \BibitemOpen
  \bibfield  {author} {\bibinfo {author} {\bibnamefont {{SXS Collaboration}}},\
  }\href@noop {} {\enquote {\bibinfo {title} {The {SXS} collaboration catalog
  of gravitational waveforms},}\ }\bibinfo {note}
  {\url{http://www.black-holes.org/waveforms}}\BibitemShut {NoStop}%
\end{thebibliography}%
\end{document}